\documentclass[a4paper,openany]{book}
\usepackage[final,dvips]{rotating,epsfig}
\usepackage{amssymb,amsbsy}
\newcommand*{\diff}[2]{\frac{\partial #1}{\partial #2}}

\newcommand*{\bbbone}{{\mathchoice \mathrm{1\mskip-4mu l} 
    \mathrm{1\mskip-4mu l} \mathrm{1\mskip-4.5mu l} \mathrm{1\mskip-5mu l}}}
\newcommand*{\vvec}{\mathbf{v}}
\newcommand*{\Fextvec}{\mathbf{F}_\mathrm{ext}}
\newcommand*{\qcvec}{\mathbf{q}_\mathrm{c}}
\newcommand*{\kappac}{\kappa_\mathrm{c}}
\newcommand*{\Mwind}{M_\mathrm{wind}}
\newcommand*{\Mdot}{\dot{M}}
\newcommand*{\mproton}{m_\mathrm{p}}
\newcommand*{\Fext}{F_\mathrm{ext}}
\newcommand*{\vsvs}{v^2_\mathrm{s}}
\newcommand*{\dvsvsdr}{v^{2\prime}_\mathrm{s}}
\newcommand*{\vs}{v_\mathrm{s}}
\newcommand*{\rc}{r_\mathrm{c}}
\newcommand*{\tauL}{\tau_\mathrm{L}}
\newcommand*{\nuL}{\nu_\mathrm{L}}
\newcommand*{\FL}{F_\mathrm{L}}
\newcommand*{\kappaL}{\kappa_\mathrm{L}}
\newcommand*{\kbolz}{k_\mathrm{B}}
\newcommand*{\davec}{\mathbf{da}}
\newcommand*{\nvec}{\mathbf{n}}
\newcommand*{\sigmaTh}{\sigma_\mathrm{Th}}
\newcommand*{\kappaTh}{\kappa_\mathrm{Th}}
\newcommand*{\rce}{r_\mathrm{ce}}
\newcommand*{\vth}{v_\mathrm{th}}
\newcommand*{\me}{m_\mathrm{e}}
\newcommand*{\LE}{L_\mathrm{E}}
\newcommand*{\gL}{g_\mathrm{L}}
\newcommand*{\gcak}{g_\mathrm{cak}}
\newcommand*{\grad}{g_\mathrm{rad}}
\newcommand*{\gradvec}{\mathbf{g}_\mathrm{rad}}
\newcommand*{\gabs}{g_\mathrm{abs}}
\newcommand*{\gTh}{g_\mathrm{Th}}
\newcommand*{\gthin}{g_\mathrm{thin}}
\newcommand*{\gthick}{g_\mathrm{thick}}
\newcommand*{\kcak}{k_\mathrm{cak}}
\newcommand*{\alphacak}{{\alpha_\mathrm{cak}}}
\newcommand*{\vrc}{v_{r\mathrm{c}}}
\newcommand*{\dvrcdr}{v_{r\mathrm{c}}^\prime}
\newcommand*{\vinf}{v_\infty}
\newcommand*{\taues}{\tau_\mathrm{es}}
\newcommand*{\vrot}{v_\mathrm{rot}}
\newcommand*{\vroto}{v_{\mathrm{rot}0}}
\newcommand*{\vrotcrit}{v_\mathrm{rot,crit}}
\newcommand*{\Omegacrit}{\Omega_\mathrm{crit}}

\newcommand*{\Ue}{U_\epsilon}
\newcommand*{\UM}{U_M}

\newcommand*{\uAc}{u_\mathrm{Ac}}
\newcommand*{\aAc}{a_\mathrm{Ac}}
\newcommand*{\yAc}{y_\mathrm{Ac}}
\newcommand*{\ys}{y_\mathrm{s}}
\newcommand*{\ysys}{y_\mathrm{s}^2}
\newcommand*{\yroto}{y_\mathrm{rot0}}
\newcommand*{\MAro}{M_{\mathrm{A}r0}}
\newcommand*{\MAr}{M_{\mathrm{A}r}}
\newcommand*{\frad}{f_\mathrm{rad}}

\newcommand*{\vAr}{v_{\mathrm{A}r}}
\newcommand*{\vAc}{v_\mathrm{Ac}}
\newcommand*{\vA}{v_\mathrm{A}}
\newcommand*{\Bro}{B_{r0}}

\newcommand*{\vro}{v_{r0}}
\newcommand*{\dr}[1]{\frac{\partial #1}{\partial r}}
\newcommand*{\du}[1]{\frac{\partial #1}{\partial u}}
\newcommand*{\dtheta}[1]{\frac{\partial #1}{\partial \theta}}
\newcommand*{\vphi}{v_\phi}
\newcommand*{\Bphi}{B_\phi}
\newcommand*{\Bphio}{B_{\phi 0}}
\newcommand*{\Bvec}{\mathbf{B}}
\newcommand*{\const}{\mathrm{const}}
\newcommand*{\rAc}{r_\mathrm{Ac}}
\newcommand*{\rs}{r_\mathrm{s}}
\newcommand*{\rf}{r_\mathrm{f}}
\newcommand*{\dyAcdu}{\dot{y}_\mathrm{Ac}}

\newcommand*{\dxidu}{\dot{\xi}}
\newcommand*{\alpharot}{\alpha_\mathrm{rot}}
\newcommand*{\alpharoteq}{\alpha_\mathrm{rot,eq}}
\newcommand*{\phiD}{\phi_\mathrm{D}}
\newcommand*{\vAbarvec}{\bar{\mathbf{v}}_\mathrm{A}}
\newcommand*{\vAbar}{\bar{v}_\mathrm{A}}
\newcommand*{\kvec}{\mathbf{k}}
\newcommand*{\xvec}{\mathbf{x}}
\newcommand*{\vbarvec}{\bar{\vvec}}
\newcommand*{\dt}[1]{\frac{\partial #1}{\partial t}}
\newcommand*{\deltarho}{\delta\rho}
\newcommand*{\deltavvec}{\delta\vvec}
\newcommand*{\deltaBvec}{\delta\Bvec}
\newcommand*{\Bbarvec}{\bar{\Bvec}}
\newcommand*{\rhobar}{\bar{\rho}}
\newcommand*{\Lsobo}{L_\mathrm{S}}
\newcommand*{\vph}{v_\mathrm{Ph}}
\newcommand*{\complexi}{\mathrm{i}}
\newcommand*{\vrbar}{\bar{v}_r}
\newcommand*{\vinfobs}{v_{\infty,\mathrm{obs}}}
\newcommand*{\Mdotobs}{\dot{M}_\mathrm{obs}}

\newcommand*{\OmegaOR}{\Omega_\mathrm{OR}}
\newcommand*{\erfc}{\mathrm{erfc}}
\newcommand*{\chib}{\chi_{\!{\ \atop \scriptstyle b}}}
\newcommand*{\chiOR}{\chi_{\!{\ \atop \scriptstyle \mathrm{OR}}}}
\newcommand*{\vpvec}{\mathbf{v}_\mathrm{p}}
\newcommand*{\dvrdr}{v_r^\prime}

\newcommand*{\vphibar}{\bar{v}_\phi}
\newcommand*{\dvphidv}{\frac{\partial v_\phi}{\partial v}}
\newcommand*{\dvphidu}{\dot{v}_\phi}
\newcommand*{\dvphideta}{\check{v}_\phi}
\newcommand*{\vp}{v_\mathrm{p}}

\newcommand*{\dvdu}{\dot{v}}
\newcommand*{\dVdeta}{\check{V}}
\newcommand*{\Bpvec}{\mathbf{B}_\mathrm{p}}

\newcommand*{\Bp}{B_\mathrm{p}}
\newcommand*{\Bpo}{B_{\mathrm{p}0}}
\newcommand*{\BpAc}{B_\mathrm{pAc}}
\newcommand*{\fradvec}{\mathbf{F}_\mathrm{rad}}
\newcommand*{\ervec}{\mathbf{e}_r}
\newcommand*{\ethetavec}{\mathbf{e}_\theta}
\newcommand*{\ez}{\mathbf{e}_z}
\newcommand*{\ephivec}{\mathbf{e}_\phi}
\newcommand*{\epvec}{\mathbf{e}_\mathrm{p}}
\newcommand*{\etvec}{\mathbf{e}_\mathrm{t}}
\newcommand*{\epr}{e_{\mathrm{p}r}}
\newcommand*{\ept}{e_{\mathrm{p}\theta}}
\newcommand*{\deprdu}{\dot{e}_{\mathrm{p}r}}
\newcommand*{\deptdu}{\dot{e}_{\mathrm{p}\theta}}
\newcommand*{\deprdeta}{\check{e}_{\mathrm{p}r}}
\newcommand*{\deptdeta}{\check{e}_{\mathrm{p}\theta}}
\newcommand*{\ddeprduu}{\ddot{e}_{\mathrm{p}r}}

\newcommand*{\dudr}{u^\prime}
\newcommand*{\dudx}{\tilde{u}}

\newcommand*{\dudth}{\breve{u}}
\newcommand*{\ddudthdu}{\dot{\breve{u}}}
\newcommand*{\detadr}{\eta^\prime}
\newcommand*{\detadx}{\tilde{\eta}}
\newcommand*{\detadth}{\breve{\eta}}
\newcommand*{\omegabarvec}{\bar{\boldsymbol{\omega}}}
\newcommand*{\omegabar}{\bar{\omega}}
\newcommand*{\domegabardeta}{\check{\bar{\omega}}}
\newcommand*{\domegadeta}{\check{\omega}}
\newcommand*{\Evec}{\mathbf{E}}
\newcommand*{\vpAc}{v_\mathrm{pAc}}
\newcommand*{\dvsvsdu}{\partial_u\vsvs}
\newcommand*{\rvec}{\mathbf{r}}
\newcommand*{\Mp}{M_\mathrm{p}}
\newcommand*{\xAc}{x_\mathrm{Ac}}
\newcommand*{\thAc}{\theta_\mathrm{Ac}}
\newcommand*{\drdu}{\dot{r}}

\newcommand*{\dthdu}{\dot{\theta}}
\newcommand*{\dthdeta}{\check{\theta}}
\newcommand*{\ddthduu}{\ddot{\theta}}
\newcommand*{\ddthdetau}{\dot{\check{\theta}}}

\newcommand*{\rhoAc}{\rho_\mathrm{Ac}}
\newcommand*{\drhodu}{\dot{\rho}}
\newcommand*{\drhodeta}{\check{\rho}}
\newcommand*{\drhoAcdeta}{\check{\rho}_\mathrm{Ac}}
\newcommand*{\arad}{a_\mathrm{rad}}

\newcommand*{\dxdu}{\dot{x}}
\newcommand*{\dxdeta}{\check{x}}
\newcommand*{\ddxduu}{\ddot{x}}
\newcommand*{\ddxdetau}{\dot{\check{x}}}
\newcommand*{\dydu}{\dot{y}}
\newcommand*{\dyxdx}{\tilde{y}_x}
\newcommand*{\dYdeta}{\check{Y}}
\newcommand*{\yrot}{y_\mathrm{rot}}
\newcommand*{\dyrotdu}{\dot{y}_\mathrm{rot}}
\newcommand*{\dyrotdeta}{\check{y}_\mathrm{rot}}
\newcommand*{\dyrotAcdeta}{\check{y}_\mathrm{rot,Ac}}
\newcommand*{\yrotAc}{y_\mathrm{rot,Ac}}
\newcommand*{\yphi}{y_\phi}
\newcommand*{\yphibar}{\bar{y}_\phi}
\newcommand*{\dyphibardu}{\dot{\bar{y}}_\phi}
\newcommand*{\dyphidu}{\dot{y}_\phi}
\newcommand*{\ddyphidetay}{\hat{\check{y}}_\phi}
\newcommand*{\dyphideta}{\check{y}_\phi}
\newcommand*{\dyphidy}{\hat{y}_\phi}
\newcommand*{\ddyphiduu}{\ddot{y}_\phi}
\newcommand*{\ddyphidyy}{\hat{\hat{y}}_\phi}
\newcommand*{\ddyphiduy}{\hat{\dot{y}}_\phi}
\newcommand*{\ddyphidetadu}{\dot{\check{y}}_\phi}
\newcommand*{\dysysdu}{\partial_u\ysys}
\newcommand*{\dysysdeta}{\partial_\eta\ysys}
\newcommand*{\ddysysdetau}{\partial_u\partial_\eta\ysys}
\newcommand*{\ddysysduu}{\partial_u^2\ysys}
\newcommand*{\AbsGradeta}{|\nabla\eta|}

\newcommand*{\dAbsGradetadu}{\partial_u|\nabla\eta|}
\newcommand*{\ddAbsGradetaduu}{\partial_u^2|\nabla\eta|}
\newcommand*{\ddAbsGradetadetau}{\partial_\eta\partial_u|\nabla\eta|}
\newcommand*{\dAbsGradetadeta}{\partial_\eta|\nabla\eta|}
\newcommand*{\dchidu}{\dot\chi}
\newcommand*{\dvarrhodu}{\dot\varrho}
\newcommand*{\dvarrhody}{\hat\varrho}
\newcommand*{\dmudu}{\dot\mu}
\newcommand*{\dmudeta}{\check\mu}
\newcommand*{\dFdeta}{\check{F}}
\newcommand*{\dadu}{\dot{a}}
\newcommand*{\dadr}{a^\prime}
\newcommand*{\dAdu}{\dot{A}}
\newcommand*{\dAzweidu}{\dot{A}_2}
\newcommand*{\dAdy}{\hat{A}}
\newcommand*{\dBdu}{\dot{B}}
\newcommand*{\dBdy}{\hat{B}}
\newcommand*{\dCdu}{\dot{C}}
\newcommand*{\dCdy}{\hat{C}}
\newcommand*{\dBdreidu}{\dot{B}_3}
\newcommand*{\dBdreidy}{\hat{B}_3}
\newcommand*{\ddudxdu}{\dot{\tilde{u}}}
\newcommand*{\ddeprdetadu}{\dot{\check{e}}_{\mathrm{p}r}}
\newcommand*{\ddetadxdu}{\dot{\tilde{\eta}}}
\newcommand*{\ddetadthdu}{\dot{\breve{\eta}}}
\newcommand*{\dvthdu}{\dot{v}_\mathrm{th}}
\newcommand*{\dat}{\Delta a_\mathrm{t}}
\newcommand*{\dbt}{\Delta b_\mathrm{t}}
\newcommand*{\dbp}{\Delta b_\mathrm{p}}
\newcommand*{\dgt}{\Delta g_\mathrm{t}}
\newcommand*{\AU}{\mathsf{AU}}
\newcommand*{\rsun}{R_\odot}
\newcommand*{\km}{\mathsf{km}}
\newcommand*{\cm}{\mathsf{cm}}
\newcommand*{\s}{\mathsf{s}}
\newcommand*{\yr}{\mathsf{yr}}
\newcommand*{\gram}{\mathsf{g}}
\newcommand*{\Msun}{M_\odot}
\newcommand*{\Lsun}{L_\odot}
\newcommand*{\Kelvin}{\mathsf{K}}
\newcommand*{\Gauss}{\mathsf{G}}
\begin{document}
%
% Titlepage:
%
\begin{titlepage}
\vspace*{2cm}
\hspace*{1.7cm}\begin{minipage}{14cm}
\begin{center}
\Huge
\textbf{New aspects in the theory of\\ 
magnetic winds\\
from massive hot stars}\\[4cm]
\normalsize
Inaugural-Dissertation\\
zur\\
Erlangung der Doktorw\"urde\\
der\\
Hohen Mathematisch-Naturwissenschaftlichen Fakult\"at\\
der\\
Rheinischen Friedrich-Wilhelms Universit\"at\\
zu\\
Bonn\\[4cm]
vorgelegt von\\[0.5ex]
\textbf{Henning Seemann}\\[0.5ex]
aus\\[0.5ex]
L\"uneburg\\[3cm]
Bonn, im M\"arz 1998
\end{center}
\end{minipage}
\vfill
\newpage
\thispagestyle{empty}
\vspace*{4cm}
\hspace*{4mm}\begin{minipage}{14cm}
\begin{center}
Angefertigt mit Genehmigung\\
der Mathematisch-Naturwissenschaftlichen Fakult\"at\\
der Universit\"at Bonn\\[8cm]
\begin{tabular}{rl}
1.~Referent: & Professor Dr.~P. L. Biermann\\
2.~Referent: & Professor Dr.~H. J. Fahr
\end{tabular}\\
\vspace{4cm}
Tag der letzten m\"undlichen Pr\"ufung: 31.3.1998\\
\end{center}
\end{minipage}
\vfill
\newpage
\vspace*{5cm}
\begin{flushright}
\thispagestyle{empty}
F\"ur alle, die mich bis hier hin gebracht haben!
\end{flushright}
\vfill
\end{titlepage}
\thispagestyle{empty}
\frontmatter
\tableofcontents
%\listoffigures
%\listoftables
%
\mainmatter
%
%-*-LaTeX-*-
% This is the 1st chapter for the PhD-thesis of Henning Seemann.
% (c) 1997-98 by Henning Seemann
%
\chapter{Introduction}
\label{Chap:introduction}
Most stars lose mass. But the importance of the stellar mass loss for the fate
of the star and the rest of the universe differs quite from star to star. Our
sun have a negligible mass los rate while the mass loss from O, B, and
Wolf-Rayet stars is very important.

O and B stars are known for a long time. While the first Wolf-Rayet stars were
discovered by Wolf \& Rayet not before 1867 \cite{Wolf:Rayet:67}. The reason
for that is that they are so few. Only about 160 Wolf-Rayet stars are known in
our galaxy today. Some more we know in the Magellanic clouds and in M33. The
most prominent spectroscopical feature of Wolf-Rayet stars are their strong
emission lines, while the spectra of most normal stars like our sun are
dominated by absorption lines.  Similar spectra can be seen in other, less
luminous objects like planetary nebula.  The common reason for the strong
emission lines is a strong, optically thick outflow of matter. In this thesis
we consider only classical Wolf-Rayet stars from the population~I when we use
the terminus ``Wolf-Rayet star''. Although planetary nebulae have a strong mass
outflow as well, the underlying physics and the stellar evolution are
different.  Wolf-Rayet stars play an especially interesting role in
astrophysics. They have the strongest stellar winds. And their winds are
strongly enhanced with heavy elements. Their winds contain nearly no hydrogen
anymore. The Wolf-Rayet stars of the WN subclass show the chemical composition
of the products of the CNO cycle in their winds. While the stars of the WC
subclass already show the products of helium burning in their winds. In both
cases we have the bare nuclei of more massive progenitor stars.  But massive
stars do not only enrich the interstellar medium with heavy
elements. Additionally the stellar winds carry momentum and kinetic
energy. These are very important to trigger e.g.\ star formation. In this
thesis we will only consider blue stars, like O, B, and Wolf-Rayet stars when
we talk about ``massive stars.'' There are massive stars in the red part of the
Hertzsprung-Russel diagram as well. But their winds contain dust and are driven
in a completely different way. Therefore they are excluded in this thesis.

Significant mass loss as a common phenomenon in massive stars is known only for
a short time, because only few massive stars like Wolf-Rayet stars show the
typical P-Cygni line profile which is the spectral indicator for strong mass
loss. A P-Cygni line is the combination of an emission line with an absorption
line connected to the blue wing of the emission line. Both lines come from the
same atomic transition. The absorption is caused in the line of sight by the
wind material moving towards the observer. This causes the blue-shift of the
absorption line. From the blue edge of the absorption line we can therefore
derive the maximum velocity in the outflowing matter. The emission line is
produced in the wind material which leaves the star roughly perpendicular to
the line of sight. Only few, weak P-Cygni lines lie in the optical
range. Therefore only few stars with very strong winds like Wolf-Rayet stars
are known for long time to have significant mass loss. Thus significant mass
loss was considered to be a rare phenomenon in stars. In 1967 Morton
\cite{Morton:67:a,Morton:67:b} used rockets to observe massive stars in the
ultraviolet (UV) range. He found that many massive stars show P-Cygni lines in
the UV-range. And so mass loss became known as a phenomenon common in massive
stars. It became also clear that the absorption of stellar light in atomic
lines plays the dominant role in driving these winds. Lucy \& Solomon
\cite{Lucy:Solomon:70} did pioneering theoretical work in this field. Based on
their results Castor, Abbott \& Klein \cite{Castor:etal:75} developed a
genius model to describe the line driving mechanism in a simple but in most
cases accurate way. We will make extensive use of their model in this thesis.

Today we have much more detailed observations of massive stars in many
frequency ranges. This gives us a modern and more detailed picture of
their winds. The wind material can today be detected in the radio
range, as well. This is more than just a confirmation of the results from
P-Cygni lines because normal wind material produces only thermal radio
emission. But many massive stars show nonthermal radio emission
\cite{Abbott:etal:86, Bieging:etal:89}. This is synchrotron emission, which is
a clear evidence for magnetic fields in these winds. Additionally an
acceleration mechanism is required to produce high energy particles for the
synchrotron emission. The most obvious mechanism is Fermi acceleration in
shocks. Shocks appear where the wind collides with the interstellar medium or
with the wind of a neighbouring star (e.g.\ in binaries). Additionally shocks
can be produced by the instability of the wind driving by radiation. Further
evidence for the existence of shocks in these winds come from the observation
of X-rays in these winds \cite{Lucy:82}. Detailed analysis of the
stellar spectra show strong evidence for wind variability as well. Thus we
can not describe the wind of massive stars as stationary, smooth outflows any
more. Modern wind models should describe in more detail how these perturbations
of the smooth wind are created, how they are amplified, and which consequences
they have for the overall wind. From these questions we can learn a lot about
these winds.

The first observations of stellar winds were made for our own sun, where due to
the short distance we can study wind phenomena much easier and in far more
detail than for other stars. L.~Biermann was the first who recognized that the
tails of the comets, which always point away from the sun, are a clear
indication for a permanent matter outflow from the sun \cite{Biermann:51,
Biermann:57}. Based on this information Parker developed the first quantitative
model for the solar wind. He showed that the hot corona of the sun
$(\sim10^6\Kelvin)$ must have a permanent matter outflow, because gravity and
the interstellar medium can not compensate the high thermal pressure. During
the years this basic model has been improved by theory and observation in many
ways. Today solar physics is a broad field in astrophysics. Most of these
improvements are too detailed to be used for winds from other stars. But one
important model, first developed for our sun, will be used in this thesis
extensively. It is the model of Weber \& Davis \cite{Weber:Davis:67} for the
solar magnetic field. In the solar wind the magnetic field plays only a
subordinate role. It merely controls the solar spin-down.  The key message of
this thesis is that the magnetic field plays an much more important role in the
wind physics of massive stars. Massive stars rotate faster and we have many
indirect evidence that they have strong magnetic fields
\cite{Biermann:Cassinelli:93}. This can enhance the wind drastically. While it
seems possible to describe the winds of O and B stars with a purely radiative
model \cite{Kudritzki:Hummer:90,Pauldrach:etal:94}, this attempt failed for
Wolf-Rayet stars.

Models which use the magnetic field as additional ingredient for the winds of
massive stars have been developed earlier \cite{Friend:MacGregor:84,
Poe:Friend:86,Poe:etal:89}. But these models suffered from various
problems. The strongest limitation to magnetic wind models is the so called
spin-down problem. The basic idea of the Weber \& Davis models is that open
field lines are fixed in the stellar envelope and extend outwards. These field
lines co-rotate with the star. Through the Lorentz force they help to push the
wind material outside. Through this process the star will lose an extra amount
of angular momentum. If the magnetic field is to strong the star will spin down
very rapidly. In this case the Weber \& Davis mechanism can not play a role in
further stages of the stars evolution. Therefore is is important to keep the
evolution of a star though its whole life in mind. It is e.g.\ not clear
whether massive stars have a magnetic field through their whole lifetime. It is
possible that young, massive stars on the main sequence have a purely radiation
driven wind, and therefore radiative wind models are so successful for these
stars. Evolved stars like Wolf-Rayet stars, whose wind are not explained yet,
may have a magnetic field, which was either generated in a later stage of the
stellar evolution or was hidden under the outer layers of the stellar envelope
blown away now. These are very speculative ideas. But we are still in an early
stage of understanding massive stars.

Another important aspect of magnetic fields in winds of massive stars is the
production of cosmic rays. We know from the observation of nonthermal radio
emission that we have shocks and a magnetic field in the winds. If the star
rotates the magnetic field will be bent into azimuthal direction far away
from the star. If we combine such a field configuration with shock running in
radial direction we get an optimal configuration for particle acceleration due
to the Fermi mechanism. The high energy particles produced here gyrate in the
magnetic field and produce the observed nonthermal radio emission. But high
energy particles, especially protons and nucleons, can also escape from the
system and contribute to the galactic cosmic ray spectrum. This is the basic
concept of Biermann \& Cassinelli \cite{Biermann:94,Biermann:Cassinelli:93},
who combined several aspects of magnetic fields in the winds of massive stars
into a self-consistent picture. This work gave rise to the initial concept of
this thesis.

In this thesis we analyze several aspects of magnetic fields in stellar
winds. The fundament for these models are the improved version of the fast
magnetic rotator model developed by Biermann \& Cassinelli
\cite{Biermann:94,Biermann:Cassinelli:93}. The first major part of this thesis
is a model for the influence of the magnetic field on the radiative instability
in the wind. Already very early it became clear that line driven winds are
unstable \cite{Lucy:Solomon:70, Milne:26}. But all previous attempts to use
these instabilities to explain the thick and fast winds, we observe in massive
stars, failed. We found that this is due to the neglect of the magnetic
field. Previous work concentrated on the effect of radiation and found that the
instability produces inward running waves \cite{Owocki:Rybicki:84}. These waves
can help to explain many details in the spectra of massive stars. But they do
not help to explain basic wind parameters like the high observed terminal wind
velocities. We found that a magnetic field turns the waves into outward running
waves \cite{Seemann:Biermann:97}, which can explain the high observed terminal
velocities with moderate magnetic fields. The latter is very important to keep
the spin-down problem under control.

In the second major part of this thesis we extend the basic wind model to take
the wind outside the equatorial plane into account. Older wind models separate
into two classes. The first class of models ignores rotation and the magnetic
field. These models treat all quantities as only radius dependent. Thus they
are one dimensional and strictly spherical. Most purely radiation driven wind
models belong to this class. The second class of models treats rotation and
often magnetic fields. They assume only rotational symmetry but restrict
themselves to the equatorial plane. Through this restriction they are
essentially one dimensional as well. Strong assumptions are necessary to
estimate the wind outside the equatorial plane. Our wind model keeps the
simplicity of an one dimensional wind model but allows to explore the wind
outside the equatorial plane. This gives us much more confidence in wind models
with rotation. The second important aspect of our extension is, that this model
can now be used to calculate winds which do not blow in the equatorial plane at
all. For such winds the old equatorial models are meaningless. This is the
large family of winds from accretion disks, which often turn into jets close to
the polar axis. Such winds appear in many astrophysical objects from young
stellar objects (YSO) to active galactic nuclei (AGN). Due to this broad
applicability we expect many interesting results from our wind model in the
future.

In Chap.~\ref{Chap:parker:wind} we start our discussion with some basic
concepts explained at the wind model of Parker. In Chap.~\ref{Chap:cak} we give
a short introduction to the theory of line driven winds. Since this thesis
concentrates on the role of the magnetic field only the basic concepts for our
own usage are given here. In Chap.~\ref{Chap:wd} the wind equations for the
equatorial plane are derived. These equations are already slightly generalized
compared to the equations of Biermann \& Cassinelli by the fact that they
incorporate the effect of a wind compressed or diluted in the equatorial
plane. In Chap.~\ref{Chap:cak:wd} we combine the results from the two previous
chapters for some initial wind models, which allow the comparison with previous
models. In Chap.~\ref{Chap:waves:shocks} we derive our model for waves in
magnetic winds from massive stars and give numerical results for a generic
model star. Chap.~\ref{Chap:fluxtube} contains our model for the wind outside
the equatorial plane. Some initial numerical results are given as well.
In Chap.~\ref{Chap:conclusions} we summarize our results in this thesis.
The appendices contain some extra numerical and mathematical material for the
non-equatorial wind model from Chap.~\ref{Chap:fluxtube}.

\vfill
%

%-*-LaTeX-*-
% This is the 2nd chapter for the PhD-thesis of Henning Seemann.
% (c) 1997-98 by Henning Seemann
%
\chapter{Introduction to stellar wind theory}
\label{Chap:parker:wind}
The first stellar wind models were developed for our sun. L. Biermann
\cite{Biermann:51,Biermann:57} first proposed a steady hydrodynamical outflow
from the sun to explain the direction of the cometary tails. Based on this idea
Parker \cite{Parker:58} showed that the sun indeed can not have a stationary
atmosphere. Due to the high temperature of the corona a stationary atmosphere
would have a high finite pressure at infinity. This pressure could not be
compensated by the interstellar medium. A continuous expansion of the solar
atmosphere is the direct consequence. Parker's solar wind is only driven by
thermal pressure. This leads to a simple model which nevertheless gives
important insight in the physics of stellar winds in general. Therefore the
important parts of Parker's model will be outlined in this chapter.

The solar wind has a velocity of approximately $300-800\,\km/\s$ at the earth's
orbit. Even very early type stars produce only a wind of up to
$3500\,\km/\s$ much smaller than the speed of
light. Therefore nonrelativistic physics is sufficient to describe these
stellar winds and will be used in this thesis.
 
For this introducing discussion to stellar winds it is sufficient to study the
average, large scale structure of the solar wind. This allows us to use the
theory of hydrodynamics, and later magnetohydrodynamics, as basis for our
models. Hydrodynamics can be used whenever the relevant scales of the model are
larger than the mean free path $\lambda$ for collisions between molecules. At
the base of the solar wind we have roughly $\lambda(3\rsun) =
4\cdot10^{-2}\rsun$. At the earth's orbit $\lambda$ has increased to
$\lambda(1\AU) = 1.5\AU$. Here and further out rigorous treatment using the
Fokker-Planck equation would be better. But this is beyond the scope of this
thesis. This problem is relaxed by the fact that the solar wind has a magnetic
field. The magnetic field forces the ionized wind particles to gyrate around
the field lines. This effect increases the probability to hit an other particle
and allows the particles to be scattered at field irregularities, so that the
hydrodynamical equilibrium is reestablished.  Furthermore the wind of early
type stars is much denser than the solar wind. This leads much shorter mean
free paths and justifies the usage of hydrodynamics as well.

\begin{figure}
\begin{center}
\begin{picture}(90,115)
%\put(0,0){\framebox(90,115){}}
\put( 5, 5){\epsfig{file=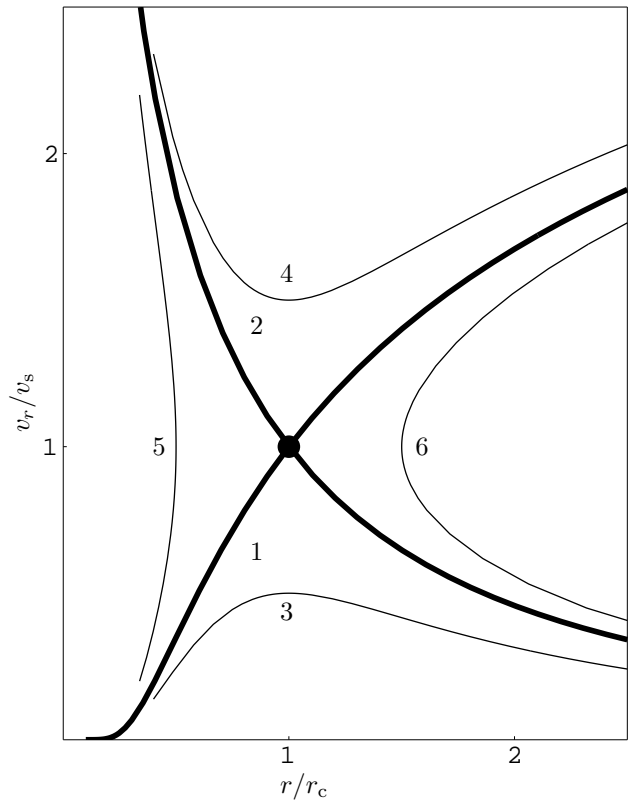}}
\put(37, 2){$r/\rc$}
\put( 1,50){\begin{turn}{90}$v_r/\vs$\end{turn}}
\put(33,33){1}
\put(33,63){2}
\put(37,25){3}
\put(37,70){4}
\put(20,47){5}
\put(55,47){6}
\end{picture}
\end{center}
\caption{\label{ParkerSolutions}The solutions of Parker's isothermal wind 
equation}
\end{figure}

The hydrodynamical wind is described by the conservation of mass
\begin{equation}\label{MassConservationHDA}
\diff{\rho}{t}+\nabla\cdot(\rho\vvec) = 0,
\end{equation}
momentum
\begin{equation}\label{EulerHDA}
\rho\diff{\vvec}{t}+\rho(\vvec\cdot\nabla)\vvec = \Fextvec,
\end{equation}
and energy
\begin{equation}\label{EnergyConservationHDA}
\rho\left(\diff{e}{t}+(\vvec\cdot\nabla)e\right) + P(\nabla\cdot\vvec) =
-\nabla\qcvec,
\end{equation}
where $\rho$ is the matter density, $t$ the time, $\vvec$ the velocity of the
bulk matter, $\Fextvec$ the sum of all external force densities, $e$ the
internal energy per unit mass, $P$ the thermal pressure, $\qcvec=-\kappac\nabla
T$ the conductive energy flux, $\kappac$ the thermal conductivity, and $T$ the
temperature. If we assume that the star does not rotate and that the wind is
stationary, the whole system reduces to one dimension
\begin{equation}\label{MassconservationHDb}
\Mdot = 4\pi\rho v_r r^2 = \mbox{const}
\end{equation}
\begin{equation}
\rho v_rv_r' = \Fext,
\end{equation}
where $'$ denotes the derivative with respect to the radius $r$. $\Mdot$ is the
mass loss rate and $v_r$ is the radial component of the wind velocity $\vvec$.
The external forces are created by radiation pressure from the stellar light,
thermal pressure, interaction of the gas with waves and shocks, and
gravitation. We will discuss radiation pressure as a major component of early
type star wind in Chap.~\ref{Chap:cak}. The creation of waves and shocks by
the instability of radiation pressure in a magnetic wind is the subject of
Chap.~\ref{Chap:waves:shocks}. Other sources for shocks and waves are e.g.\
discussed in \cite{Castor:86}. The influence of waves and shocks on
the average wind through Eq.~\ref{EulerHDA} was analysed by Koninx
\cite{Koninx:PhD}. The gravitational potential derives from Poissons equation
\begin{equation}\label{DefGravitationA}
\nabla^2\Phi = -4\pi G \rho,
\end{equation}
where $G$ is the gravitational constant. The contribution of the wind mass 
\begin{equation}\label{WindMass}
\Mwind=\int\rho\,dV \approx \Mdot \frac{r_\infty}{v_r}
\end{equation}
is negligible compared to the stellar mass, where $r_\infty$ is the radius of
the heliosphere. For the sun we find $\Mwind \approx 10^{-12}\Msun,$ where
$\Msun$ is the solar mass. All normal stars are also reasonably spherically
symmetric. We can therefore approximate the gravitational potential by
\begin{equation}\label{Gravitationb}
\Phi = -\frac{GM}{r},
\end{equation}
where $M$ is the stellar mass. We now express the thermal pressure $P$ by
the isothermal speed of sound $\vs$
\begin{eqnarray}
P     &=& \vsvs\rho\\
\vsvs &=& \frac{\kbolz T}{m},
\end{eqnarray}
where $m$ is the average mass of the wind particles.  If we use gravitation and
thermal pressure as the only relevant forces, as Parker did
\begin{equation}\label{ParkerForces}
\Fext = -\rho\Phi'-P',
\end{equation}
we can write the momentum equation as
\begin{equation}\label{ParkerWind}
v_r v_r' = \frac{f(r)}{g(r,v_r)} = 
         \frac{\frac{2\vsvs}{r}-\dvsvsdr-\frac{GM}{r^2}}
              {1-\frac{\vsvs}{v_r^2}}.
\end{equation}
In the general case this ordinary differential equation has to be
solved simultaneously with an equation for $T(r)$ which can be derived
from Eq.~\ref{EnergyConservationHDA}. This can be found in
\cite{Parker:58,Mihalas:Book}.  For our discussion it is sufficient to
analyze the isothermal approximation.  In this case $g$ in
Eq.~\ref{ParkerWind} depends only on $v_r$. For $v_r>\vs$, $g$ is larger
than zero and otherwise smaller than zero. $f(r)$ changes its sign at the
Parker radius
\begin{equation}\label{ParkerRadius}
\rc = \frac{GM}{2\vsvs}.
\end{equation}
This leads to a classical family of solutions. The members with $v_r>0$ are
sketched in Fig.~\ref{ParkerSolutions}. If a solution reaches $|v_r|=\vs$ at a
radius $r\not=\rc$, $v_r'$ becomes infinite. This leads to double valued
solutions like solutions 5\&6. These are unphysical for a stationary wind. If a
solution passes through the critical point $|v_r|=\vs,\ r=\rc$, l'Hospital's
rule gives 
\begin{equation}
v_r'(\rc) = \pm\frac{2\vs^3}{GM},
\end{equation}
which leads to two solutions. Solution 1 is subsonic at the solar surface and
continously accelerates up to supersonic velocities. This solution was proposed
by Parker for the solar wind. Solution 2 describes a wind which has for some
reason a very high velocity at the solar surface. This does not fit the
observation there. The same is true for solution 4, which is supersonic at all
radii. Solution 3 did indeed fit all observations in 1958. It was proposed by
Chamberlain \cite{Chamberlain:61} as solar \textsl{breeze.} Finally the
question was solved in favour of the transonic \textsl{winds}\/ by observations
from satellites, which found a supersonic plasma at the earth's orbit. The
solutions with negative velocities can be interpreted as accretion scenarios
\cite{Johnson:Axford:71, Mathews:Baker:71}.

Critical points are very useful for the solution of wind equation, because they
contain extra information. Equation~\ref{ParkerWind} is a single ordinary
differential equation of first order. For such an equation it is normally
necessary to specify $v_r$ at one point in order to select a unique solution
from the infinite family of possible solutions. Once we know from physical
arguments that the solution has to go through the critical point, only a small
finite number of solutions remain. This becomes important in the case of
stellar winds, where less parameters are accessible for direct observation than
in the case of our sun.

The isothermal model has the advantage that neglecting the energy balance
simplifies the mathematics. Physically it means that we implicitly introduce
additional energy sources and sinks to maintain the constant temperature. Close
to the sun, where the temperature changes only by one order of magnitude
between the corona and the earth's orbit, this is a reasonable
approximation. When we go further out the wind has to cool down adiabatically
finally. In the isothermal case Eq.~\ref{ParkerWind} can be readily
integrated to
\begin{equation}
\left(\frac{v_r}{\vs}\right)^2-2\ln\left(\frac{v_r}{\vs}\right) = 
4\ln\left(\frac{r}{\rc}\right)+4\frac{\rc}{r}+C.
\end{equation}
For $r\rightarrow\infty$ and $v_r\gg\vs$ this leads to $v_r =
2\vs\ln^{0.5}(r/\rc) \rightarrow \infty$, because we add an infinite amount of
thermal energy to maintain the constant temperature at infinity.

Real stellar winds do not extend to infinity, but they collide with the
interstellar medium. This happens when the ram pressure equals
the pressure of the interstellar medium. Therefore all numerical models for
stellar wind compute the wind only up to a large but finite radius. This radius
should be chosen so that it is larger than any radius relevant in the wind. In
Chap.~\ref{Chap:wd} we will use $u\sim r^{-1}$ as spatial coordinate. In this
case it is mathematically possible to extend the computation to infinity in the
mathematical sense. This can not be done using the isothermal
approximation. But since the wind is terminated in the real world by the
collision with the interstellar medium anyway, we will compute our wind models
only to a large radius using the isothermal approximation.

In this chapter we established the fundamental framework for our stellar wind
models. We will use nonrelativistic hydrodynamics or magnetohydrodynamics (MHD)
in the stationary or quasi stationary form to describe stellar winds. This
thesis emphasizes the role of the magnetic field. Thus we treat the wind
as isothermal and do not go into the details of radiation transfer theory.

%-*-LaTeX-*-
% This is the 3rd chapter for the PhD-thesis of Henning Seemann.
% (c) 1997-98 by Henning Seemann
%
\chapter{Stellar radiation and winds}
\label{Chap:cak}
\section{Introduction}
Photons carry momentum and energy. Both can be exchanged with matter. Therefore
it seems quite obvious that the stellar light can play an important role in
stellar wind physics. Unfortunately, the detailed treatment of this process is
very complicated.

We start our discussion with some basic definitions from radiation theory
following the description of Mihalas \cite{Mihalas:Book} and Falcke
\cite{Falcke:Msc}. The specific radiation intensity
\begin{equation}
I_\nu=\frac{dE_\mathrm{rad}}{d\nu\,dt\,\nvec\cdot\davec\,d\Omega}
\end{equation}
is the differential energy $dE_\mathrm{rad}$ per unit frequency $\nu$ and time
$t$, which is emitted from an area of size $da$ oriented in direction $\davec$
into an solid angle $d\Omega$ in direction $\nvec$. If the radiation with
solid angle angle $d\Omega$ and direction  $\nvec$ passes through
the area $\davec$ along a distance $ds$, and the volume $dV=\davec\,ds$ is
filled with matter of density $\rho$, the energy
\begin{equation}
dE_\mathrm{abs} = I_\nu\rho\kappa_\nu\,d\Omega\,d\nu\,dt\,\nvec\cdot\davec\,ds
\end{equation}
is absorbed. The opacity $\kappa_\nu$ can be equivalently expressed by the
total cross section for absorption
\begin{equation} 
\sigma_\nu = m\kappa_\nu.
\end{equation}
The momentum transfer $\mathbf{dp}=\nvec\,dp$ per unit volume and due to the
absorption can be expressed by
\begin{equation}
dp = \frac{dE_\mathrm{abs}}{c\,dV\,dt}.
\end{equation}
This assumes that all photons which are emitted from the matter, including
scattered photons, are emitted isotropically. Therefore their net momentum is
zero and can be neglected in the momentum transfer between radiation and
matter. This absorption approximation is sufficient for our models. More
elaborate radiation transfer models, which include non-isotropic emission, have
been calculated by Gayley et al.\ \cite{Gayley:etal:95} and others. To obtain
the total radiative acceleration we have to integrate $dp$ over frequency and
the angle of incoming radiation:
\begin{eqnarray}
\grad &=& \frac{1}{\rho}\int\!\int dp \\
      &=& \int\!\int I_\nu\kappa_\nu\nvec\,d\Omega\,d\nu\\
\label{DefgradA}
      &=& c^{-1}\int_0^\infty F_\nu\kappa_\nu\,d\nu
\end{eqnarray}
$\sigma$ and $\kappa$ are complicated functions of temperature, density,
chemical composition, frequency, and the intensity $I_\nu$. Since the radiation
and matter influences each other while the radiation passes through many layers
of material, we end up with an enormous nonlocal set of equations. Solving this
problem rigorously leads to good quantitative results for the radiation
transfer but also needs an enormous effort \cite{Kudritzki:Hummer:90,
Pauldrach:etal:94}. This is beyond the scope of this thesis. Therefore it is
necessary to simplify the problem to a level which is reasonably close to
physical reality but simple enough. For the treatment of hot
star winds at this level there are two important models for the interaction of
radiation and wind: Thompson scattering and line absorption.

In Sect.~\ref{Sec:Thompson} we discuss Thompson scattering as a very simple
mechanism for the interaction between matter and radiation. In
Sect.~\ref{Chap:cak:cak} we discuss how photon absorption in atomic lines can
be described in a simple model. And finally we discuss in
Sect.~\ref{Sec:cak:cakp} the critical point in solutions for these radiation
driven winds.
\section{The Thompson wind}
\label{Sec:Thompson}
Due to the high temperature in the wind of hot stars and due to the strong
ultraviolet radiation field, the matter in the wind is highly ionized. This
leads to a large number of free electrons. Free electrons are, compared to
atoms and molecules, very simple objects. Their total cross section for
scattering of radiation is
\begin{equation}
\sigmaTh = \frac{8\pi}{3}\rce^2 = 6.65\ 10^{-25}\cm^2
\end{equation}
\begin{equation}
\rce = \frac{e^2}{\me c^2}
\end{equation}
where $\rce$ is the classical electron radius. The angular dependence of the
differential cross section and the backscattering of the photons can be
neglected for our treatment (absorption approximation). To calculate the
opacity we need to know the number of free electrons per unit mass. Since the
wind can be assumed to be fully ionized this depends only on the ratio of
protons (and electrons) to nucleons. For hydrogen this is 1, for helium and
heavier elements it is 0.5. This leads to
\begin{equation}
\kappaTh = (1+X)\frac{\sigmaTh}{2\mproton}\approx 0.2(1+X)\frac{\cm^2}{\gram},
\end{equation}
where $X$ is the number fraction of hydrogen in the wind. Since $\kappaTh$ is
independent of $\nu$, we can integrate Eq.~\ref{DefgradA} directly. If we
assume that the wind is optically thin, so that the flux $F_\nu$ depends only
on the geometry, we get
\begin{equation}
\gTh = \frac{\kappaTh L}{4\pi r^2 c},
\end{equation}
where
\begin{equation}
L = \int_0^\infty \!\!\int F_\nu\,d\Omega\,d\nu = 4\pi r^2 F
\end{equation}
is the total luminosity of the star. Here we assumed that the star is point
like. This simplifies the integration over the angle of the incoming
photons. Models which take the finite size of the star into account had been
calculated for the CAK-model \cite{Friend:Abbott:86,Pauldrach:etal:86}. The
Thompson radiation pressure has the same radial dependence than gravity. So if
we add Thompson scattering to the external force balance
(Eq.~\ref{ParkerForces}) we get
\begin{eqnarray}
\Fext  &=& \rho(g_\mathrm{grav}+\gTh)+F_\mathrm{pressure}\\
       &=& -\frac{GM\rho}{r^2}(1-\Gamma)-P'\\
\Gamma &=& \frac{L}{\LE} = \frac{\kappaTh L}{4\pi GMc}.
\end{eqnarray}
$\Gamma$ is the well known Eddington factor, which describes the modification
to gravity by Thompson scattering. If $\Gamma$ would exceed 1 the star would
disintegrate, because the radiation pressure overcomes gravity. This is known
as the Eddington limit. Since the Thompson model is so simple it can be
integrated without extra effort in every wind model. 
\section{The theory of Castor, Abbott \& Klein}
\label{Chap:cak:cak}
In the last section we saw that a sufficient high opacity can drive any
wind. But the Thompson opacity is not high enough to explain the strong winds
we see in many hot stars. Already in the twenties Johnson \cite{Johnson:25} and
Milne \cite{Milne:26,Milne:27} proposed absorption in spectral lines as an
additional mechanism to drive outflows from stars.  The line opacity,
especially in resonance lines, is much higher than the continuum opacity. But
this is much more difficult to treat due to the complicated interaction between
the radiation intensity, as function of frequency and radius, and the opacity
depending on individual excitation levels of the atoms.  If the wind is very
thin (i.e. the radiation is not strongly attenuated - even in lines) we could
used an opacity averaged in frequency like the Rosseland opacity to calculate a
`gray' wind model. This would just replace the low Thompson opacity by the
higher Rosseland opacity. But in the wind of hot stars many lines
become optically thick.  The strong dependency of the line opacity and the
radiation intensity on frequency requires to take the Doppler effect for
accelerating winds into account.  Sobolev \cite{Sobolev:57} did pioneering work
to find a simple formalism to describe this mechanism.  Lucy and Solomon
\cite{Lucy:Solomon:70} were the first who applied this idea to hot stars
shortly after Morton \cite{Morton:67:a,Morton:67:b} had observed P-Cygni
profiles in the UV-spectra of such stars. But their model used only a few
resonance lines and therefore gave much smaller mass loss rates than
observed. Castor Abbott, and Klein \cite{Castor:74,Castor:etal:75} (CAK) solved
this problem by including a large number of lines. Nevertheless they managed to
keep the theory simple by encapsulating the complicated atomic physics in a
simple fit formula. This allows us to use this formula in this project.

We do not need the full details of this theory. Therefore we follow here the
derivation of Owocki \cite{Owocki:90}. The CAK-theory is a hydrodynamical
theory. It assumes that the momentum which is absorbed by some heavy elements
is distributed among the whole wind material. This distinguishes it from the
early work by Johnson and Milne, who assumed the heavy elements escape from the
star while the hydrogen remains bounded. The second assumption is that the
temperature and any nonthermal population of the atomic levels are known in
advance. This fixes the opacity of the wind material.  The CAK-theory takes
into account, that optically thick lines can shadow themselves from the stellar
light, and that this effect can be suppressed by the Doppler-shift in the
expanding wind.

We start the derivation with a single line of opacity
\begin{eqnarray}
\kappa_\nu &=& \kappaL \phi(x)\\
x          &=& \frac{\nu-\nuL}{\nuL} \frac{c}{\vth}\\
\vth       &=& \sqrt{\frac{2\kbolz T}{\mproton}},
\end{eqnarray}
where $x$ is the deviation from the line frequency $\nuL$ in units of the
Doppler shift due to the thermal motion $\vth$. Here $\kappaL$ is the line
opacity, $\phi(x)$ is the normalized line profile function, $c$ is the speed of
light, $\kbolz$ is the Boltzmann constant, and $\mproton$ is the mass of the
proton. The deviation of the mean mass $m$ of the wind particles from the
proton mass is later incorporated in the CAK parameter $\alphacak$ and
$\kcak$. If we neglect now the momentum of the emitted photons, we can express
the radiative acceleration of this line resulting from a flux $F_\nu$ by the
momentum of the absorbed photons
\begin{equation}\label{gLA}
\gL(r,\nuL,\kappaL)\approx\gabs(r,\kappaL)=
        \int_0^\infty \kappaL\,\phi\!\left(x-\frac{v_r(r)}{\vth}
        \right)\frac{F_\nu}{c} \frac{\nuL\vth}{c}e^{-\tau(x,r)}\,dx,
\end{equation}
where we have taken the attenuation of the flux $F_\nu$ by the optical depth 
\begin{eqnarray}\label{gradtau}
\tau(x,r) &=& \int_R^r\kappaL\rho(\tilde{r})\,\phi\!\left(x-
              \frac{v_r(\tilde{r})}{\vth}\right)\,d\tilde{r}\\ 
          &=& \label{gradtaub}
              \int^x_{x-\frac{v_r(r)}{\vth}}\kappaL\,\rho\,\phi(\tilde{x})
              \frac{\vth}{v_r'}\,d\tilde{x}
\end{eqnarray} 
into account, where $R$ is the radius of the stellar surface $(v_r(R)\approx0)$
and $\tilde{x} = x-(v_r(\tilde{r})/\vth)$ the frequency in the comoving frame.
The line profile function $\phi(x)$ is sharply peaked at $x=0$ and therefore
only the small spatial region of the wind with $|x-v_r(\tilde{r})/\vth|\lesssim
1$ will contribute to the integral in Eq.~\ref{gradtaub}. The thickness of this
region is given by the Sobolev length
\begin{equation}\label{defLsobo}
\Lsobo = \frac{\vth}{v_r'(r)}
\end{equation}
with the velocity gradient of the wind
\begin{equation}
v_r' = \frac{dv_r}{dr}.
\end{equation}
The important contribution of Sobolev at this point was the observation that
$\rho(\tilde{r})/v_r'(\tilde{r})$ is approximately constant over this
region of the wind. Since we have a single line wind, where the same line
causes the flux attenuation and the wind acceleration, both effects take place
at approximately the same radius $r$. We can therefore approximate
$\rho(\tilde{r})/v_r'(\tilde{r})$ by $\rho(r)/v_r'(r)$.  This
allows us to draw this term out of the integral
\begin{equation}
\tau(r,x) \approx \tauL\int_{x-\frac{v_r(r)}{\vth}}^\infty
                    \phi(\tilde{x})\,d\tilde{x},
\end{equation}
where we have introduced the Sobolev optical depth of a single line
\begin{equation}
\tauL(r) = \frac{\kappaL\rho\vth}{v_r'}.
\end{equation}
For convenience we shift now to the comoving frame at radius $r$ by $\hat{x} =
x-v_r(r)/\vth$ to solve
Eq.~\ref{gLA} analytically
\begin{eqnarray}
\gL(r,\nuL,\kappaL)   
         &\approx& \frac{\FL\kappaL\vth\nuL}{c^2\tauL}\int_{-\infty}^\infty
                   \tauL\,\phi(\hat{x})\times\nonumber\\
         &       & \exp\left(-\tauL\int^\infty_{\hat{x}}
                   \phi(\tilde{x})\,d\tilde{x}\right)\,d\hat{x}\\
         &\approx& \frac{\FL\kappaL\nuL\vth}{c^2}
                   \frac{1-e^{-\tauL}}{\tauL}.\label{gLB}
\end{eqnarray}
In the case of an optically thin line $\tauL \ll 1$ we find
\begin{equation}
\gthin = \frac{\FL\kappaL\nuL\vth}{c^2}.
\end{equation}
In the optically thick case $\tauL \gg 1$ we find 
\begin{equation}\label{gthickDef}
\gthick = \frac{\gthin}{\tauL} = \frac{\FL\nuL}{c^2\rho}v_r'.
\end{equation}
This expression is independent of $\kappaL$ because the thick line will absorb
all the photons in its frequency range anyway. But $\gthick$ depends on the
velocity gradient $v_r'$, because $v_r'$ describes how efficiently
the line can escape its own shadow and therefore can absorb new photons at a
different Doppler shifted frequency.

The major improvement of Castor, Abbott, and Klein was to take many lines into
account by summing Eq.~\ref{gLA} for many lines. This includes the
assumption that different lines do not overlap in frequency even taking the
Doppler shift due to the wind velocity into account. For a real line list this
computation
can only be done numerically. It is then possible to encapsulate the atomic
physics of the line list in a function $M(t)$ called the \textsl{force
multiplier.} CAK found that this force multiplier can then be approximated by
\begin{equation}\label{CAKMt}
M(t) = \kcak t^{-\alphacak},
\end{equation}
where.
\begin{equation}
t = \frac{\kappaTh \rho \vth}{v_r'}
\end{equation}
is the line independent Sobolev optical depth. CAK \cite{Castor:etal:75} found
$\kcak=1/30$ and $\alphacak=0.56$. Later Abbot \cite{Abbott:82} tabulated
improved values for $\kcak$ and $\alphacak$.

The results given in Eq.~\ref{CAKMt} can be obtained analytically if we assume
that the lines have a flux-weighted number distribution that has a power
law in opacity given by
\begin{equation}
N(\nuL,\kappaL)=\frac{F}{\nuL \FL}\frac{1}{\kappa_0}
                \left(\frac{\kappaL}{\kappa_0}\right)^{\alphacak-2},
\end{equation}
where $F$ is the total radiation flux and $\kappa_0$ is a normalisation
constant. We can now incorporate this expression in Eq.~\ref{gLA} and
integrate over opacity.
\begin{eqnarray}\label{gCAKA}
\gcak(r,v_r,v_r') 
         &=      & \int_0^\infty N(\nuL,\kappaL)\gL(r,\nuL,\kappaL)\,d\kappaL\\
         &\approx& \frac{F\vth\kappa_0}{c^2}
                   \left(\frac{v_r'}{\kappa_0\rho\vth}\right)^\alphacak
                   \times\nonumber\\
         &       & \int_{-\infty}^\infty\int_0^\infty
                   \tauL^{\alphacak-1}e^{-\tau(\hat{x},r)}\,d\tauL
                   \phi(\hat{x})\,d\hat{x}\\
         &\approx& \frac{F\vth\kappa_0}{c^2}
                   \left(\frac{v_r'}{\kappa_0\rho\vth}\right)^\alphacak
                   \Gamma(\alphacak)\times\nonumber\\
         &       & \int_{-\infty}^\infty
                   \left(\int^\infty_{\hat{x}}\phi(\tilde{x})
                   \,d\tilde{x}\right)^{-\alphacak} \phi(\hat{x})\,d\hat{x}\\
         &\approx& \frac{F\vth\kappa_0}{c^2}
                   \frac{\Gamma(\alphacak)}{1-\alphacak}
                   \left(\frac{v_r'}{\kappa_0\rho\vth}\right)^\alphacak\\
         &\approx& \frac{F\kappaTh}{c}M(t),
\end{eqnarray}
where $\Gamma(\alphacak)$ is the complete Gamma function.
Here we find the relation between $\kappa_0$ and $\kcak$:
\begin{equation}
\kcak = \frac{\vth}{c}\left(\frac{\kappa_0}{\kappaTh}\right)^{1-\alphacak}
        \frac{\Gamma(\alphacak)}{1-\alphacak}
\end{equation}

If we include $\gcak$ and $\gTh$ in the wind equation we get different
analytical conditions for a complete physical solution than in the case of a
Parker wind. The reason for this is that $\gcak$ depends additionally on the
local wind acceleration. To analyze the solutions we rewrite the wind
equation (Eq.~\ref{ParkerWind}) to
\begin{equation}\label{CAKWind}
A(r,v_r)v_r'+B(r) = C(r,v_r) v_r^{\prime\alphacak}
\end{equation}
\begin{eqnarray}
\label{CAKTermA}
A &=& v_r-\frac{\vsvs}{v_r}\\
\label{CAKTermB}
B &=& \frac{GM}{r^2}(1-\Gamma)-2\frac{\vsvs}{r}+\vs^{2\prime}\\
\label{CAKTermCa}
C &=& \frac{\kappaTh L}{4\pi r^2 c}\kcak
      \left(\frac{1}{\kappaTh\rho\vth}\right)^\alphacak\\
\label{CAKTermCb}
  &=& \frac{GM\Gamma\kcak}{r^2}
      \left(\frac{4\pi r^2v_r}{\kappaTh\Mdot\vth}\right)^\alphacak.
\end{eqnarray}
Equation~\ref{CAKWind} is an implicit differential equation, which in general
can not be solved analytically. But this is one of the minor numerical problems
in CAK wind models.
\begin{figure}
\begin{center}
\begin{picture}(120,140)
%\put(0,0){\framebox(120,140){}}
\put(20,135){A)}
\put(65,135){B)}
\put(20,107){C)}
\put(65,107){D)}
\put(20,79){E)}
\put(65,79){F)}
\put(20,51){G)}
\put(65,51){H)}
\put(15,-2){\epsfig{file=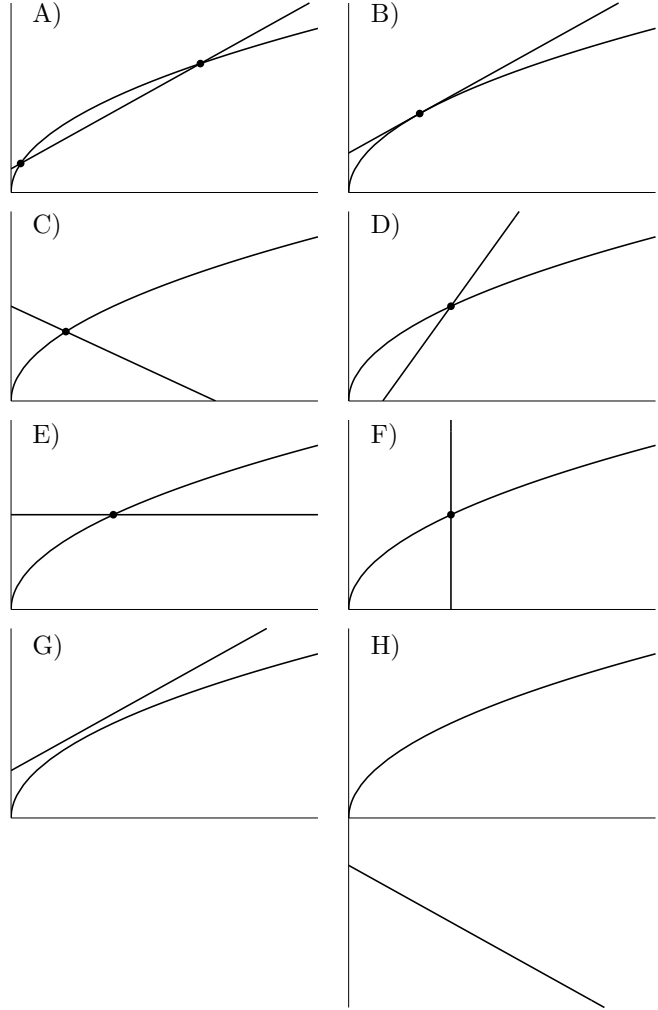,width=9cm}}
\end{picture}
\end{center}
\caption[The solutions for the CAK wind]{\label{CAKSolutions}The possible
solutions of Eq.~\ref{CAKWind}: The left hand side of Eq.~\ref{CAKWind} is
linear in $v_r'$: $A\,v_r'\,+B$, while the right hand side:
$C\,v_r^{\prime\alphacak}$ depends on $v_r'$ roughly like $\sqrt{v_r'}$ because
$\alphacak$ is between 0 and 1. But $A$ and $B$ can have either sign
(Eqs.~\ref{CAKTermA}\&\ref{CAKTermB}) depending on the values of $r$ and
$v_r(r)$. From Eq.~\ref{CAKTermCb} we see that $C$ is always positive.  In
Figs.~A--H we sketch the left hand side and the right hand side of
Eq.~\ref{CAKWind} (ordinate) as function of $v_r'$ (abscissa). $r$ and $v_r$
are kept constant in each figure. A solution for Eq.~\ref{CAKWind} is given at
every value of $v_r'$ where the two curves intersect.  In Fig.~A
Eq.~\ref{CAKWind} has two solutions. In Figs.~B--F Eq.~\ref{CAKWind} has one
solution. And in Figs.~G\&H Eq.~\ref{CAKWind} has no solution. In general the
coefficients $A$ and $B$ are not constant in a single wind solution. Thus for
different radii different of the figures sketched here may apply in the same
wind solution. For a physical wind solution $r$ and $v_r$ and thus $A$ and $B$
are continuous everywhere. Therefore the transition between these different
figures has to be continuous as well. A nonmagnetic wind passes from the base
of the wind to infinity through the following sequence: C--E--A--B(here is the
CAK critical point)--A--D. See Sect.~\ref{Sec:cak:cakp} for details.  The case
of Fig.~F appears only if magnetic fields are included
(Chap.~\ref{Chap:cak:wd}). In this case $A$ and $B$ are infinite at the
Alfv\'enic point.}
\end{figure}
In the case of an isothermal wind we can simplify these equations by
substituting $u=-1/r$, $w=v_r^2/2$, and $\dot{w}= (dw/du)$: 
\begin{equation}\label{CAKWindB}
\bar{A}(w)\dot{w}+\bar{B}(u)=\bar{C}\dot{w}^\alphacak
\end{equation}
\begin{eqnarray}
\bar{A}(w) &=& 1-\frac{\vsvs}{2w}\\
\bar{B}(u) &=& GM(1-\Gamma)+2\vsvs u\\
\bar{C}    &=& GM\Gamma\kcak
                 \left(\frac{4\pi}{\kappaTh\Mdot\vth}\right)^\alphacak.
\end{eqnarray}
In the case of a cold wind without thermal pressure ($\vs=0$)
Eq.~\ref{CAKWindB} becomes independent of $r$. Therefore $\dot{w}=(dw/du)$ is
independent of $r$ as well. This leads to the often used velocity law
\begin{eqnarray}
v_r(r) &=& \vinf\sqrt{1-\frac{R}{r}}\\
\vinf  &=& \sqrt\frac{2\dot{w}}{R},
\end{eqnarray}
where $R$ is the stellar radius. Since in this case $A=1$ and $B=GM(1-\Gamma)$
are larger than zero, we can have two (Fig.~\ref{CAKSolutions}.A), one
(Fig.~\ref{CAKSolutions}.B), or no (Fig.~\ref{CAKSolutions}.G) solutions for
$\dot{\omega}$ depending on the value of $C(\Mdot)$. The case of one solution
is special, because it has the minimum value for $C(\Mdot)$ and therefore the
maximum value of $\Mdot$. The unique value for $\dot{w}$ in this case leads to
\begin{eqnarray}
\label{MdotCAK}
\Mdot_\mathrm{max} &=& \frac{4\pi G M}{\kappaTh\vth}\alphacak
  (\kcak\Gamma)^{1/\alphacak}\left(\frac{1-\alphacak}{1-\Gamma}
  \right)^{(1-\alphacak)/\alphacak}\\
\label{vinfCAK}
\vinf(\Mdot_\mathrm{max}) &=& \sqrt{\frac{\alphacak}{1-\alphacak}
   \frac{2GM(1-\Gamma)}{R}}.
\end{eqnarray}
This case is often referred to as the analytical CAK-solution. Nevertheless this
situation is not satisfying, because solutions with lower mass loss rates are
not excluded by any physical argument.
\section{The CAK point}
\label{Sec:cak:cakp}
For a unique solution with a unique value for $\Mdot$ we have to take the
thermal pressure into account. Since $\alphacak$ is between 0 and 1,
Eq.~\ref{CAKSolutions} can have no, one, or two solutions for $v_r'$
depending on the values of $A$, $B$. The possible cases are sketched in
Fig.~\ref{CAKSolutions}. If $C$ is zero we have the Parker or Thompson wind as
limiting case.  At $r=\infty$ we have a negative value for $B$. This requires
$A>0$ as plotted in Fig.~\ref{CAKSolutions}.D and therefore a supersonic
wind. Close to the star the gravity term in Eq.~\ref{CAKTermB} will dominate
and cause a positive value for $B$. This transition from
Fig.~\ref{CAKSolutions}.D to Fig.~\ref{CAKSolutions}.A takes place in the
supersonic part of the wind ($A>0)$. Otherwise we would end up with the
situation of Fig.~\ref{CAKSolutions}.H where no solution for $v_r'$ can be
found. In the case of Fig.~\ref{CAKSolutions}.A we have two possible
solutions. But the continuous transition from Fig.~\ref{CAKSolutions}.D allows
only the larger value of $v_r'$.

If we use the CAK-model in the subsonic part of the wind we get a negative
value for $A$. Therefore $B$ must be positive and we have the unique solution
for $v_r'$, which is sketched in Fig.~\ref{CAKSolutions}.C. At the sonic
point ($A=0$, Fig.~\ref{CAKSolutions}.E) the solution switches to
Fig.~\ref{CAKSolutions}.A. But now the proper solution for $v_r'$ is the
lower value, because only this matches continously the unique solution at the
sonic point.

If we want a continuous steady state solution, which extends from the base of
the wind $(v_r\ll\vs)$ to $r=\infty$, we have to connect the solution branches
discussed above in a proper way. This can only happen if the solution passes
through a point where Fig.~\ref{CAKSolutions}.B applies. Only here we can shift
$v_r'$ continously from the lower value in Fig.~\ref{CAKSolutions}.A to
Fig.~\ref{CAKSolutions}.B and then to the higher value in
Fig.~\ref{CAKSolutions}.A. This point is the \emph{critical CAK-point.} It
fixes the whole solution. The best way to solve Eq.~\ref{CAKWind} numerically
is to find the critical CAK-point and to integrate Eq.~\ref{CAKWind} from there
to the base of the wind and to $r=\infty$.

The critical CAK-point is specified by four quantities: $\Mdot,$ $\rc,$
$\vrc,$ and $\dvrcdr$. Therefore we need four conditions to find the critical
CAK-point. The first one is Eq.~\ref{CAKWind}. From Fig.~\ref{CAKSolutions}.B
we see that both sides of Eq.~\ref{CAKWind} have the same inclination when they
are plotted as function of $v_r'$. This happens only in
Fig.~\ref{CAKSolutions}.B and is therefore the second condition:
\begin{equation}\label{CAKpointB}
A(\rc,\vrc) = \alphacak C(\rc,\vrc)v_{r\mathrm{c}}^{\prime\alphacak}
\end{equation}
A physical solution for $v_r$ should be smooth everywhere. This allows us to
differentiate Eq.~\ref{CAKWind} totally with respect to $r$:
\begin{eqnarray}
\lefteqn{
\left(\diff{A}{r}+\diff{A}{v_r}v_r'\right)v_r'+Av_r''+
\diff{B}{r} = }\nonumber\\
&&\left(\diff{C}{r}+\diff{C}{v_r}v_r'\right)v_r^{\prime\alphacak} +
\alphacak C v_r^{\prime(\alphacak-1)}v_r''
\end{eqnarray}
At the critical CAK-point we can use Eq.~\ref{CAKpointB} to eliminate
$v_r''$. This leads to the third condition:
\begin{equation}\label{CAKpointC}
\left.\left(\diff{A}{r}+\diff{A}{v_r}v_r'\right)v_r'+
\left(\diff{B}{r}+\diff{B}{v_r}v_r'\right) = \right|_{\rc,\vrc}
\left(\diff{C}{r}+\diff{C}{v_r}v_r'\right)v_r^{\prime\alphacak}
\end{equation}
The forth condition is not located at the critical point but at the base of the
wind, where our solution should fit the conditions of the photosphere. We can
specify the optical depth or the wind velocity there.  The
Eq.\ \ref{CAKWind}, \ref{CAKpointB} \& \ref{CAKpointC} can be solved
analytically for $\rc,$ $\vrc,$ and $\dvrcdr$.  CAK solved for $\Mdot$ instead
of $\rc$. In any case we have to assume a value for the forth unknown, find
the critical CAK-point, integrate Eq.~\ref{CAKWind}, and
finally improve our assumption about the forth unknown using the boundary
condition at the base of the wind. This procedure converges sufficiently fast.

The CAK model gives us a simple description for the momentum transfer from the
radiation to the wind. But this includes no description for the energy transfer
between radiation and wind. Therefore it is not possible to solve
simultaneously the energy equation for the local temperature. Instead
we have to assume a temperature distribution before we can solve
Eq.~\ref{CAKWind} for the velocity distribution. A separate model can then be
used to improve the temperature distribution as CAK did \cite{Castor:etal:75,
Castor:74:b}. This iteration converges fast, because the velocity distribution
depends only weakly on the temperature distribution.

When we ask for a wind solution with constant but nonzero temperature, we can
apply our discussion about the critical CAK-point. So we would expect a unique
solution. But if we write down the third critical point condition for $w$
\begin{equation}
\diff{\bar{A}}{w}\dot{w}^2+\diff{\bar{B}}{u} = 0
\end{equation}
\begin{eqnarray}
\diff{\bar{A}}{w} &=& \frac{\vsvs}{2w^2} > 0\\
\diff{\bar{B}}{u} &=& 2\vsvs >0
\end{eqnarray}
we see that $\dot{w}^2$ must be negative. This would lead to complex and
therefore unphysical solutions for Eq.~\ref{CAKWindB}.

The basic theory, described here, had been improved in many ways. This includes
the improved fit formula for the force multiplier by Abbott \cite{Abbott:82},
the correction for the finite size of the star \cite{Friend:Abbott:86,
Pauldrach:etal:86}, the inclusion of multiple photon scattering
\cite{Lucy:Abbott:93, Gayley:etal:95}, and the detailed line balance models of
Kudritzki \cite{Kudritzki:Hummer:90, Pauldrach:etal:94}.  But in this thesis we
want to emphasize the effect of magnetic fields. Therefore we will restrict
ourself to this basic formalism.
\vfill
%

%-*-LaTeX-*-
% This is the 4th chapter for the PhD-thesis of Henning Seemann.
% (c) 1997-98 by Henning Seemann
%
\chapter{Magnetic fields in stellar winds}
\label{Chap:wd}
\section{Introduction}
Already Parker \cite{Parker:58} argued about the role of the magnetic field in
the solar wind. But the solar magnetic field is weak. Its contribution to the
momentum balance is small compared to the thermal pressure of the hot
corona. Therefore he neglected the Lorentz force as we did in Chapter
\ref{Chap:parker:wind}. But he saw the importance of the magnetic field for the
angular momentum balance of the sun.

Weber \& Davis \cite{Weber:Davis:67} were the first who replaced the
hydrodynamical treatment of Parker by a magnetohydrodynamic (MHD) treatment
to analyze the importance of the magnetic field for the solar wind. The second
important ingredient for their model was the rotation of the sun, which was
neglected by Parker as well. Without rotation the material will stream out
radially along radial, open field lines. In this case there will be no net
force and we end up with the physics of Parker's model.

Biermann \& Cassinelli \cite{Biermann:Cassinelli:93, Biermann:94} developed an
improved version of the Weber \& Davis model, which treats the magnetic field
in the equatorial plane without any further simplifications beyond the basic
assumptions we made already as well. This allows for an oblique magnetic field
at the stellar surface, which occurs in fast magnetic rotators. This model is
the basis for our analysis of magnetic stellar winds. We present here the
derivation in a generalized version. Biermann \& Cassinelli and Weber \& Davis
assumed that meridional diameter of the fluxsheet is proportional to $r$. This
leads e.g.\ to radial magnetic field proportional to $r^{-2}$. We introduce an
arbitrary function $a(r)$, which allows us to describe the derivation from this
$r^{-2}$ law due to the influence of a wind compressed disk or a wind, which is
diluted in the equatorial plane by a non-equatorial wind collimated towards the
poles.

In Sect.~\ref{Sec:wdmodel} we derive our wind equations from the fundamental
equations of magnetohydrodynamics. In Sect.~\ref{Sec:simple} we discuss briefly
simplified versions of our equations. And finally we discuss in
Sect.~\ref{Sec:spindown} the spin-down problem, which gives limitations for the
model parameters.
\section{The model}
\label{Sec:wdmodel}
We start our derivation from the magneto-hydrodynamical equations for an ideal,
perfectly conducting gas in a stationary situation.

In the case of the Parker wind we had a perfect spherical symmetry, which
reduced the problem directly to one spacial dimension. The one dimensional
solution was valid for all areas in the wind. For a rotating star we have only
one axis of symmetry. Therefore we can reduce our problem strictly only by one
dimension. For a stellar wind it is most useful to introduce spherical
coordinates parallel to the rotation axis and then to neglect the dependence of
any physical quantity on the azimuthal angle $\phi$. For winds from other
objects (e.g. disks) a cylindrical coordinate system might be more useful.  In
any case we should solve the MHD equations in the two dimensions of the
meridional plane. In this plane we can not expect more symmetry because the MHD
equations mix the radial symmetry of gravitation and radiation with the
cylindrical symmetry of the centrifugal force. A rigid two dimensional
treatment is beyond the scope of this project. But we can reduce the problem to
one dimension if we assume a priori a certain geometry of the flow in the
meridional plane. This concept is called the fluxsheet approximation. We have
to assume the path of the fluxsheet in the meridional plane and its cross
section, varying along the fluxsheet. Then we need only to find the various
physical quantities along the fluxsheet. This is a one dimensional problem. In
this chapter we choose the equatorial plane as fluxsheet. From observation we
know, that our sun has a wind in the equatorial plane. And the principal
symmetry between northern and southern hemisphere of a star makes this scenario
quite probable for other stars as well. The only other reasonable configuration
in the equatorial plane is an accretion disk. We will discuss non-equatorial
winds in Chap.~\ref{Chap:fluxtube}.
\begin{figure}
\begin{center}
\begin{picture}(90,90)
%\put(0,0){\framebox(90,90){}}
\put(0,0){\epsfig{file=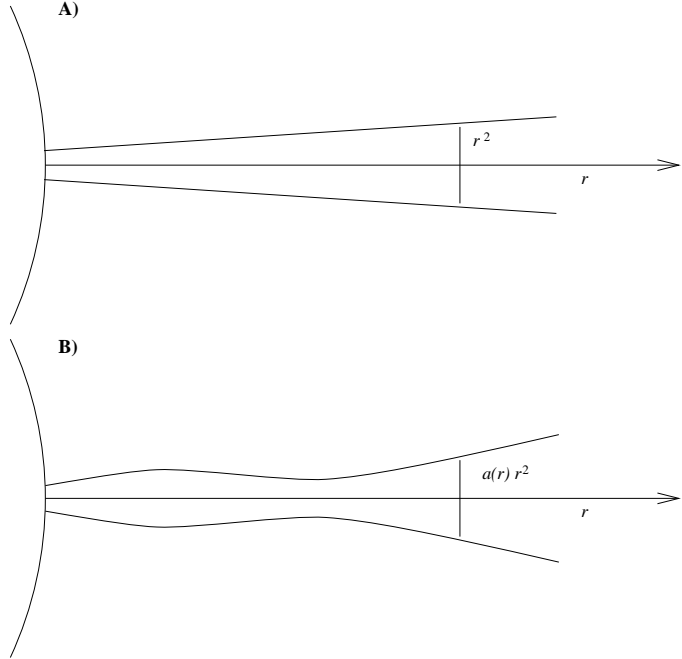,width=9cm}}
\end{picture}
\end{center}
\caption[The cross section factor $a(r)$]{\label{AreaFactor}An equatorial
fluxsheet in the $(r,\theta)$ plane. Fig.~\textbf{A)} shows the situation in
previous models. The thickness of the fluxsheet in the meridional plane is
proportional to $r$. Due to the rotational symmetry the same is true in the
equatorial plane. The cross section of the fluxsheet is therefore proportional
to $r^2$. Fig~\textbf{B)} The dimensionless function $a(r)$ describes the
deviation of the fluxsheet cross section from $r^2$.}
\end{figure}

We start from the equations of magnetohydrodynamics (MHD) for a nonviscous,
perfectly conducting, and quasi stationary gas:
\begin{eqnarray}
\label{MassConservationMHDc}
\nabla\cdot(\rho\vvec)         &=& 0 \\
\label{SheetWinda}
\rho(\vvec\cdot\nabla)\vvec    &=& \Fextvec-\frac{1}{4\pi}
                                   \Bvec\times(\nabla\times\Bvec)\\
\label{fluxfreezinga}
\nabla\times(\vvec\times\Bvec) &=& 0.
\end{eqnarray}
The external forces
\begin{eqnarray}
\Fextvec &=& \Fext\ervec\\
\label{Fextdef}
         &=& -\rho\nabla\Phi-\nabla P+\fradvec
\end{eqnarray}
include gravity, radiation pressure, and thermal pressure as discussed in the
previous chapters. Since we restrict our discussion to the equatorial plane
$(\theta=\pi/2)$ we can set $v_\theta$ and $B_\theta$ to zero. But there
derivatives with respect to $\theta$ might not be zero. The momentum equation
then simplifies to
\begin{eqnarray}\label{momentumphia}
\rho\frac{v_r}{r}\dr{r\vphi}
&=& \frac{1}{4\pi}\left[(\nabla\times\Bvec)\times\Bvec\right]_\phi=
    \frac{B_r}{4\pi r}\dr{r\Bphi}\\
\label{momentumra}
\left(v_r\dr{v_r}\right)\left[1-\frac{\vsvs}{v_r^2}\right]
&=& -\frac{GM}{r^2} + \frac{\vsvs}{r}\left(2 + \frac{r}{a}\dr{a}\right)-
    \frac{\partial\vsvs}{\partial r}\nonumber\\
& & +{}\grad+\frac{\vphi^2}{r} -\frac{1}{8\pi\rho r^2}
    \dr{}\left(r \Bphi\right)^2.
\end{eqnarray}
In the fluxsheet concept the conservation laws for magnetic flux and matter are
given by
\begin{eqnarray}
\label{FluxConservationa}
F_B   &=& B_r r^2 a(r)  = \const\\
\label{MassConservationMHDa}
\Mdot &=& 4\pi\rho r^2 v_r a(r) = \const.
\end{eqnarray}
The definition of $\Mdot$ can be misleading. It is in analogy to the definition
for a spherical wind, where we have $a(r) \equiv 1$. In the case of a
cylindrical wind with rotation we must calculate the wind for all latitudes to
get an accurate value for the total mass loss of the star. But if we assume
that the wind is mostly spherical at the base of the wind $(r=R)$, and if we
normalize $a$ to $a(R)=1$, Eq.~\ref{MassConservationMHDa} will give a good
estimate for the total mass loss rate. If we assume that the mass loss is
reduced at the poles due to the missing rotation there,
Eq.~\ref{MassConservationMHDa} will give an upper limit for the real mass loss
rate. Which case is true, depends on the rotation rate of the star.

The flux freezing condition Eq.~\ref{fluxfreezinga} for the perfectly
conducting wind in the equatorial plane leads to
\begin{equation}
0=\dr{}\left[r\left(\vphi B_r-v_r\Bphi\right)\right]-
  \dtheta{}\left(v_\theta\Bphi-\vphi B_\theta\right).
\end{equation}
The symmetry between northern and southern hemisphere requires that all
quantities except $v_\theta$ and $B_\theta$ have a vanishing derivative with
respect to $\theta$ in the equatorial plane. Now we replace the
$\theta$-derivatives of $v_\theta$ and $B_\theta$ using $\nabla\cdot\Bvec=0$
and Eqs.~\ref{MassConservationMHDc}, \ref{FluxConservationa} \&
\ref{MassConservationMHDa}
\begin{equation}
0=\dr{}\left[r\left(\vphi B_r-v_r\Bphi\right)\right]-
  \frac{\dadr}{a}\left(v_r\Bphi-\vphi B_r\right),
\end{equation}
where we used the $^\prime$ to mark the derivative with respect to $r$.  This
equation can be directly integrated and leads to
\begin{eqnarray}
a r(v_r\Bphi-\vphi B_r) &=& \const\\
                        &=& -\Omega a_0 R^2 \Bro\\
\label{fluxfreezingb}
                        &=& -\Omega F_B.
\end{eqnarray}
Here we introduce the concept of the 'base of the wind'. At the base of the
wind we have a negligible radial wind velocity. It is the transition region
between the static envelope and the dynamically expanding wind of the star.  We
use $R$ for the radius of the base of the wind and denote other physical
quantities at the base of the wind with the index $_0$. For stars with a weak,
optically thin wind, like our sun and most main sequence stars, the base of the
wind is identical with the photosphere defined by an optical depth of one for
electron scattering. This is e.g.\ not true for Wolf-Rayet stars. We denote the
rotation rate of the star with $\Omega$. It is defined by the rotation of the
quasi stationary magnetic field configuration. At the base of the wind the
radial velocity $\vro$ can be neglected compared to the co-rotation velocity
$\vroto=\Omega R$. If at the base of the wind the radial component $\Bro$ and
the azimuthal component $\Bphio$ of the magnetic field are of the same order,
the azimuthal component of the wind velocity $\vphi$ will be nearly identical
to the co-rotation velocity at the base of the wind. In this case $\Omega$ will
be the rotation velocity of the stellar surface as well. We will later find
this confirmed in our numerical models.  It is important to note at this place,
that $\Omega$ may depend on latitude. This is e.g.\ the case for our sun. From
theoretical models we know as well, that the rotation rate can change
drastically inside the star.

Due to the rotational symmetry of our star the angular momentum component
parallel to the rotation axis is a conserved quantity.
Equation~\ref{momentumphia} describes the transport of angular momentum in the
wind. Using the conservation laws for magnetic flux and matter we can integrate
Eq.~\ref{momentumphia} to yield the angular momentum loss of the star per unit
wind mass
\begin{equation}\label{LjdefA}
L_j = r\vphi-\frac{B_r}{4\pi\rho v_r}r\Bphi = \const.
\end{equation}
This includes the angular momentum in the wind matter and the angular momentum
in the bent magnetic field. We parametrize $L_j$ by
\begin{equation}\label{LjdefB}
L_j=\epsilon\Omega R^2,
\end{equation}
where $\Omega R^2$ is the angular momentum transport per unit mass by matter
at the base of the wind. The angular momentum transport of the magnetic field
at the base of the wind is included through the dimensionless factor
$\epsilon$. For all reasonable configurations $\epsilon$ is larger than 1. We
discuss this later in detail.

The radial velocity can be expressed relative to the radial Alfv\'enic
velocity by
\begin{equation}
\MAr=\frac{v_r}{\vAr}=\frac{v_r}{\frac{B_r}{\sqrt{4\pi\rho}}}.
\end{equation}
This leads to
\begin{equation}\label{auxeqa}
\frac{1}{\MAr^2}=
\frac{B_r^2}{4\pi\rho v_r^2}\frac{a^2r^4}{a^2r^4}=
\frac{F_B^2}{\Mdot}\frac{1}{v_r ar^2}=
\frac{1}{\MAro^2}\frac{\vro a_0 R^2}{v_r ar^2}
\end{equation}
and
\begin{eqnarray}
\label{auxeqc}
\frac{1}{\MAr^2}\frac{r^2}{R^2}\frac{1}{\epsilon}
&=&\frac{F_B^2}{\Mdot}\frac{\Omega}{L_j v_r a}\\
\label{auxeqd}
&=&\frac{1}{\MAro^2}\frac{\vro}{v_r}\frac{1}{\epsilon}\frac{a_0}{a}.
\end{eqnarray}
With Eq.~\ref{fluxfreezingb},\ref{LjdefA}\&\ref{auxeqc} we can now express
$\vphi$ by
\begin{eqnarray}
\vphi &=& \frac{L_j}{r}
            \frac{1-\frac{F_B^2}{\Mdot}\frac{\Omega}{L_j v_r a}}
                 {1-\frac{F_B^2}{\Mdot}\frac{1}{v_r a r^2}}\\
      &=& \frac{L_j}{r}
            \frac{1-\frac{1}{\MAr^2}\frac{r^2}{R^2}\frac{1}{\epsilon}}
                 {1-\frac{1}{\MAr^2}}\\
      &=& (\Omega r)\frac{\UM-\Ue}{\UM},
\end{eqnarray}
where we have introduced the auxiliary quantities
\begin{eqnarray}
\Ue &=& 1-\epsilon\left(\frac{R}{r}\right)^2\\
\UM &=& 1-\frac{1}{\MAr^2}\\
    &=& 1-\frac{1}{\MAro^2}\frac{\vro}{v_r}
          \left(\frac{R}{r}\right)^2\frac{a_0}{a}.
\end{eqnarray}
Now we have expressed $\vphi$ in terms of $r$ and $v_r$. Before we can solve
Eq.~\ref{momentumra} as an ordinary differential equation in $v_r(r)$ we have
to do the same for 
\begin{eqnarray}
\Bphi &=& \frac{1}{v_r r}\left[r\vphi B_r-\frac{\Omega F_B}{a}\right]\\
      &=& -\frac{\Omega F_B}{v_r r}\frac{1}{\UM}\left[\frac{\UM}{a}-
            \frac{r B_r}{\Omega F_B}\frac{L_j}{r}
            \left(1-\frac{F_B^2}{\Mdot}\frac{\Omega}{L_j v_r}\frac{1}{a}
            \right)\right]\\
      &=& -\frac{\Omega F_B}{v_r r}\frac{1}{\UM}\frac{1}{a}
            \left[1-\frac{L_j}{\Omega r^2}+
            \frac{F_B^2}{\Mdot v_r a}\left(\frac{1}{r^2}-\frac{1}{r^2}\right)
            \right]\\
\label{Bphia}
      &=& -\frac{\Omega F_B}{v_r r a}\frac{\Ue}{\UM}.
\end{eqnarray}
The Lorentz force term in Eq.~\ref{momentumra} can be split into two parts
by
\begin{eqnarray}
-\frac{1}{8\pi\rho r^2}\dr{}(r\Bphi)^2
&=& -\frac{a v_r}{2\Mdot}\dr{}(r\Bphi)^2\\
&=& \underbrace{-\frac{a}{v_r}\dr{}\left(\frac{v_r^2}{2}
    \frac{(r\Bphi)^2}{\Mdot}\right)}_{II}+\nonumber\\
&&  \underbrace{\frac{a(r\Bphi)^2}{v_r\Mdot}}_I\left(v_r\dr{v_r}\right).
\end{eqnarray}
Each part can now be expressed in terms of $r$, $v_r$, and $\dvrdr$.
\begin{eqnarray}
I  &=& \frac{a}{v_r \Mdot}
         \left(\frac{\Omega F_B}{av_r}\frac{\Ue}{\UM}\right)^2\\
   &=& \left(\frac{\Omega r}{v_r}\right)^2\frac{a_0}{a}\frac{1}{\MAro^2}
         \frac{\vro}{v_r}
         \frac{R^2}{r^2}\left(\frac{\Ue}{\UM}\right)^2\\
   &=& \frac{1}{\MAro^2}\frac{a_0}{a}\left(\frac{\Omega R}{\vro}\right)^2
         \left(\frac{\vro}{v_r}\right)^3\left(\frac{\Ue}{\UM}\right)^2
\end{eqnarray}
\begin{eqnarray}
II &=& -\frac{a\Omega^2F_B^2}{2\Mdot v_r}
         \dr{}\left(\frac{\Ue}{a\UM}\right)^2\\
   &=& -\frac{aa_0}{\MAro^2}\left(\frac{\Omega R^2}{\vro}\right)^2
         \frac{\vro^3}{v_r}\left(\frac{\Ue}{a\UM}\right)
         \left(\frac{\Ue^\prime a\UM-\Ue(a\UM)^\prime}{a^2\UM^2}\right)\\
   &=& -\frac{1}{\MAro^2}\frac{a_0}{a}\left(\frac{\Omega R^2}{\vro}\right)^2
         \frac{\vro^3}{v_r}\frac{\Ue}{\UM^3}
         \left[2\epsilon\frac{R^2}{r^3}\UM-\right.\nonumber\\
   & & \left.\Ue\left(\frac{a^\prime}{a}+
         \frac{2}{\MAro^2}\frac{\vro}{v_r}\frac{R^2}{r^3}\frac{a_0}{a}\right)
         -\Ue\frac{1}{\MAro^2}\frac{\vro v_r^\prime}{v_r^2}\frac{a_0}{a}
         \right]\\
   &=& -\frac{1}{\MAro^2}\frac{a_0}{a}\left(\frac{\Omega R^2}{\vro}\right)^2
         \frac{\vro}{v_r}\frac{\vro^2}{r}\left.\frac{\Ue}{\UM^3}
         \right[2(1-\Ue)\UM-\nonumber\\
   & & \left.2\Ue(1-\UM)-\Ue\frac{ra^\prime}{a}-
         \Ue(1-\UM)\frac{r v_r^\prime}{v_r}\right]\\
   &=& \frac{1}{\MAro^2}\frac{a_0}{a}\left(\frac{\Omega R^2}{\vro}\right)^2
         \frac{\vro}{v_r}\frac{\vro^2}{r}\frac{\Ue}{\UM^2}
         \left(2\frac{\Ue-\UM}{\UM}+\frac{\Ue}{\UM}\frac{ra^\prime}{a}
         \right)+\nonumber\\
   & & \underbrace{\frac{1}{\MAro^2}\frac{a_0}{a}
         \left(\frac{\Omega R^2}{\vro}\right)^2
         \left(\frac{\vro}{v_r}\right)^3\frac{\Ue^2}{\UM^3}
         (1-\UM)}_{III}\left(v_r\dr{v_r}\right)
\end{eqnarray}
Here we used again the $^\prime$ to mark the derivative with respect to $r$.
It useful to collect terms depending on $v_r^\prime$, because this makes it
later easier to find $v_r^\prime$ as function of $r$ and $v_r$.
\begin{eqnarray}
I+III &=& \frac{1}{\MAro^2}\frac{a_0}{a}\left(\frac{\Omega R}{\vro}\right)^2
         \left(\frac{\vro}{v_r}\right)^3\left(\frac{\Ue}{\UM}\right)^2
         \left[1+\frac{1-\UM}{\UM}\right]\\
     &=& \frac{1}{\MAro^2}\frac{a_0}{a}\left(\frac{\Omega R}{\vro}\right)^2
         \left(\frac{\vro}{v_r}\right)^3 \frac{\Ue^2}{\UM^3}         
\end{eqnarray}
As final result for the Lorentz force term we get then
\begin{eqnarray}
\lefteqn{-\frac{1}{8\pi\rho r^2}\dr{}\left(r\Bphi\right)^2=}\\
&&  \left(v_r\dr{v_r}\right)\left[\frac{1}{\MAro^2}\frac{a_0}{a}
    \left(\frac{\Omega R}{\vro}\right)^2\left(\frac{\vro}{v_r}\right)^3 
    \frac{\Ue^2}{\UM^3}\right]-\\
&&  \frac{1}{\MAro^2}\frac{a_0}{a}
    \left(\frac{\Omega R}{\vro}\right)^2\frac{\vro}{v_r}\frac{\vro^2}{r}
    \frac{\Ue}{\UM^2}\left(2\frac{\UM-\Ue}{\UM}-
    \frac{\Ue}{\UM}\frac{ra^\prime}{a}\right).
\end{eqnarray}
This can now be combined with the term for the centrifugal force
\begin{equation}
\frac{\vphi^2}{r} =
  \left(\frac{\Omega R}{\vro}\right)^2\frac{\vro^2}{R}\frac{r}{R}
  \left(\frac{\UM-\Ue}{\UM}\right)^2
\end{equation}
to
\begin{eqnarray}
\lefteqn{\frac{\vphi^2}{r}-\frac{1}{8\pi\rho r^2}\dr{}\left(r\Bphi\right)^2=}
\nonumber\\
&&  \left(v_r\dr{v_r}\right)\left[\frac{1}{\MAro^2}\frac{a_0}{a}
    \left(\frac{\Omega R}{\vro}\right)^2\left(\frac{\vro}{v_r}\right)^3 
    \frac{\Ue^2}{\UM^3}\right]+\nonumber\\
&&  \left(\frac{\Omega R}{\vro}\right)^2 \frac{\vro^2}{r}
    \frac{r^2}{R^2}\frac{\UM-\Ue}{\UM^2}\left(\UM-\Ue-
    \frac{2}{\MAro^2}\frac{\vro}{v_r}\frac{R^2}{r^2}\frac{a_0}{a}
    \frac{\Ue}{\UM}\right)+\nonumber\\
&&  \frac{1}{\MAro^2}\frac{a_0}{a}
    \left(\frac{\Omega R}{\vro}\right)^2\frac{\vro}{v_r}\frac{\vro^2}{r}
    \frac{\Ue^2}{\UM^3}\frac{ra^\prime}{a}\\
&=& \left(v_r\dr{v_r}\right)\left[\frac{1}{\MAro^2}
    \left(\frac{\Omega R}{\vro}\right)^2\left(\frac{\vro}{v_r}\right)^3 
    \frac{\Ue^2}{\UM^3}\frac{a_0}{a}\right]-\nonumber\\
&&  \left(\frac{\Omega R}{\vro}\right)^2 \frac{\vro^2}{r}
    \frac{R^2}{r^2}\frac{1}{\UM^2}
    \left(\epsilon-\frac{1}{\MAro^2}\frac{\vro}{v_r}\frac{a_0}{a}\right)
    \left[\left(2\frac{\Ue}{\UM}+1\right)
    \frac{1}{\MAro}\frac{\vro}{v_r}\frac{a_0}{a}-\epsilon\right]+\nonumber\\
&&  \left(\frac{\Omega R}{\vro}\right)^2\frac{1}{\MAro^2}
    \frac{\vro}{v_r}\frac{\vro^2}{r}
    \frac{\Ue^2}{\UM^3}\frac{a_0ra^\prime}{a^2}
\end{eqnarray}
Now we can write the radial momentum equation (Eq.~\ref{momentumra}) as an
ordinary differential equation in $v_r(r)$, which allows the desired one
dimensional solution:
\begin{eqnarray}\label{momentumrb}
\lefteqn{-\frac{r^2}{GM}\left(v_r\dr{v_r}\right)\left[1-\frac{\vsvs}{v_r^2}
        -\frac{1}{\MAro^2}\left(\frac{\vro}{v_r}\right)^3\left(\frac{\Omega
        R}{\vro}\right)^2\frac{\Ue^2}{\UM^3}\frac{a_0}{a}\right]=}\nonumber\\
&& 1 - \frac{\vsvs r}{GM}\left(2+\frac{r\dadr}{a}\right) + \frac{r^2}{GM}
     \dr{\vsvs}-\frac{r^2 \grad}{GM}+\nonumber\\
&& \frac{\Omega^2R^3}{GM}\frac{R}{r}\frac{1}{\UM^2}
    \left(\epsilon-\frac{1}{\MAro^2}\frac{\vro}{v_r}\frac{a_0}{a}\right)
    \left[\left(2\frac{\Ue}{\UM}+1\right)
    \frac{1}{\MAro}\frac{\vro}{v_r}\frac{a_0}{a}-\epsilon\right]-\nonumber\\
&&  \frac{\Omega^2 R^3}{GM}\frac{r}{R}\frac{1}{\MAro^2}\frac{\vro}{v_r}
    \frac{\Ue^2}{\UM^3}\frac{a_0ra^\prime}{a^2}
\end{eqnarray}
In this chapter we neglect the radiative acceleration although we included the
radiation term for reference into Eq.~\ref{momentumrb}. Here we want to
concentrate on the effect of the magnetic field. The CAK model for the line
acceleration acceleration introduces a radiative term which is nonlinear in
$v_r'$. We discuss the consequences of such a term in the next chapter.

For the numerical solution of this equation we introduce dimensionless
variables
\begin{eqnarray}
\label{dimlessa}
y    &=& \frac{v_r}{v_*}\\
u    &=& \frac{r_*}{r}\\
\label{dimlessc}
GM &=& r_*v_*^2.
\end{eqnarray}
Using $u\sim r^{-1}$ instead of $u\sim r$ has the advantage that the region
close to the stellar surface, where the wind is strongly accelerated on a short
length scale is enlarged. This region is the interesting region, where most
things happen. the enlargement in the $u$-scale will ensure proper solutions
for the wind equation in this region. On the other side we have the region far
away from the star, where the wind is streaming outward with a nearly constant
velocity. This region is physically and numerically uninteresting. In the
$u$-scale this region is shrunk. And therefore not much numerical effort will
be spent to solve the wind equation there. With a proper temperature
stratification this region can even extend to $r=\infty$
$(u=0)$. Chap.~\ref{Chap:cak:wd} shows several plots with wind solutions as
functions of $u$, which show nicely that $u$ is the appropriate coordinate to
compute and to plot wind solutions up to large radii.  Our choice for the
length and velocity scale has still one degree of freedom. We can still choose
an arbitrary value for $r_*$ or $v_*$. This of course has no influence on the
physics. But an improper choice can trouble the numerics. One good choice for
fast magnetic rotator models (strong magnetic field, fast rotation) is the
Michel velocity
\begin{equation}
v_\mathrm{M} = \frac{\vro}{\MAro^{2/3}}
               \left(\frac{\Omega R}{\vro}\right)^{2/3}.
\end{equation}
We choose for simplicity just the stellar radius as length scale
\begin{equation}
r_*=R.
\end{equation}
This length scale is reasonable for all wind models -- even for nonrotating or
nonmagnetic wind models.
The dimensionless version of Eq.~\ref{momentumrb} is
\begin{eqnarray}\label{momentumrc}
\lefteqn{y\dydu\left[1-\frac{\ys^2}{y^2}-\frac{\yroto^2}{y^2}\frac{Q}{y}
         \frac{\Ue^2}{\UM^3}\frac{a_0}{a}\right]=}\nonumber\\
&& 1-\frac{\ys^2}{u}\left(2 - \frac{u\dadu}{a}\right)
     -\du{\ys^2} - \frac{\arad}{u^2} +\nonumber\\
&& \frac{\lambda}{\UM^2}\frac{u}{u_0}\left[\epsilon-\frac{Q}{y}\frac{a_0}{a}
     \right]\left[\left(2\frac{\Ue}{\UM}+1\right)\frac{Q}{y}\frac{a_0}{a}-
     \epsilon\right]+\nonumber\\
&& \frac{\lambda}{\UM^2}\frac{u_0}{u}\frac{\Ue^2}{\UM}\frac{Q}{y}
     \frac{a_0u\dadu}{a^2},
\end{eqnarray}
where we have used the dimensionless quantities:     
\begin{eqnarray}
\arad   &=& \frac{r_*^2\grad}{GM}\\
\lambda &=& \frac{\Omega^2 R^3}{GM}\\
\yroto  &=& \frac{\Omega R}{v_*}\\
Q       &=& \frac{y_0}{\MAro^2}\\
\ys     &=& \frac{\vs}{v_*}\\
\dadu   &=& \diff{a}{u}\\
\dydu   &=& \diff{y}{u}
\end{eqnarray}
In analogy to Eq.~\ref{CAKWind} we can write Eq.~\ref{momentumrc} as
\begin{equation}\label{momentumrd}
A(u,y)\dydu+B(u,y) = 0.
\end{equation}
Equation~\ref{momentumrc} is the fluxsheet generalization of the results of
Biermann \& Cassinelli \cite{Biermann:Cassinelli:93} and Biermann
\cite{Biermann:94}. 
\section{Simplified versions}
\label{Sec:simple}
For this model it is important that both, magnetic field and rotation, are
present. If we would turn the rotation off $(\lambda=\yroto=0)$
Eq.~\ref{momentumrb} directly simplifies to the equation of the Parker
wind. The open magnetic field lines would be stretched out radially by the
wind.

If we take rotation into account but assume that the star has no relevant
magnetic field, the conservation of angular momentum (Eq.~\ref{LjdefA}) would
simplify to
\begin{equation}\label{vphinoB}
r\vphi = {\rm const}
\hspace{5mm}\Rightarrow \hspace{5mm} 
\vphi = \frac{v_{\phi 0}R}{r} = \frac{\Omega R^2}{r}.
\end{equation}
For Eq.~\ref{momentumrc} we find then
\begin{equation}
y\dydu\left[1-\frac{\ys^2}{y^2}\right] = 1-\frac{\ys^2}{u}\left(2 - 
\frac{u\dadu}{a}\right)
-\du{\ys^2} - \frac{\arad}{u^2}+\lambda\frac{u}{u_0}.
\end{equation}
This scenario in combination with a CAK radiation force has been analyzed by
Pauldrach et al. \cite{Pauldrach:etal:86}.
\section{The critical points}
\label{Sec:wd:critp}
As in the case of the Parker or CAK wind it is important to understand the
critical points of Eq.~\ref{momentumrc} to find proper physical solutions.
In this model we have not one but three critical points. Under certain
circumstances the solution need not pass through all these points. But
nevertheless it is important to understand them all.

The most important critical point is the Alfv\'enic point. At this point with
$r=\rAc$ the radial wind velocity equals the radial Alfv\'enic velocity
$(\MAr=1)$. And therefore $\UM$ will be zero at this point and change its
sign. If we require that $\Bphi$ is finite and continuous, $\Ue$ must be
zero and change its sign there as well (Eq.~\ref{Bphia}).
For the radius of the Alfv\'enic point and the radial velocity there we find
\begin{eqnarray}
\label{defuAc}
\uAc &=& \frac{u_0}{\sqrt{\epsilon}}\\
\label{defyAc}
\yAc &=& \frac{Q}{\epsilon}.
\end{eqnarray}
As in Chap.~\ref{Chap:parker:wind} we can try to use l'Hospital's rule again
to find $\dydu$ at the critical point:
\begin{equation}\label{dyAcequ}
\left.\frac{\Ue}{\UM}\right|_{r=\rAc}
  = \left.\frac{\dot{U}_\epsilon}{\dot{U}_M}\right|_{u=\uAc}
  = \left.\frac{\frac{\dadu}{a}-\frac{2}{u}}
    {\frac{\dadu}{a}+{\dydu}{y}-\frac{2}{u}}\right|_{u=\uAc}
\end{equation}  
In the case of the Weber \& Davis wind we do not find a finite number of
possible values for $\dyAcdu$. Therefore we do not have a simple X-type
critical point here like in the Parker wind. This local analysis allows a range
of values for $\dyAcdu$. We find
in general
\begin{equation}
\frac{\Ue}{\UM}=-\frac{\Bphi}{B_r}\frac{v_r}{\Omega r}.
\end{equation}
Depending of the values of $v_r$ and $\Bvec$, $\Ue/\UM$ can have different
values. This leads to different values for $\dyAcdu$ according to
Eq.~\ref{dyAcequ}.

Since $\Ue/\UM$ is finite, and $\UM$ equals 0, at $\rAc$ the $\Ue^2/\UM^3$ term
in $A(u,y)$ will dominate for $r\approx\rAc$. For $r\lesssim\rAc$ we have
$A>0$. But if the velocity at the base of the wind is small enough,
$A(u_0,y_0)$ will be less than zero due to the then dominating thermal pressure
term. For $r\gtrsim\rAc$ we have $A<0$. And for sufficient high velocities at
$r=\infty$ $A$ will be positive due to the inertia term. So we have two radii
$R<\rs<\rAc$ and $\rAc<\rf<\infty$ where $A$ will be zero. At these radii we
have a X-type singularity as in the Parker wind. We denote these critical
points and their analogies in the case of an magnetic CAK wind as the inner and
the outer critical points. The condition $A=0$ leads to
\begin{eqnarray}
0 &=& v_r-\frac{\vsvs}{v_r}-\frac{v_r}{\MAro^2}\left(\frac{\vro}{v_r}\right)^3
      \left(\frac{\Omega R}{\vro}\right)^2\frac{\Ue^2}{\UM^3}\frac{a_0}{a}\\
      &=& v_r-\frac{\vsvs}{v_r}-\frac{\Bphi^2}{4\pi\rho v_r}\frac{1}{\UM}\\ &=&
      \UM\left(1-\frac{\vsvs}{v_r^2}\right)-
      \left(\frac{\vA^2}{v_r^2}-\frac{\vAr^2}{v_r^2}\right)\\
\label{DispersionWD}
  &=& v_r^4 - (\vA^2+\vsvs)v_r^2+\vsvs\vAr^2,
\end{eqnarray}
where we have used the total Alfv\'enic velocity
\begin{equation} 
\vA = \frac{B_r^2+\Bphi^2}{4\pi\rho}.
\end{equation}
Equation~\ref{DispersionWD} is the well known dispersion relation for slow and
fast magnetosonic waves \cite{Jackson:Book}. Therefore at the slow and fast
critical point the radial wind velocity $v_r$ equals the local phase
velocity for slow respectively fast magnetosonic waves. There is a deep
physical connection to the problem of causality in a stationary physical
model. Weber \& Davis \cite[Figs.~1\&2]{Weber:Davis:67} described the topology
for the solution Eq.~\ref{momentumrd} and the role of the critical points. In
the next chapter we combine the Weber \& Davis model in the form presented here
with the CAK model. There we will discuss the critical points and the
solution topology in detail.
\section{The spin-down problem}
\label{Sec:spindown}
One important aspect in the theory of magnetic winds is the angular momentum
balance. Every star which rotates and loses mass will lose angular momentum
as well, because the rotating surface layer of the star carries a certain
fraction of the stellar angular momentum. When this layer evaporates into the
wind it will take its angular momentum along. Due to a magnetic field or
viscosity the angular momentum loss may be enhanced. We expressed this by the
factor $\epsilon$ in our formula for the angular momentum loss per unit mass
(Eq.~\ref{LjdefB}). A normal star without an enhanced angular momentum loss
has $\epsilon=1$. The angular momentum loss rate\footnote{$\dot{L}=\partial
L/\partial t$ is beside $\Mdot$ the second exception from the rule that the dot
$\dot{ }$ marks a derivative with respect to $u$.} of a star is
\begin{equation}\label{LdotdefA}
\dot{L} = \int L_j \diff{\Mdot}{\omega}\,d\omega,
\end{equation}
where $L_j$ is the angular momentum loss per unit mass (Eq.~\ref{LjdefA}) and
$\omega$ is the solid angle. In order to evaluate Eq.~\ref{LdotdefA} properly
we would have to calculate a whole 3-dimensional wind model. This is is beyond
this thesis. So we have to assume the star is sufficiently spherically
symmetric at its surface, so that we can use the results from our equatorial
wind model. This assumption is only good if the rotation rate is sufficiently
low. Since this is not the case for most of our models, the results of this
section should only be used carefully as a rough estimate. We find under these
assumptions from Eq.~\ref{LdotdefA}
\begin{equation}\label{LdotdefB}
\dot{L} = \frac{2}{3}\epsilon\Mdot R^2\Omega = \frac{2}{3}\Mdot\rAc^2\Omega.
\end{equation}
The interesting point now is to compare the timescales for mass loss
\begin{equation}
\tau_M=\frac{M}{\Mdot}
\end{equation}
and angular momentum loss 
\begin{equation}\label{tauLdefA}
\tau_L=\frac{L}{\dot{L}}.
\end{equation}
We know from observation and stellar evolution models that massive stars lose
a large fraction of their mass before they explode as a supernova. I.e.\ their
mass loss timescale is comparable to their lifetime. If rotation plays an
important role through the whole life of a massive star, its angular momentum
loss timescale should not be much smaller. Otherwise the star will spin-down
already in an early stage of its evolution. The problem now is to estimate $L$,
the total angular momentum of our star. Today no models for the inner structure
of rapidly rotating stars with significant magnetic fields are available. The
most simple by extremely unphysical assumption is a rigidly (!) rotating star
of constant (!) density.  In this case we have $L=(5/2)MR^2\Omega$ and
therefore
\begin{equation}\label{tauLdefB}
\tau_L=\frac{3}{5}\frac{\tau_M}{\epsilon}.
\end{equation}
This result suggests that every star has an angular momentum loss timescale
smaller than the mass loss timescale. The naive interpretation would be that
even a star without magnetic field $(\epsilon=1)$ would spin down rapidly. But
of course, a star without angular momentum transport between different layers
of material will not spin down at all, if we assume that it does not expand or
shrink. If the angular momentum is conserved in every layer then every layer
will keep to overall rotation rate. Due to all the simplifications
Eq.~\ref{tauLdefB} allows to criticize wind models only if $\tau_M$ and $\tau_L$
differ by more than one or two orders of magnitude.  A final answer for the
spin-down problem requires a unified model which describes the evolution of the
star and its wind for all latitudes in detail over the whole lifetime. This is
beyond todays possibilities.

Therefore our wind models are reasonable as long as they have a small value of
$\epsilon$. But what is the physical meaning of $\epsilon$? It is the factor by
which the angular momentum loss is enhanced compared with the case of no
magnetic field. From our discussion in the previous section we see that the
radius of the Alfv\'enic point is given by $\rAc^2=\epsilon R^2$. If we make
the quite physical assumption that the magnetic field transports angular
momentum from the star away, we will always have $\epsilon>1$ and therefore
$\rAc>R$. A star with $\epsilon$ less than 1 would lose less angular momentum
than a star without magnetic field. If we assume that at the stellar surface
(or somewhere just below) we have a layer of stellar material without
significant outward motion $(v_r=0)$, we will have strict co-rotation there due
to the flux freezing. When this layer evaporates into the wind, the magnetic
field would have to transport angular momentum back into the star to keep the
angular momentum loss less than in the case of a nonmagnetic star. Since the
nonmagnetic star rotates with a constant rotation rate, a magnetic star with
$\epsilon$ less than 1 would spin up like a rotating firework.
\begin{figure}
\begin{center}
\begin{picture}(120,85)
%\put(0,0){\framebox(120,85){}}
\put(0,0){\epsfig{file=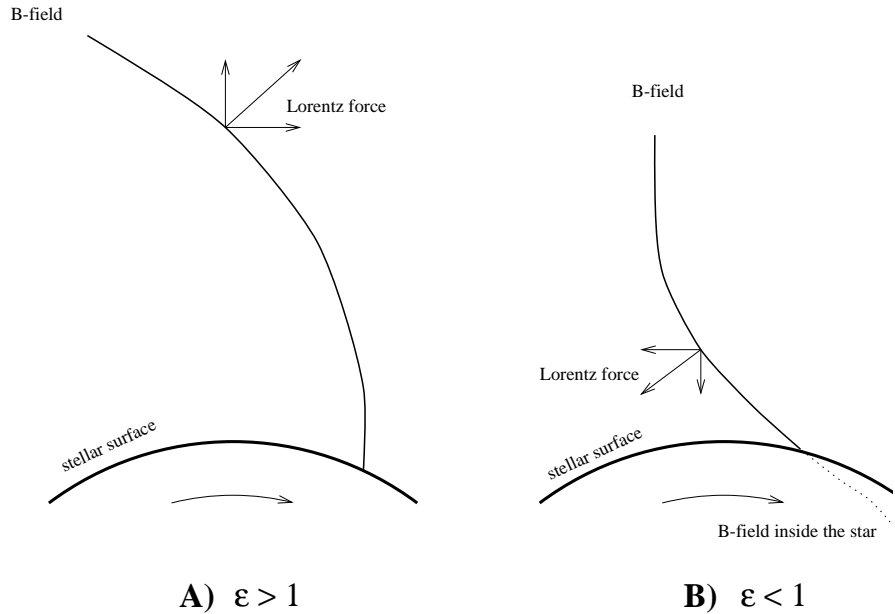,width=12cm}}
\end{picture}
\end{center}
\caption[Wind models with $\epsilon>1$ and $\epsilon<1$.]{\label{f:epsilon}
shows the magnetic field configuration in the equatorial plane of wind models
with $\epsilon>1$ (A) and $\epsilon<1$ (B). }
\end{figure}
Figure~\ref{f:epsilon} shows shows the magnetic field configuration in the
equatorial plane of wind models with $\epsilon>1$ (A) and $\epsilon<1$ (B).  In
the first case (A) the magnetic field line pushes the wind material outside and
in the direction of the rotation. The first effect enhances the wind
outflow. The second effect transfers extra angular momentum from the magnetic
field to the wind matter. This extra angular momentum comes via the curvature
of the field line from the star. In the second case the magnetic field line is
mostly tangential at the stellar surface. The wind accelerated by radiation
pressure pushes against the field line. This bends the field line
clockwise. Through this curvature angular momentum is transferred from the wind
material via the field line back to the star. Therefore we have $\epsilon$ less
than 1. The consequence of the curved field line is a Lorentz force which
decelerates the wind in radial and azimuthal direction. If this effect is very
strong the wind can even start to rotate counterclockwise. In any case
scenarios of type B have a decelerating Lorentz force close to the stellar
surface. The force balance close to the surface fixes the mass loss
rate. Therefore wind models of type B will have a lower mass loss rate than the
corresponding model without magnetic field. It is not even clear whether a
configuration of type B is stable, because the field line has to be bent
clockwise inside the star (dotted line in Fig.~\ref{f:epsilon}.B). Otherwise it
would hit the clockwise bent stellar surface again. This would violate our
assumption of rotational symmetry. We therefore conclude that $\epsilon$ must
be larger than 1 in physical wind models with magnetic fields. And we will
restrict our further analysis to that case.  \vfill
%
 
%-*-LaTeX-*-
% This is the 5th chapter for the PhD-thesis of Henning Seemann.
% (c) 1997-1998 by Henning Seemann
%
\chapter{Magnetic CAK-winds}
\label{Chap:cak:wd}
\section{Introduction}
In this chapter we discuss various numerical solutions for the equatorial wind
model developed in the previous chapters. Where possible we compare our results
with the literature. The combined wind equation for line driven winds with
magnetic fields is
\begin{equation}\label{CAKWDWind}
A(u,y)\dydu+B(u,y) = C(u,y) \dydu^\alphacak.
\end{equation}
We know this form of the wind equation already from our discussion of
nonmagnetic, line driven winds in Chap.~\ref{Chap:cak}
(Eq.~\ref{CAKWind}). But now we use the dimensionless quantities
(Eqs.~\ref{dimlessa}--\ref{dimlessc}) and extend the coefficients $A$ and $B$
by the magnetic terms derived in Chap.~\ref{Chap:wd} (Eq.~\ref{momentumrc}):
\begin{eqnarray}
A(u,y) &=& y\left[1-\frac{\ys^2}{y^2}-\frac{\yroto^2}{y^2}\frac{Q}{y}
         \frac{\Ue^2}{\UM^3}\frac{a_0}{a}\right]\\
B(u,y) &=& 1-\Gamma-\frac{\ys^2}{u}\left(2 - \frac{u\dadu}{a}\right)
       -\du{\ys^2} +\nonumber\\
&& \frac{\lambda}{\UM^2}\frac{u}{u_0}\left[\epsilon-\frac{Q}{y}\frac{a_0}{a}
     \right]\left[\left(2\frac{\Ue}{\UM}+1\right)\frac{Q}{y}\frac{a_0}{a}-
     \epsilon\right]+\nonumber\\
&& \frac{\lambda}{\UM^2}\frac{u_0}{u}\frac{\Ue^2}{\UM}\frac{Q}{y}
     \frac{a_0u\dadu}{a^2}\\
\label{CAKWDWindC}
C(u,y) &=& \Gamma\kcak\left(\frac{4\pi GM}{\kappaTh\Mdot\vth}
           ay\right)^\alphacak.
\end{eqnarray}
Eq.~\ref{CAKWDWindC} distinguishes from Eq.~\ref{CAKTermCb} beside the overall
factor $r^2/GM$ by the geometry factor $a(u)^\alphacak$. This factor comes from
the equation of continuity, which we used between Eq.~\ref{CAKTermCa} and
Eq.~\ref{CAKTermCb}. The geometry factor $a(u)$ was not introduced before
chapter ~\ref{Chap:wd}.

A solution of this wind equation can pass through up to three critical
points. In Sec.~\ref{Sec:wd:critp} we discussed the three critical points of
a magnetic wind without line acceleration (i.e. $C=0$). The Alfv\'enic critical
point, we discussed there, exists in a line driven, magnetic wind as well. And
our arguments, that $A$ will pass through zero somewhere inside and outside the
Alfv\'enic point, are still valid. But since the we have now the line driving
term $C\dydu^\alphacak$ in our Eq.~\ref{CAKWDWind}, we can not use the old
Parker argument that $A(u,y)=0$ implies $B(u,y)=0$. Therefore we do not have
the classical X-type critical point at the radii where $A(u,y)$ is zero. Rather
$A$ and $B$ change their signs in the subtle way we described for the critical
point of nonmagnetic CAK-wind in Sec.~\ref{Sec:cak:cakp}. So we have two
critical points of the CAK-type.

In order to compute a numerical model we have to fix several model parameters.
The star is described by its mass $M$, its luminosity $L$, its radius $R$, its
rotation rate $\Omega$, and its surface radial magnetic field strength
$\Bro$. For the wind we have to specify the CAK-parameters $\alphacak$ and
$\kcak$, the temperature profile $T(r)$, the electron scattering opacity
$\kappaTh$, and the meridional shape of the wind described here by the area
function $a(r)$. When we put all these numbers into
Eqs.~\ref{CAKWDWind}--\ref{CAKWDWindC} we find that two further quantities
remain to be fixed. E.g.\ we can specify additionally the mass loss rate
$\Mdot$ and the radius of the Alfv\'enic point $\rAc$. These two parameters are
very crucial because they allow to relate our numerical models to observation
and to control the stellar angular momentum loss (Eq.~\ref{LdotdefB}), which is
important for the stellar evolution. Therefore it would be nice, if these
parameters were rather results than input parameters of our models. This can be
realized, if we require that our wind solution has to pass through all three
critical points. Through this additional assumption $\Mdot$ and $\rAc$ become
eigenvalues of our wind equation. We find these eigenvalues using the shooting
method. For given values of $\Mdot$ and $\rAc$ we can find the two CAK-type
points by solving twice the nonlinear set of equations
\ref{CAKWDWind},\ref{CAKpointB}\&\ref{CAKpointC}. From these points we can
integrate Eq.~\ref{CAKWDWind} to the Alfv\'enic point
(Eqs.~\ref{defuAc}\&\ref{defyAc}), to the stellar surface, and to infinity.  In
Sec.~\ref{Sec:wd:critp} we argued that the acceleration $\dydu$ is not fixed a
priori at the Alfv\'enic critical point. But, of course, $\dydu$ should be
continuous at the Alfv\'enic point. Therefore we should find the same value of
$\dydu(\uAc)$ when we integrate Eq.~\ref{CAKWDWind} from the inner and from the
outer CAK-point to the Alfv\'enic point. This is the first condition for the
eigenvalues $\Mdot$ and $\rAc$. The second condition is related to the stellar
surface. We defined a stellar radius $R$ as the radius at the base of the
wind. In our models we require a certain, small, initial wind velocity $y_0$ at
this radius. This condition is very reasonable from the theoretical point of
view, because a theoretical wind model should describe, how the wind material
is accelerated from the low, subsonic velocity of the quasi static, stellar
atmosphere to the high observed terminal velocities.  Alternatively we could
require a certain optical depth (e.g. $\tau_0=1$) at the stellar radius $R$.
For stars with an optical thin wind like O and B stars these two alternative
definitions of $R$ are practical identical. For stars with an optical thick
wind like Wolf-Rayet stars the optical depth at the subsonic base of the wind
is not known. If we would start our wind model for a Wolf-Rayet star at
$\tau=1$, we would describe only the outer, high velocity part of the
wind. Such a model could not be considered as a complete wind
model. Additionally specifying $\tau_0$ requires more numerical effort than
specifying $y_0$. Therefore we prefer to specify $y_0$.

In Sec.~\ref{Sec:FMG} we compare our results with the model of Friend \&
MacGregor \cite{Friend:MacGregor:84}, who published a model with
a different wind equation. In Sec.~\ref{Sec:twopoints} we our model to the case
of only two critical points.  This is important for our study of winds outside
the equatorial plane in Chap.~\ref{Chap:fluxtube}.
\section{The wind models of Friend \& MacGregor revisited}
\label{Sec:FMG}
We start our discussion with wind models which have no compression or expansion
of the wind in meridional direction, i.e.\ the cross section function $a(r)$ as
defined in Eqs.~\ref{FluxConservationa}\&\ref{MassConservationMHDa} is
constant. Additionally we assume that the wind passes through all three
critical points.  Such a model can be compared to the model published by Friend
\& MacGregor \cite{Friend:MacGregor:84}. Their model is based on the CAK model
for the radiative acceleration and the Weber Davis model for the magnetic
acceleration (their Eqs.~1--21). Up to this point their model is identical with
our model. But from these equations they derive a final differential equation
for $v_r$ (their Eqs.~22--26) which differs in the magnetic terms from our
final equation (Eqs.~\ref{CAKWDWind}--\ref{CAKWDWindC}). And this is not only a
question of different notation. Our final equation is derived strictly from the
common basic assumptions (their Eqs.~1--21) without any further assumptions or
approximations. Due to the necessary brevity in an article Friend \& MacGregor
give no further information how they obtain their final equation from the basic
equations. Therefore we can not analyze the differences between the final
equation of Friend \& MacGregor and our final equation in further details. To
study the differences of our models in the numerical level we recalculated
their numerical models with our software using our final wind equation. They
calculated 16 models with varying magnetic field and rotation rate for the O6ef
star $\lambda$ Cephei. For this star they used the parameters: $M=50\Msun$,
$L=6.76\times 10^5\Lsun$, $R=19.7\rsun$. The sound speed $\vs=30\,\km/\s$ is
constant in the wind. This leads to a wind temperature of about
$65000\,\Kelvin$. For the radiative acceleration they claim to use the original
CAK \cite{Castor:etal:75} values $\alphacak=0.7$ and $\kcak=1/30$. From the
statement that the CAK terminal velocity (their Eq.~30, our Eq.~\ref{vinfCAK})
is $1240\,\km/\s$ we can derive a value of $\Gamma=0.319$ for the Eddington
factor. Therefore we find for the electron scattering opacity per unit mass
$\kappaTh=0.309\,\cm^2/\gram$ (called $\sigma_e$ by Friend \& MacGregor). For
the CAK mass loss rate (their Eq.~31, our Eq.~\ref{MdotCAK}) they quote a value
of $5.2\times 10^{-6}\Msun/\yr$. For their parameters we find a value of
$9.74\times 10^{-7}\Msun/\yr$. We find the same discrepancy in the mass loss
rates of the numerical models. We can fix this by assuming a modified CAK
parameter $\kcak=0.107$ instead of $\kcak=1/30$. Finally we can deduce from
their Tab.~2.B that their reference radius for the terminal velocity of their
numerical models is 200 times the Alfv\'enic radius. For all numerical models
we state the rotation rate in units of the critical rotation rate
\begin{equation}
\alpharot = \frac{\vrot}{\vrotcrit} = \frac{\Omega}{\Omegacrit} = 
\frac{\Omega R}{\sqrt{\frac{GM}{R}(1-\Gamma)}}.
\end{equation}
$\alpharot$ gives a good intuitive feeling, whether the star rotates fast or
slow. In the case $\alpharot=1$ the centrifugal and radiative (Thompson
scattering) forces will compensate gravity at the equator. An eruptive mass
loss would be the consequence. This case is called the $\Omega$-limit by Langer
\& Heger \cite{Langer:Heger:97}. They claim that stars can really reach this
limit. 

\begin{figure}[ht]
\begin{center}
\begin{picture}(120,75)
%\put(0,0){\framebox(120,75){}}
\put(0,2){\epsfig{figure=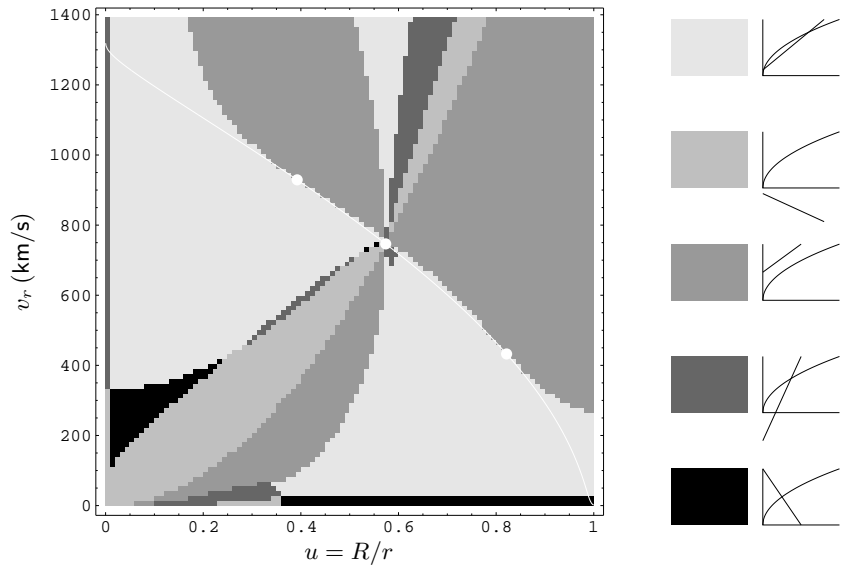,width=12cm}}
\put(44,1){\small$u=R/r$}
\put(5,34){\begin{sideways}\small$v_r\,(\km/\s)$\end{sideways}}
\end{picture}
\caption{\label{f:FM:m01:a}shows the radial wind velocity $v_r$ (white curve)
as function of $u=R/r$ for our version of Friend \& MacGregor's model 1 (weak
$\Bvec$ field \& slow rotation). The three critical points (from left to right:
the outer, the Alfv\'enic, and the inner critical point) are marked by white
dots. Additionally we plot, using different gray scales, for every point in the
$(u,v_r)$ plane the local solution logic for Eq.~\ref{CAKWDWind}, which is the
magnetic version of Eq.~\ref{CAKWind}. The ``local solution logic'' is
explained for the nonmagnetic case in Sect.~\ref{Sec:cak:cakp} and
Fig.~\ref{CAKSolutions}. The different cases in Fig.~\ref{CAKSolutions} are
connected to the different grayscales in this plot by the legend on the right
hand side.  In the magnetic case we have two critical points of the
(nonmagnetic) CAK-type: The inner and outer critical points. At the Alfv\'enic
point Fig.~\ref{CAKSolutions}.F applies. The wind solution shown in this plot
passes from the stellar surface $(u=1)$ to infinity $(u=0)$ through the
following sequence (c.f.\ Fig~\ref{CAKSolutions}: C--E--A--B(here is the inner
critical point)--A--D--F(here is the Alfv\'enic critical point)--C--A--B(here
is the outer critical point)--A--D. The situation around the Alfv\'enic point
can be seen better in Fig.~\ref{f:FM:m13:a}.}
\end{center}
\end{figure}

\begin{figure}[p]
\begin{center}
\begin{picture}(120,75)
%\put(0,0){\framebox(120,75){}}
\put(0,2){\epsfig{figure=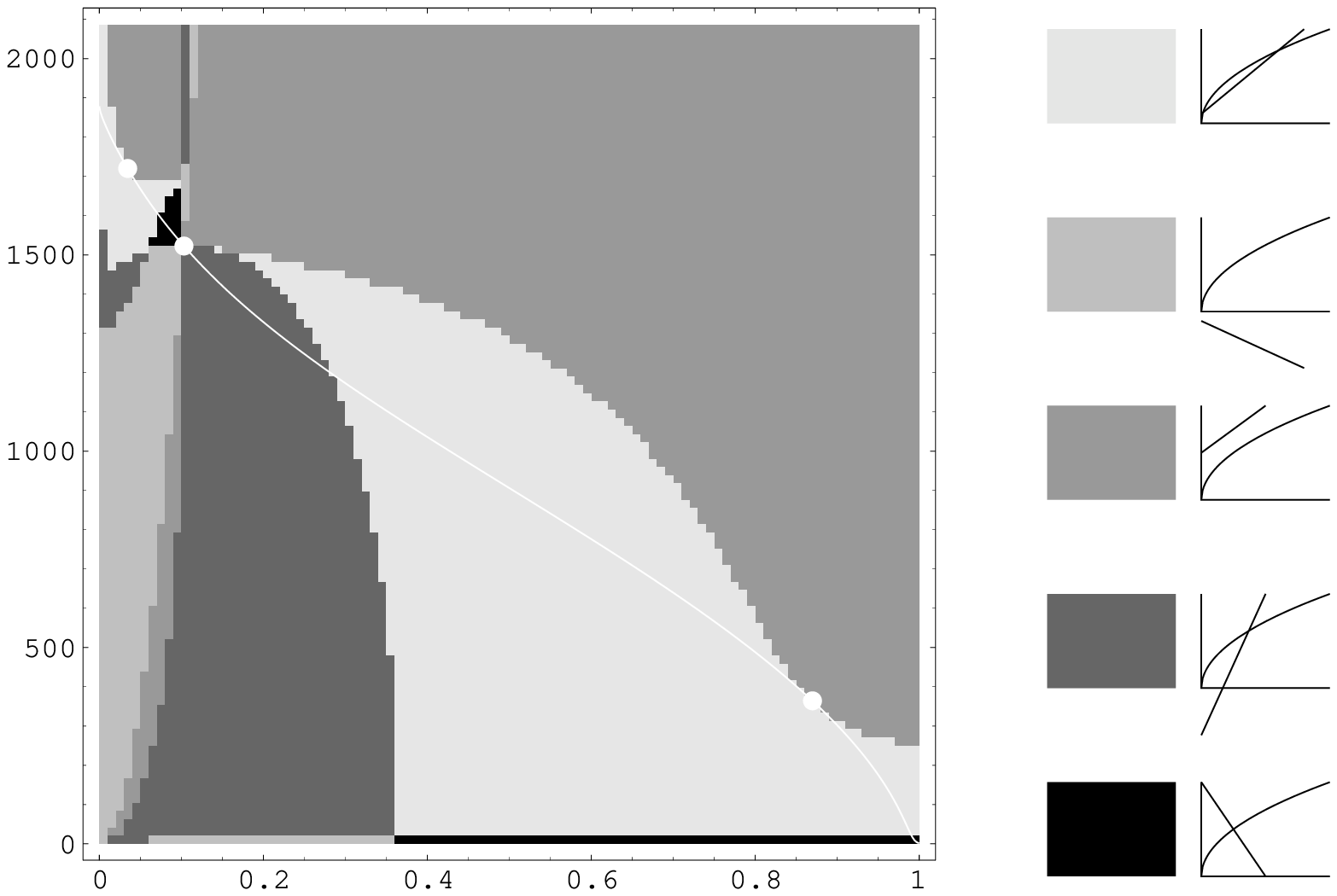,width=12cm}}
\put(44,1){\small$u=R/r$}
\put(5,34){\begin{sideways}\small$v_r\,(\km/\s)$\end{sideways}}
\end{picture}
\caption{\label{f:FM:m04:a}shows the same as Fig.~\ref{f:FM:m01:a} but for our
version of Friend \& MacGregor's model 4 (Strong $\Bvec$ field \& slow
rotation). For the technical explanation of this plot check
Fig.~\ref{f:FM:m01:a}.}
\end{center}
\end{figure}

\begin{figure}[p]
\begin{center}
\begin{picture}(120,75)
%\put(0,0){\framebox(120,75){}}
\put(0,2){\epsfig{figure=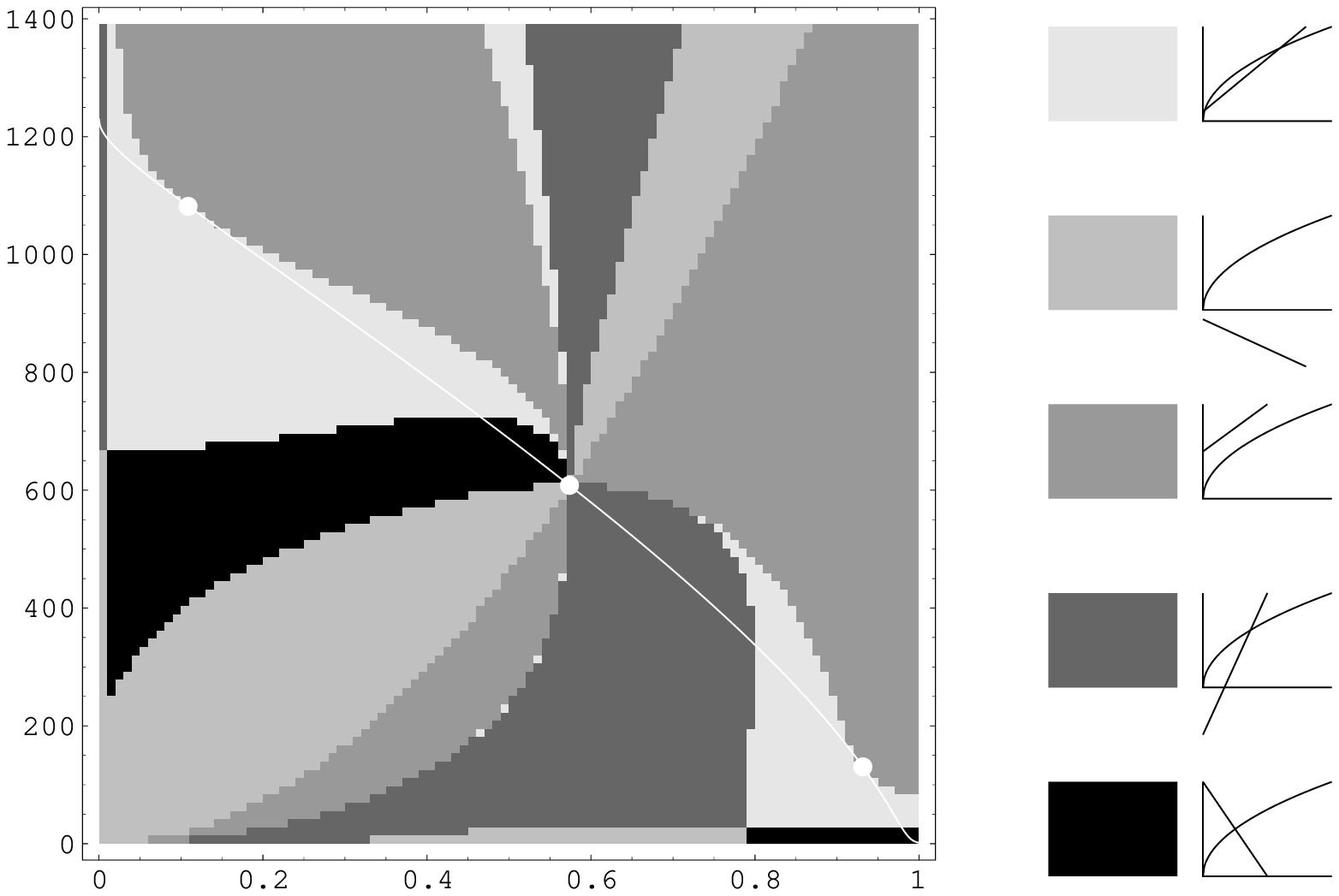,width=12cm}}
\put(44,1){\small$u=R/r$}
\put(5,34){\begin{sideways}\small$v_r\,(\km/\s)$\end{sideways}}
\end{picture}
\caption{\label{f:FM:m13:a}shows the same as Fig.~\ref{f:FM:m01:a} but for our
version of Friend \& MacGregor's model 13 (weak $\Bvec$ field \& fast
rotation). For the technical explanation of this plot check
Fig.~\ref{f:FM:m01:a}.}  
\end{center}
\end{figure}

\begin{figure}[p]
\begin{center}
\begin{picture}(120,75)
%\put(0,0){\framebox(120,75){}}
\put(0,2){\epsfig{figure=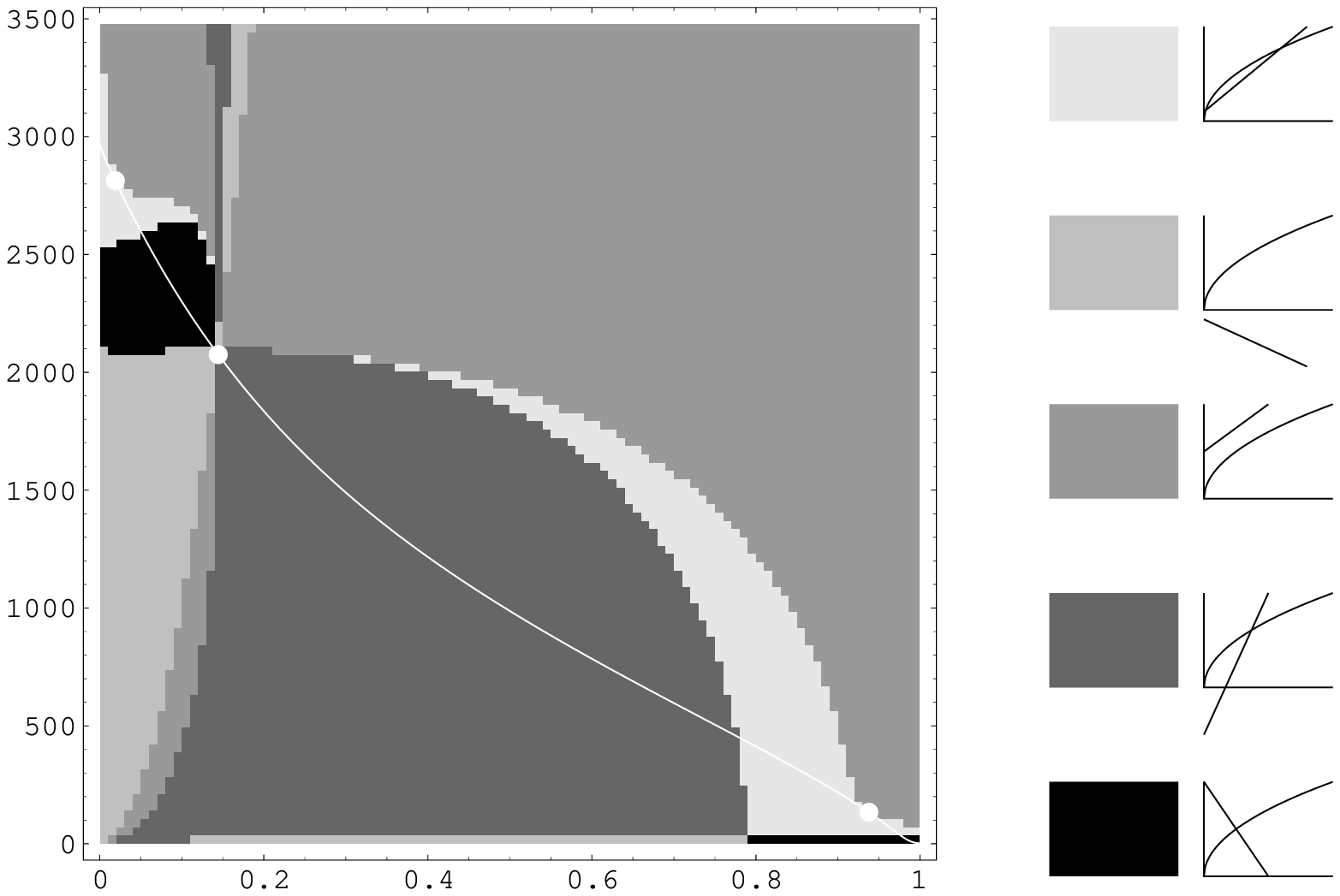,width=12cm}}
\put(44,1){\small$u=R/r$}
\put(5,34){\begin{sideways}\small$v_r\,(\km/\s)$\end{sideways}}
\end{picture}
\caption{\label{f:FM:m16:a}shows the same as Fig.~\ref{f:FM:m01:a} but for our
version of Friend \& MacGregor's model 16 (Strong $\Bvec$ field \& fast
rotation). For the technical explanation of this plot check
Fig.~\ref{f:FM:m01:a}.}  
\end{center}
\end{figure}

In Table \ref{Tab:FMResults} we display analogously to Friend \& MacGregor's
Tab.~1 our results for the 16 wind models extended by our values for the wind
efficiency and the ratio between terminal velocity and the total Alfv\'enic
velocity.
\begin{table}
\caption[Friend \& MacGregor's \cite{Friend:MacGregor:84} models recalculated]{
\label{Tab:FMResults}Wind models for $\lambda$ Cephei analogous to the models 
of Friend \& MacGregor \cite{Friend:MacGregor:84}.}
\begin{center}
\begin{tabular*}{13cm}{*{11}{r@{\extracolsep{\fill}\hspace{1mm}}}}
\hline
\hline
\rule[-3mm]{0mm}{8mm}
no. & $\Bpo\atop(\Gauss)$ & $\alpharot$ & $\frac{r_{c1}}{R}$ & $\frac{\rAc}{R}$
& $\frac{r_{c2}}{R}$ & $\vAc\atop(\km/\s)$ &
$\vinf\atop(\km/\s)$ & ${10^6\,\Mdot}\atop(\Msun/\yr)$ &
$\frac{\Mdot\vinf}{L/c}$ & $\frac{v_\infty}{v_{\mathrm{A}\infty}}$\\
\hline
% Dies sind meine Ergebnisse in Analogie zu Friend & MacGregor Apj 282 Tab 1.
%no   Bp0    arot    rcn     rc     rcf    vac   vinf   Mdot     eff   y/yA
 1 &  200 & 0.218 & 1.22 & 1.74 &  2.54 &  746 & 1305 & 5.26 & 0.507 & 7.90\\
 2 &  400 & 0.218 & 1.19 & 2.97 &  4.54 & 1024 & 1395 & 5.28 & 0.543 & 4.38\\
 3 &  800 & 0.218 & 1.17 & 5.32 & 10.29 & 1255 & 1553 & 5.31 & 0.606 & 2.58\\
 4 & 1600 & 0.218 & 1.15 & 9.70 & 28.98 & 1523 & 1871 & 5.33 & 0.732 & 1.71\\
 5 &  200 & 0.430 & 1.11 & 1.74 &  3.79 &  719 & 1315 & 5.50 & 0.534 & 4.14\\
 6 &  400 & 0.430 & 1.10 & 2.84 &  6.43 & 1054 & 1530 & 5.60 & 0.631 & 2.63\\
 7 &  800 & 0.430 & 1.09 & 4.87 & 14.77 & 1410 & 1890 & 5.71 & 0.792 & 1.82\\
 8 & 1600 & 0.430 & 1.08 & 8.34 & 40.42 & 1896 & 2534 & 5.79 & 1.076 & 1.42\\
 9 &  200 & 0.610 & 1.08 & 1.74 &  6.20 &  660 & 1271 & 5.98 & 0.562 & 2.90\\
10 &  400 & 0.610 & 1.08 & 2.73 &  8.90 & 1022 & 1577 & 6.28 & 0.730 & 2.05\\
11 &  800 & 0.610 & 1.07 & 4.47 & 18.97 & 1458 & 2070 & 6.54 & 0.995 & 1.57\\
12 & 1600 & 0.610 & 1.07 & 7.40 & 48.11 & 2070 & 2894 & 6.73 & 1.428 & 1.32\\
13 &  200 & 0.697 & 1.07 & 1.74 &  9.14 &  608 & 1213 & 6.45 & 0.581 & 2.46\\
14 &  400 & 0.697 & 1.07 & 2.66 & 11.15 &  969 & 1551 & 6.96 & 0.796 & 1.84\\
15 &  800 & 0.697 & 1.07 & 4.26 & 21.91 & 1426 & 2086 & 7.39 & 1.132 & 1.48\\
16 & 1600 & 0.697 & 1.07 & 6.92 & 52.60 & 2075 & 2961 & 7.67 & 1.667 & 1.28\\
\hline
\end{tabular*}
\end{center}
\end{table}
We find larger differences only for the radius of the slow $(\approx7\%)$ and
fast $(\approx20\%)$ critical points. In the first case this is due to our
slightly different inner boundary condition. We require a fixed radial velocity
$(\vro=\vs/20)$ at the photosphere. Friend \& MacGregor required an optical
depth for electron scattering of $\taues=1$ at the base of the wind. In the
second case the differences are due to the bad numerical conditioning for
finding the far critical point. All other values in Tab. \ref{Tab:FMResults}
differ less than 3\% between Friend \& MacGregor's and our calculations. This
difference is too small to distinguish any physical from numerical
effects. Therefore we have to conclude that we found the same wind solutions as
Friend \& MacGregor although we used a different wind equation. Their
additional assumptions or simplifications, which lead to their Eqs.~22--26,
appear to be well chosen. For a deeper physical interpretation of these results
we refer the reader to the paper of Friend \& MacGregor, since we do not want
to reprint all their arguments here in detail. In order to check our inner
boundary condition we calculated models with different values of $\vro$ up to
$3\vs$. We found that this has only a very small influence on the wind
models. The model parameter like $\Mdot$ and $\vinf$ changed not more than
10\%. This shows that our boundary condition at the base of the wind is not
critical for our results. For most models they changed only a few percent. This
weak dependence on the inner boundary condition explains also why we got the
same results as Friend and MacGregor although they use a completely different
inner boundary condition. Figs.~\ref{f:FM:m01:a}--\ref{f:FM:m16:a} show the
velocity profiles for four of our models plotted as function of $R/r$ (the
white curve). These figures show the different regions in the $(u=R/r,v_r)$
plane, which have zero, one, or two solutions for Eq.~\ref{CAKWDWind}. The
position of the three critical points are marked by white points. Tracing the
solution from the stellar surface $(R/r=1)$ to infinity $(R/r=0)$ we can see
how $A(u,y)$ and $B(u,y)$ evolve through the wind and why the solution has to
pass through the two CAK-type critical points, if we require, that the wind
starts at a subsonic velocity and extends to arbitrary large radii.  

The interesting point in this section is that the approximations made by Friend
\& MacGregor leading to their version of the wind equation seems to be quite
appropriate. It is not clear whether this is still true in the case of higher
rotation rates.  We will use higher rotation rates in Chap.~\ref{Chap:fluxtube}
for our non-equatorial models.

\begin{figure}[p]
\begin{center}
\begin{picture}(120,75)
%\put(0,0){\framebox(120,75){}}
\put(0,2){\epsfig{figure=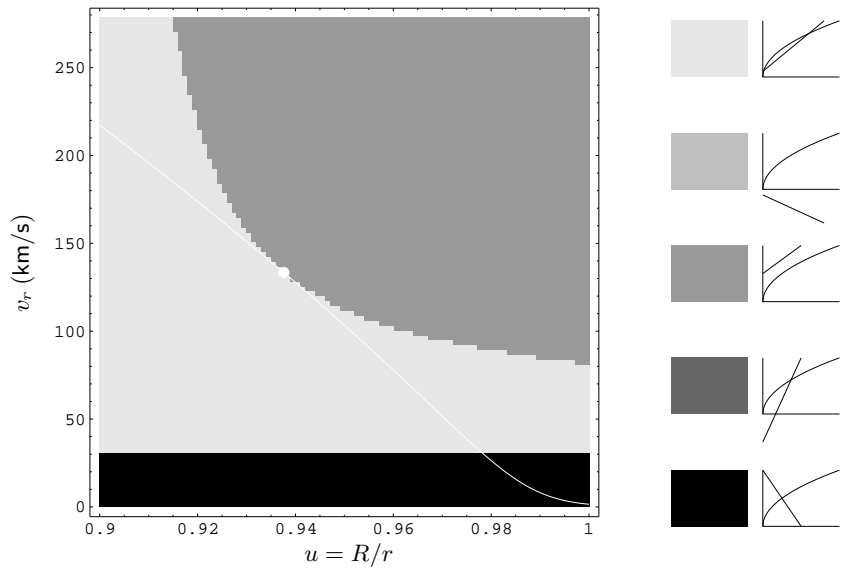,width=12cm}}
\put(44,1){\small$u=R/r$}
\put(5,34){\begin{sideways}\small$v_r\,(\km/\s)$\end{sideways}}
\end{picture}
\caption{\label{f:FM:m16:b}shows the inner part of our wind model 16
(Fig.~\ref{f:FM:m16:a}). The region with two solutions extends down to the
stellar surface at $u=1$. The wind solution can only extend into the subsonic
part, after it has passed through the inner critical point. For the technical
explanation of this plot check Fig.~\ref{f:FM:m01:a}.}
\end{center}
\end{figure}
\begin{figure}[t]
\begin{center}
\begin{picture}(120,75)
%\put(0,0){\framebox(120,75){}}
\put(0,2){\epsfig{figure=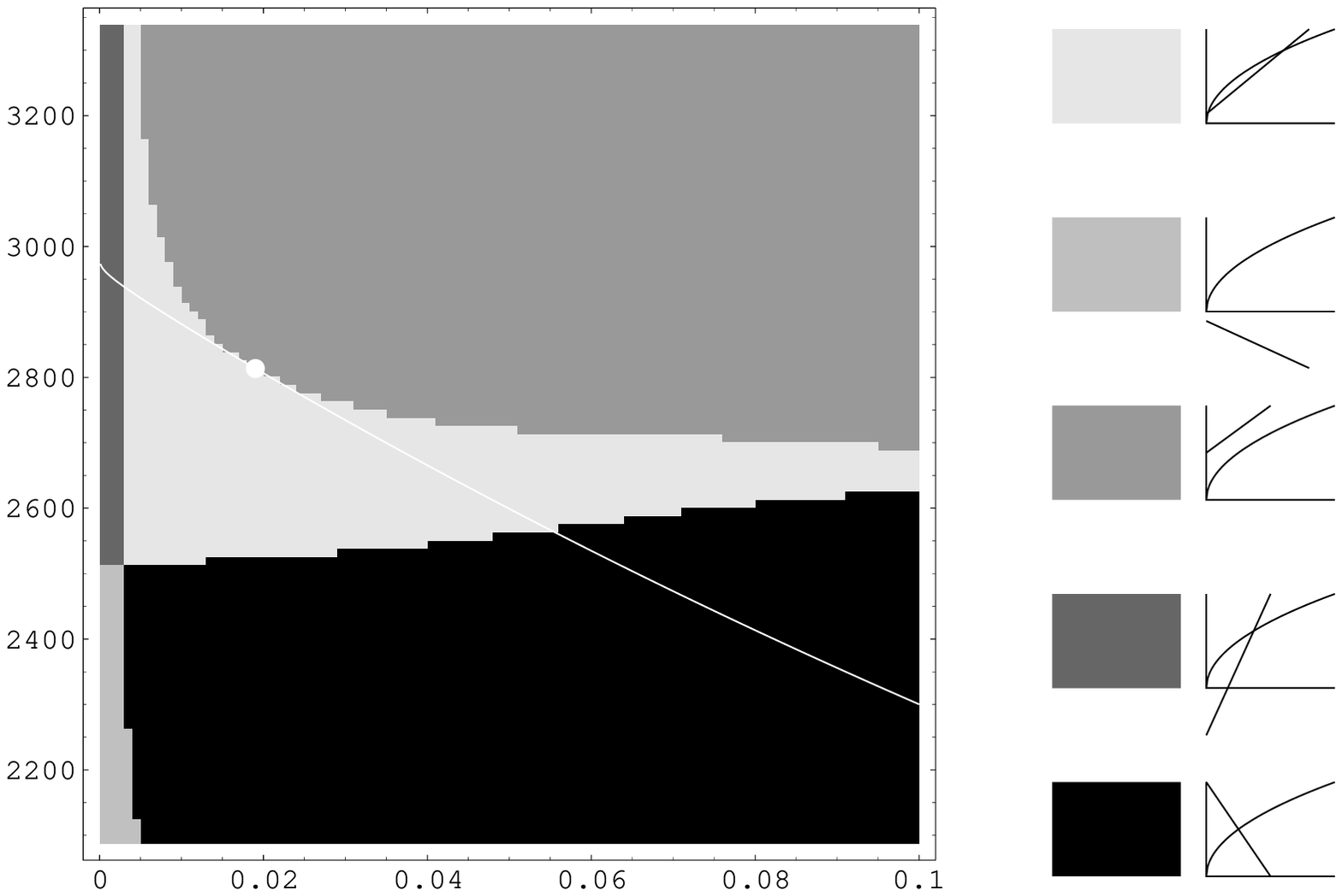,width=12cm}}
\put(44,1){\small$u=R/r$}
\put(5,34){\begin{sideways}\small$v_r\,(\km/\s)$\end{sideways}}
\end{picture}
\caption{\label{f:FM:m16:d}shows the outer part of our wind model 16
(Fig.~\ref{f:FM:m16:a}). The wind solution can only extend beyond
$u\approx0.05$ $(r\approx200R)$ after it has passed though the outer critical
point. For the technical explanation of this plot check
Fig.~\ref{f:FM:m01:a}.}
\end{center}
\end{figure}
\section{Models with only two critical points}
\label{Sec:twopoints}
When we look at Fig.~\ref{f:FM:m16:b} we see that the region where
Eq.~\ref{CAKWDWind} has two solutions extends down to the stellar surface --
but only with a supersonic wind velocity. In Sec.~\ref{Sec:wd:critp} we
argued that the wind has to pass through the inner critical point, because only
if we switch the solution branch at the inner critical point the solution can
extend into the subsonic part. But it is not clear whether our wind equation is
still valid in the subsonic regime. Effects like turbulence, optical thickness,
pulsation, meridional motions or other might dominate the wind there. In this
case we might get wrong results if we try to describe the wind down to the
subsonic regime with our model. Nevertheless if we restrict our calculations to
the supersonic regime we still have to start close to the star with a wind
velocity small compared to the terminal velocity. Otherwise our model is
irrelevant because it does not answer the fundamental questions of wind
physics: (1) How does the wind material escape from the gravity of the star?
And (2) how is it accelerated to the high observed terminal velocities?

To analyze this question we modified our program, so that the wind solutions
pass only through the outer and the Alfv\'enic critical points. Through this
step we lose one of our conditions for the eigenvalues of
Eq.~\ref{CAKWDWind}. Now we have to specify either the acceleration at the
Alfv\'enic point or the wind velocity at the stellar surface in order to
integrate Eq.~\ref{CAKWDWind} between the stellar surface and the Alfv\'enic
point. Therefore we do not find a unique solution any more. Rather we find a
set of solutions with $\Mdot$ as a function of $\rAc$ or vice versa.

We calculated solutions for the four corner models of Friend \& MacGregor
(models 1,4,13,16).  Hereby we used an initial wind velocity at the stellar
surface of $v_{r0}=1.3\,\vs=39\,\km/\s$. In Tab.~\ref{Tab:FM:noinner} we list
for the four models the solutions with the smallest and the largest radius of
the Alfv\'enic point $(\rAc)$. As reference level for the terminal velocity
$\vinf$ we chose $100R$. This allows us to compare these results with the
models in the next paragraph.
\begin{table}[t]
\caption[Wind models without inner critical point]{
\label{Tab:FM:noinner}Wind models for $\lambda$ Cephei which do not pass
through the inner critical point}
\begin{center}
\begin{tabular*}{13cm}{*{10}{r@{\extracolsep{\fill}\hspace{1mm}}}}
\hline
\hline
\rule[-3mm]{0mm}{8mm}
no. & $\Bpo\atop(\Gauss)$ & $\alpharot$ &
$\frac{r_{c2}}{R}$ & $\frac{\rAc}{R}$ & $\vAc\atop(\km/\s)$ &
$\vinf\atop(\km/\s)$ & ${10^6\Mdot}\atop{(\Msun/\yr)}$ &
$\frac{\Mdot\vinf}{L/c}$ & $\frac{v_\infty}{v_{\mathrm{A}\infty}}$\\
\hline
%no    Bp0    arot    rcf     rc    vac    vinf   Mdot     eff    y/yA
 1a &  200 & 0.218 &  1.60 & 1.54 & 1024 & 1751 & 4.92 & 0.624 & 11.89\\
 1b &  200 & 0.218 &  2.26 & 1.67 &  820 & 1379 & 5.24 & 0.520 &  8.58\\
 4a & 1600 & 0.218 &  9.09 & 8.16 & 4080 & 4384 & 2.81 & 0.894 &  4.45\\
 4b & 1600 & 0.218 & 24.26 & 9.43 & 1684 & 1977 & 5.09 & 0.720 &  1.81\\
13a &  200 & 0.697 &  1.50 & 1.41 & 1611 & 2985 & 3.73 & 0.814 &  7.28\\
13b &  200 & 0.697 &  5.04 & 1.64 &  753 & 1380 & 5.90 & 0.594 &  2.91\\
16a & 1600 & 0.697 & 49.05 & 7.09 & 2220 & 3030 & 6.84 & 1.513 &  1.30\\
16b & 1600 & 0.697 & 25.66 & 7.84 & 3685 & 4367 & 3.37 & 1.075 &  1.56\\
\hline
\end{tabular*}
\end{center}
\end{table}
\begin{figure}[p]
\begin{center}
\begin{picture}(120,70)
%\put(0,0){\framebox(120,75){}}
\put(0,2){\epsfig{figure=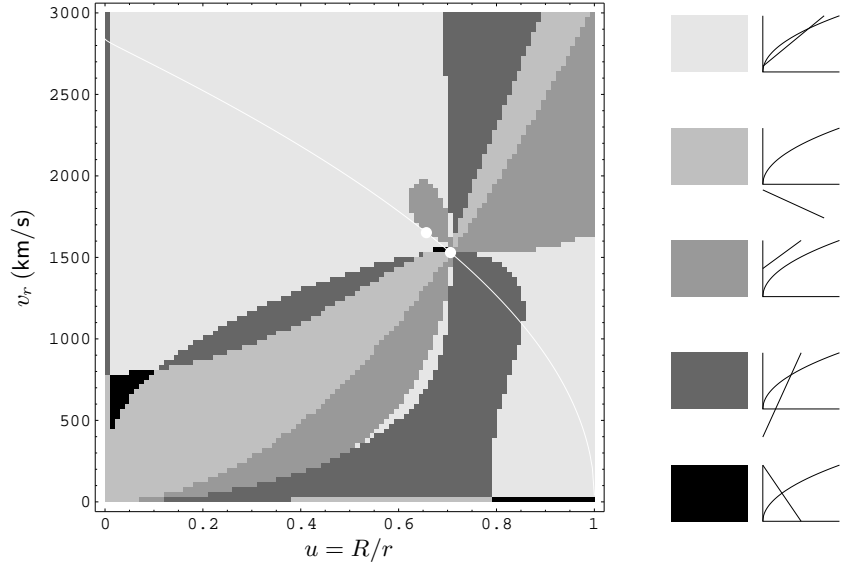,width=12cm}}
\put(44,1){\small$u=R/r$}
\put(5,34){\begin{sideways}\small$v_r\,(\km/\s)$\end{sideways}}
\end{picture}
\caption{\label{f:FM:m13:b}shows our wind model 13b. This solution has no inner
critical point. Therefore the sequence of the solution logic from the stellar
surface $(u=1)$ to infinity $(u=0)$ is (c.f.\ Fig.~\ref{CAKSolutions}):
A--D--F(here is the Alfv\'enic critical point)--C--A--B(here is the outer
critical point)--A--D. For the technical explanation of this plot check
Fig.~\ref{f:FM:m01:a}.}
\end{center}
\end{figure}
\begin{figure}[p]
\begin{center}
\begin{picture}(120,70)
%\put(0,0){\framebox(120,75){}}
\put(0,2){\epsfig{figure=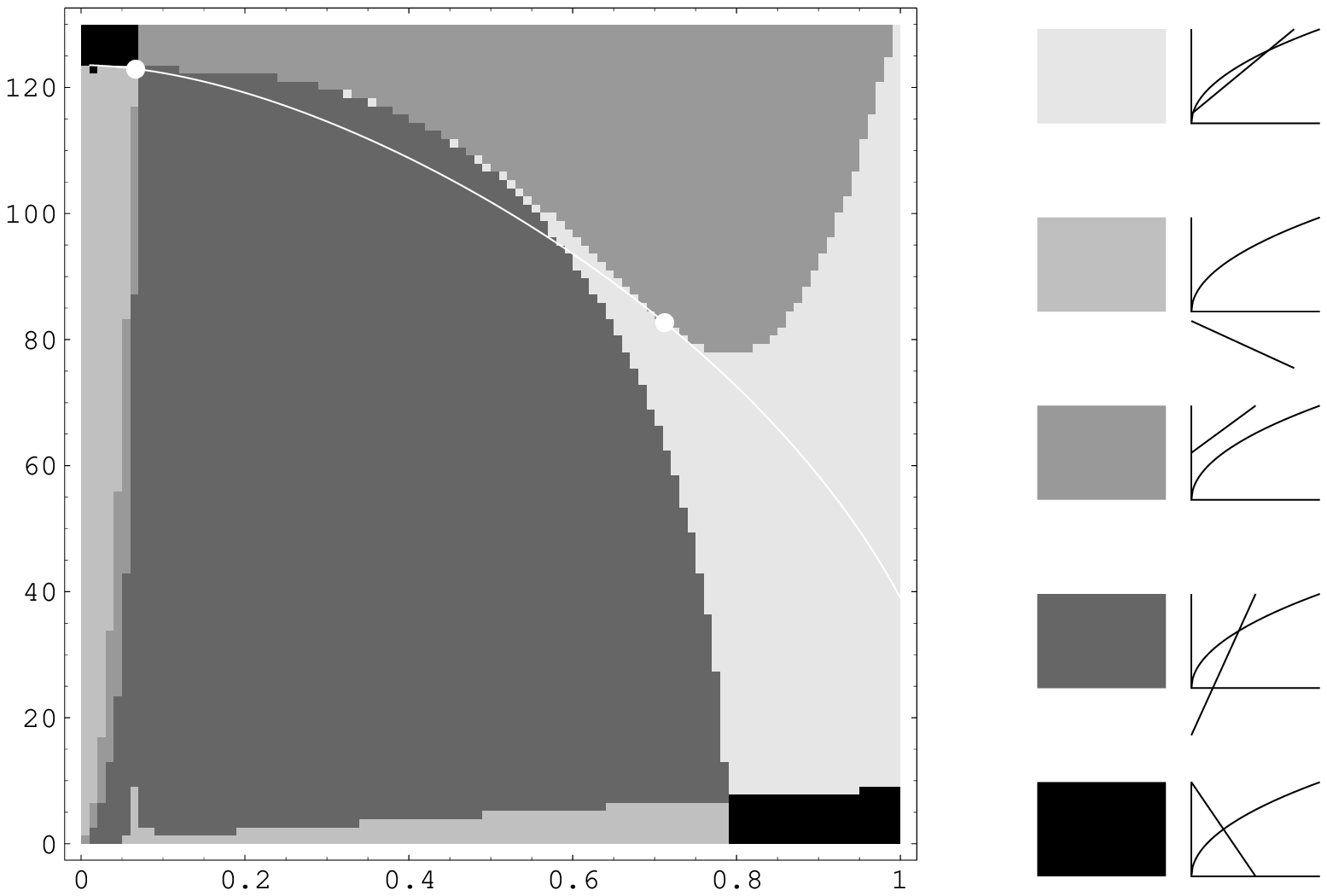,width=12cm}}
\put(44,1){\small$u=R/r$}
\put(5,34){\begin{sideways}\small$v_r\,(\km/\s)$\end{sideways}}
\end{picture}
\caption{\label{f:FM:m13:c}shows our wind model 13c. This solution has no outer
critical point. Therefore the sequence of the solution logic from the stellar
surface $(u=1)$ to infinity $(u=0)$ is (c.f.\ Fig.~\ref{CAKSolutions}):
A--B(here is the inner critical point)--A--D--F(here is the Alfv\'enic critical
point)--C--A. For the technical explanation of this plot check
Fig.~\ref{f:FM:m01:a}.}
\end{center}
\end{figure}
For the models 1,4\&13 we find that the Alfv\'enic radius $\rAc$ is reduced
compared to the models with 3 critical points. But for model 16 we find only
solutions with an enlarged Alfv\'enic radius. All other quantities vary up to
an factor of two. This shows that even for fixed magnetic field strength and
rotation rate magnetic wind models have a strong capability to fit
observational constraints. Additionally we see that a proper treatment of the
wind physics at the stellar surface is crucial for quantitative wind models.
We did not expect this in the last section where we found that models with 3
critical points are insensitive for the initial wind velocity $\vro$.
Therefore it is important to improve the wind physics near to the stellar
surface. Figure~\ref{f:FM:m13:b} shows our wind solution 13a. The interesting
aspect about solution 13a is that it has a significantly reduced angular
momentum loss $(\epsilon=2)$ compared to the model with 3 critical points
no.~13 $(\epsilon=3)$. Additionally model 13a has a much steeper velocity law
and a much higher terminal velocity than model 13. This can also help to fit
observations. In Fig.~\ref{f:FM:m13:b} we show the velocity profile for our
wind model~13a.

It is also possible to question whether the outer critical point exists. It
will certainly exists when we have a finite temperature at an sufficiently
large radius. But is is not necessary that our wind solution extends
to infinity in the mathematical sense. This would not be physical as we
discussed in Chap.~\ref{Chap:parker:wind}. We can see in
Fig.~\ref{f:FM:m16:d} that the wind solution has to pass through the outer
critical point before it reaches at $r\approx200R$ a region, where again only
local solution of Eq.~\ref{CAKWDWind} exists. This was the argument which
required the outer critical point. But we will discuss in the next chapter a
model for linear waves in magnetic stellar winds. We will find there, that
these waves are highly instable and will grow to linear shocks
rapidly. Although the latter effect demands for a numerical model for
nonlinear waves, we can seriously expect that the wind at large radii will not
be a continuous, smooth, stationary flow, as assumed in the wind model used in
this chapter, but rather a chaotic system of shocks and nonlinear
waves. Therefore it might be that the smooth wind model used in this chapter is
not a proper description for the stellar wind at large radii -- even for the
time averaged properties of the wind. If we restrict our wind model to smaller
radii with $r<200R$, the need for the outer critical point is
relieved. Therefore we calculate now a set of wind models which have an inner
but no outer critical point. As in the previous case we have therefore one free
parameter in the model. Analogous to Tab.~\ref{Tab:FM:noinner}
Tab.~\ref{Tab:FM:noouter} shows our results for wind models without an outer
critical point. Hereby we restricted ourself to a maximum of $15R$ for the
radius of the Alfv\'enic critical point $\rAc$. Especially for wind model~13
such a high value for $\rAc$ is already unphysical.
\begin{table}
\caption[Wind models without outer critical point]{
\label{Tab:FM:noouter}Wind models for $\lambda$ Cephei which do not pass
through the outer critical point}
\begin{center}
\begin{tabular}{rrrrrrrrrr}
\hline
\hline
\rule[-3mm]{0mm}{8mm}
no. & $\Bpo\atop(\Gauss)$ & $\alpharot$ &
$\frac{r_{c1}}{R}$ & $\frac{\rAc}{R}$ & $\vAc\atop(\km/\s)$ &
$\vinf\atop(\km/\s)$ & ${10^6\Mdot}\atop{(\Msun/\yr)}$ &
$\frac{\Mdot\vinf}{L/c}$ & $\frac{v_\infty}{v_{\mathrm{A}\infty}}$\\
\hline
%no    Bp0    arot    rcn       rc    vac   vinf   Mdot     eff   y/yA
 1c &  200 & 0.218 &  1.38 &  1.79 &  717 & 1046 & 5.22 & 0.399 & 5.63\\
 1d &  200 & 0.218 &  1.88 &  2.13 &  569 &  759 & 4.63 & 0.257 & 3.29\\
 4c & 1600 & 0.218 &  1.14 &  9.87 & 1473 & 1566 & 5.32 & 0.609 & 1.30\\
 4d & 1600 & 0.218 &  3.99 & 15.00 &  724 &  731 & 4.69 & 0.250 & 0.39\\
13c &  200 & 0.697 &  1.05 &  1.83 &  598 &  883 & 5.97 & 0.385 & 1.47\\
13d &  200 & 0.697 &  1.40 & 15.00 &  123 &  124 & 0.43 & 0.004 & 0.02\\
16c & 1600 & 0.697 &  1.04 &  7.27 & 1978 & 2267 & 7.30 & 1.208 & 0.83\\
16d & 1600 & 0.697 &  1.07 & 15.00 &  645 &  648 & 5.27 & 0.249 & 0.11\\
\hline
\end{tabular}
\end{center}
\end{table}
Figure~\ref{f:FM:m13:c} shows the extreme case of solution~13c. The wind
solution starts at the stellar surface well above the region of no wind
solution. Since our solutions pass through the inner critical point, we could
have chosen a subsonic wind velocity a the stellar surface. But in order to
compare these solutions with our solutions from the last paragraph we chose the
same initial wind velocity of $\vro=39\,\km/\s$.  Our solutions pass through
the inner and the Alfv\'enic critical point, and stops at 100 stellar
radii. This is well below the region where the passage through the outer
critical point becomes important $(r\approx300R)$.

\begin{figure}[p]
\begin{center}
\begin{picture}(120,77)
%\put(0,0){\framebox(120,77){}}
\put(10,3){\epsfig{figure=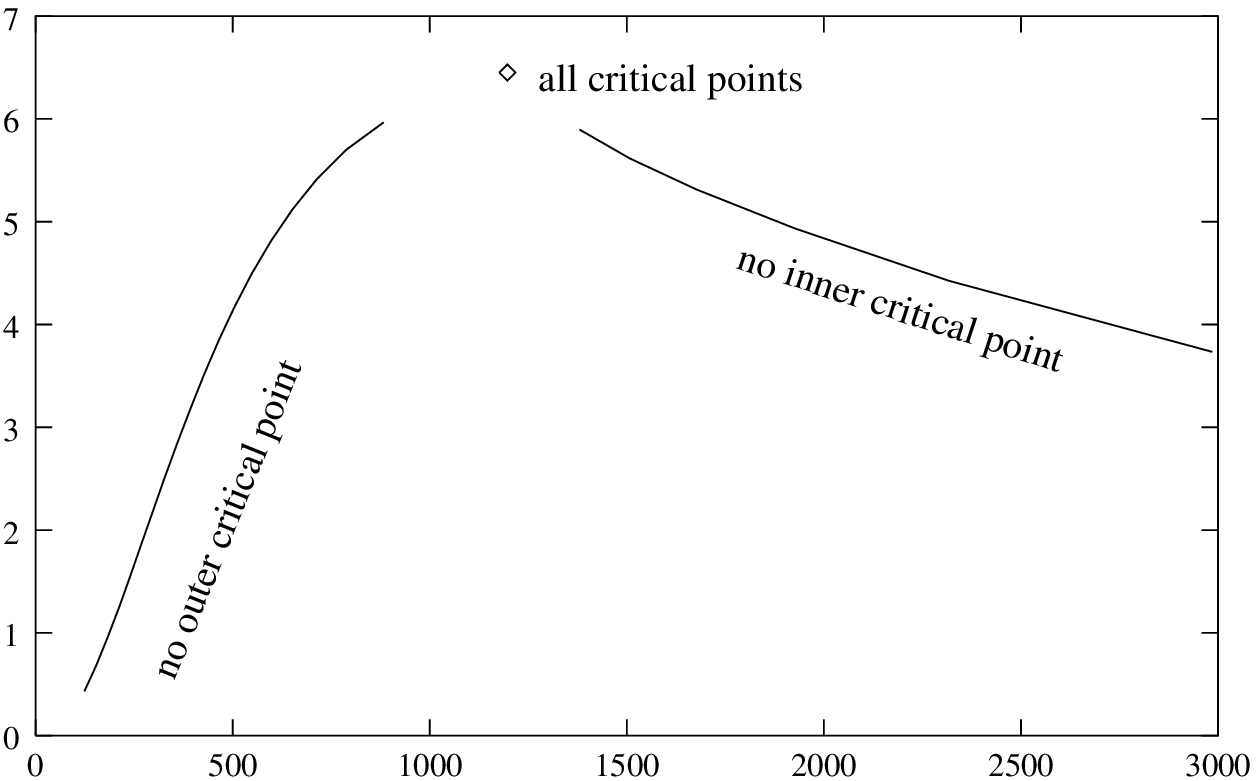,width=11cm}}
\put(58,1){\small$\vinf\,(\km/s)$}
\put(1,34){\begin{sideways}\small$\Mdot\,(10^{-6}\Msun/\yr)$\end{sideways}}
\end{picture}
\caption{shows our wind models 13 the mass loss rate $\Mdot$ vs.\ the terminal
velocity $\vinf$.\label{f:FM:m13:d}}
\end{center}
\end{figure}

\begin{figure}[p]
\begin{center}
\begin{picture}(120,77)
%\put(0,0){\framebox(120,77){}}
\put(6,3){\epsfig{figure=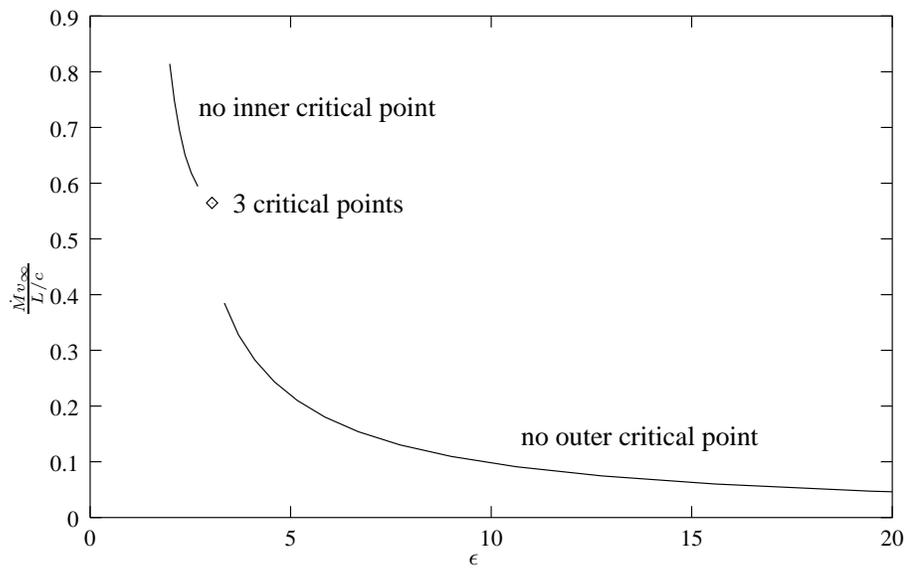,width=114mm}}
\put(62,1){\small$\epsilon$}
\put(1,34){\begin{sideways}\small$\frac{\Mdot\vinf}{L/c}$\end{sideways}}
\end{picture}
\caption{shows the wind efficiency $(\Mdot\vinf)/(L/c)$ vs.\ the angular
momentum enhancement factor $\epsilon$.\label{f:FM:m13:e}}
\end{center}
\end{figure}

Since a picture can tell more than 1000 words we show finally for our wind
model~13 the mass loss rate $\Mdot$ vs.\ the terminal velocity $\vinf$
(Fig.~\ref{f:FM:m13:d}) and the wind efficiency $(\Mdot\vinf)/(L/c)$ vs.\ the
angular momentum enhancement factor $\epsilon$ (Fig.~\ref{f:FM:m13:e}). The
wind model with three critical points seems to form a situation between the two
cases with only two critical points. From the view point of avoiding the
spin-down problem in a strong wind models without an inner critical point are
very interesting. They allow an increased wind efficiency together with a
reduced angular momentum loss. This was made possible by the fact that we
modified our model close to the star. Our modification to assume that the wind
is already slightly supersonic at the base of the wind is arbitrary. Somehow
the wind has to become supersonic. This problem could be solved by using a
different theory for the wind close to the stellar surface. Such a theory could
e.g.\ incorporate the consequences of an optically thick wind. Further out we
should smoothly switch to our wind model taking advantage of the high wind
efficiency and the low angular momentum loss of our wind solutions without an
inner critical point. Therefore such an alternative model must include a
consistent transition from the inner to the outer regions of the wind.

\enlargethispage{8mm}
We did not make yet use of the capability of our theory to describe winds which
are compressed or diluted in the equatorial plane. We will discuss this aspect
in a more generalized framework in Chap.~\ref{Chap:fluxtube}.
%

%-*-LaTeX-*-
% This is the 6th chapter for the PhD-thesis of Henning Seemann.
% (c) 1997-98 by Henning Seemann
%
\chapter{Waves and shocks in stellar winds}
\label{Chap:waves:shocks}
This chapter is based on an article \cite{Seemann:Biermann:97} by the author
and his academic supervisor Prof.\ P. L. Biermann.
\section{Introduction}
In the previous chapters we asked for a stationary solution of the wind
problem. This assumption leads to a strong simplification of the model. Such a
simplified model is a good starting point for understanding a physical system
by describing its time averaged features.  Therefore the first models which
were applied to the solar wind were stationary (c.f.\
Chaps.~\ref{Chap:parker:wind}\&\ref{Chap:wd}), as well. But from observations
we know today that the solar wind is far from being stationary. The strong
fluctuations in the solar corona heat the corona to the high temperatures,
which are necessary to drive the solar wind by thermal pressure. Therefore even
the time averaged features of the solar wind can not be understood without
taking fluctuations into account. Therefore it is quite possible that the wind
of hot stars can not be described properly without taking fluctuations into
account.

The analysis of wind fluctuations breaks into three parts: (1) How are
fluctuations generated? And how do they develop in the stellar wind? (2)
Which consequences do they have for the overall model of the wind? And (3)
which influence does fluctuations have on observations?

Wind fluctuations can have two consequences for observation: (1) Is it possible
to observe fluctuations directly by analyzing the time dependence of
observational parameters like line profiles? (2) How do fluctuations influence
the derivation of physical parameters like mass loss rate and terminal velocity
from observational parameters like spectra and fluxes? Since this thesis is a
theoretical orientated work, we do not want to go to deep into observational
details here. Observed fluctuations can have two reasons: (1) Periodic
fluctuations are probably connected to the stellar rotation. Some localized
feature on the rotating stellar surface perturbes the wind in the line of sight
periodically. This scenario is similar to a pulsar. Such fluctuations have
e.g.\ be observed by Gagn\'e et al.\ \cite{Gagne:etal:97}. These observations
allow to measure the stellar rotation rate, which is crucial for our
models. Gagn\'e et al.\ call $\theta^1$~Orionis an `oblique magnetic rotator'
with a rotation period of 15.422 days. This star therefore rotates with 3\% of
its critical rotation rate. (2) The second type of fluctuations have a
stochastic characteristic. This type of fluctuations are presumably created in
the wind by some kind of plasma instability. We will discuss such instabilities
in this chapter. Further evidence for fluctuations are the observed X-ray and
nonthermal radio emission. Lucy \cite{Lucy:82} proposed that the observed
X-rays are emitted by very hot plasma, which is heated up by shocks. Shocks are
required as well to explain the nonthermal radio emission observed by Bieging
et al.\ \cite{Bieging:etal:89} from OB stars and by Abbott et al.\
\cite{Abbott:etal:86} from Wolf-Rayet stars. This nonthermal radio emission is
in fact synchrotron emission by relativistic electrons. Therefore this gives
clear evidence for non-negligible magnetic fields on these stars. The
relativistic electrons can be produced by Fermi acceleration in shocks. In
binary systems these shocks could be created by the colliding winds of the two
components. For single stars intrinsic wind instabilities are the only possible
source for the required shocks.

Fluctuations in the wind influence the observation of basic wind parameters
like the terminal velocity and the mass loss rate.  The {\it observed} terminal
velocity $\vinf$ is inferred from the blue edge of P-Cygni lines. This is in
fact the maximal velocity reached by a significant fraction of the wind
material. In the case of an permanently accelerating stationary wind this is
approximately the theoretical terminal velocity $\vinf=v(r=\infty)$.  But if we
have fluctuations in the wind where regions of temporarily enhanced wind
velocity have simultaneously an enhanced density, we will see the enhanced
velocity as the blue edge of the P-Cygni lines. In the linear theory such
simultaneously enhancements are described by outward running waves.  This
enhanced velocity is in the case of linear waves the unperturbed $\vinf$ plus
the velocity amplitude of the waves or, if {\it outward} running waves have
already steepened into outward running shocks, the unperturbed $\vinf$ roughly
plus the speed of the shock front. The latter scenario was used by Lucy to
explain the X-ray emission from these stars.  

The observed values for the mass loss rate $\Mdot$ are influenced by waves in
two ways: (i) $\Mdot$ is inferred from observed values of the density $\rho$ by
$\Mdot \sim \rho r^2 \vinf$. An overestimated $\vinf$ therefore leads to an
overestimated $\Mdot$. (ii) The values for $\rho$ are inferred from observed
radio or UV fluxes. A matter distribution perturbed by waves or shocks
generates higher fluxes, which lead to higher values for $\rho$ if the
perturbations are not taken into account properly (Abbott et al.\
\cite{Abbott:etal:81} and Hillier \cite{Hillier:91}). The nonthermal
contribution to the radio emission additionally increases the estimates for
$\rho$, if it is not properly separated from the thermal emission.
Fluctuations can also be an additional source and transport mechanism for
momentum. When waves are amplified by radiation they extract extra momentum
from the radiation and therefore increase the overall efficiency of radiation
pressure. When these linear waves later dissipate or steepen into shocks, they
transfer their momentum to the wind. This effect was analyzed by Koninx
\cite{Koninx:PhD}. Koninx model and our discussion about the derivation of
observed values for $\Mdot$ and $\vinf$ are valid only if the wind is dominated
by \textit{outward} running waves.

The fluctuations in the solar wind are generated in the convective envelope of
the sun. From there they propagate into the wind. At the base of the wind they
heat up the plasma to the very high temperatures $(\sim 10^6\Kelvin)$ of the
corona. Hot stars do not have a convective envelope. Their envelope is mostly
radiative. Such a radiative envelope will not produce strong fluctuations and
therefore presumably no corona. Therefore we need another mechanism to produce
fluctuations. A good candidate is the radiative line driving force. This force
is the dominating driving mechanism for the unperturbed wind. Any instability
in this force could create significant fluctuations. The possibility for
such an instability was already mentioned by Castor, Abbott and Klein in their
famous paper about line driven winds \cite{Castor:etal:75}.

This idea was later elaborated by several authors. MacGregor et al.\
\cite{MacGregor:etal:79} developed a model for short, optical thin waves. They
found that these waves are unstable and grow rapidly. This result was opposed
by Abbott \cite{Abbott:80}, who found stable waves assuming long
wavelengths. Owocki and Rybicki \cite{Owocki:Rybicki:84} unified these
contradicting models by a more elaborate, but still analytical linear model for
all wavelengths. This model showed that waves with short wavelength $(\lambda
\lesssim \Lsobo)$ will grow rapidly. They refined their model later in a series
of papers \cite{Owocki:Rybicki:85, Owocki:Rybicki:86, Rybicki:etal:90,
Owocki:Rybicki:91}. The basic results is that in the thin part of the wind
fluctuations are strongly amplified. The fluctuations are dominated by inward
running waves steepening into shells of increased velocity but decreased
density. These fluctuations are advected outward by the average wind.  These
papers all do a linear analysis of the fluctuations. Krolik \& Raymond
\cite{Krolik:Raymond:85} and MacFarlane \& Cassinelli
\cite{MacFarlane:Cassinelli:89} analyzed the structure of a single outward
running shock shell in a nonmagnetic wind and found that the shock shell is
driven by the radiation field to a velocity much higher than the velocity of
sound and therefore much higher than the phase velocity for outward running
linear waves found by Owocki \& Rybicki \cite{Owocki:Rybicki:84}. But later
Owocki et al.\ \cite{Owocki:etal:88} found in a nonlinear time dependent
calculation, that waves in the stellar wind are dominantly running inward and
steepen into inward running, reverse shocks. This confirmed their previous
result from linear analysis.

The aim of this chapter is to show in the limit of a linear analysis that the
objections of Owocki et al.\ \cite{Owocki:etal:88} do not apply, if a strong
magnetic field and rotation are present. Therefore we connect the ideas of a
rotating magnetic field, developed in the previous chapters, and of
unstable waves (Owocki \& Rybicki \cite{Owocki:Rybicki:84} in the wind of
a massive star and do a linear stability analysis. We show that the magnetic
field increases the number of wave modes and changes their properties
significantly. High phase velocities can be achieved due to the high Alfv\'en
velocity. Therefore inward running waves will not be advected outward and will
not steepen into reverse shocks anymore, which dominate at large radii.
Furthermore outward and inward running waves have the same growth timescale in
the short wavelength regime where both modes are unstable. So we can expect
outward running waves far from the star and forward shocks in the nonlinear
regime. 

In Sect.~\ref{Sec:ORmodel} we outline the model of Owocki \& Rybicki
\cite{Owocki:Rybicki:84} for the radiative instability. In
Sect.~\ref{Sec:waveeq} we derive the dispersion relation for radiatively
amplified waves in the presence of a magnetic field. In
Sect.~\ref{Sec:unpermodels} we discuss our unperturbed wind models. In
Sect.~\ref{Sec:numres} we discuss the waves we found for the wind models of
Sect.~\ref{Sec:waveeq}. In Sect.~\ref{Sec:consec} we discuss the observational
consequences of our model. And in Sect.~\ref{Sec:waveconclu} we describe some
conclusions.
\section{The radiative Instability model of Owocki \& Rybicki}
\label{Sec:ORmodel}
In this section we outline briefly the model for the radiative instability of
Owocki \& Rybicki \cite{Owocki:Rybicki:84}, which we are going to use for our
linear analysis of radiative instabilities in magnetic winds. A more detailed
treatment including several refinements can be found in the original paper and
in the subsequent series of papers \cite{Owocki:Rybicki:85, Owocki:Rybicki:86,
Rybicki:etal:90, Owocki:Rybicki:91}. We start our analysis with the radiative
acceleration by a single line as described by Eq.~\ref{gLA} in
Sect.~\ref{Chap:cak:cak}. If we assume that the average wind is perturbed by a
small, linear fluctuation, we can describe the perturbed line
force\footnote{Variables marked with a\ $\bar{\ }$\ refer in this chapter to
the unperturbed wind.}  $\gL=\bar\gL+\delta\gL$ to first order in the perturbed
optical depth $\tau=\bar\tau+\delta\tau$ and the line profile
$\phi(\hat{x})=\bar\phi(\hat{x})-\phi^\prime(\hat{x})\delta v_r(r)/\vth$
perturbed by the Doppler shift due to the velocity fluctuation. But we assume
that the temperature remains constant.
\begin{equation}
\delta\gL(r,\kappaL) = -\gthin\int_{-\infty}^\infty\left(
                          \phi(\hat{x})\delta\tau(\hat{x},r)+
                          \phi^\prime(\hat{x})
                          \frac{\delta v_r(r)}{\vth}\right)
                          e^{-\tau(\hat{x},r)}\,d\hat{x}
\end{equation}
Now we can integrate by parts the second term of the integrand using the
Sobolev approximation for the quantities of the \textit{unperturbed} wind.
\begin{equation}
\delta\gL(r,\kappaL) \approx
    -\gthin\int_{-\infty}^\infty\phi(\hat{x})\left(\delta\tau(\hat{x},r)
      -\frac{\delta v_r^\prime(r)}{v_r^\prime(r)}\right)
      e^{-\tau(\hat{x},r)}\,d\hat{x}
\end{equation}
We assumed that the line opacity $\kappaL$ is not affected by the
perturbation. From Eq.~\ref{gradtau} we can derive the perturbation of the
optical depth
\begin{eqnarray}
\delta\tau(\hat{x},r) &=& 
                \int^r_R\left(\kappaL\delta\rho(\tilde{r})\phi(\tilde{x})-
                 \rho(\tilde{r})\phi^\prime(\tilde{x})
                 \frac{\delta v_r(r)}{\vth}\right)\,
                 d\tilde{r}\\
           &\approx&   
               \int^r_R \kappaL\rho\left(\frac{\delta\rho}{\rho}-
                \frac{\delta v_r^\prime(\tilde{r})}{\bar v_r^\prime(\tilde{r})}
                 \right)\phi(\tilde{x})\,d\tilde{r}.\\
           &\approx&
               \bar\tauL\int^\infty_{\hat{x}}
                 \frac{\delta\tauL}{\bar\tauL}\phi(\tilde{x})\,d\tilde{x}
\end{eqnarray}
Here we used the same trick as for $\delta\gL$ to get rid of the $\phi^\prime$
term. In the WKB approximation, that gradients of the mean flow are small, we
can use
\begin{equation}
\frac{\delta\tauL}{\bar\tauL}=\frac{\delta\rho}{\bar\rho}-
   \frac{\delta v_r^\prime(\tilde{r})}{\bar v_r^\prime(\tilde{r})}\approx
   -\frac{\delta v_r^\prime(\tilde{r})}{\bar v_r^\prime(\tilde{r})}.
\end{equation}
In our linear analysis we can describe the perturbation by a sinusoidal wave
$\delta v_r(r) \approx \delta v_re^{ikr}$. In the limit of a strong line
$(\tauL \gg 1)$ we can find a simple expression for $\delta\gL$.
\begin{equation}
\frac{\delta\gL}{\delta v_r} = \frac{ik\Lsobo\gthin\bar\tauL}{\vth}
  \int_{-\infty}^\infty \phi(\hat{x})e^{-\tau(\hat{x},r)}
  \int_{\hat{x}}^\infty \phi(\tilde{x}) e^{ik\tilde{r}}\,
  d\tilde{x}\,d\hat{x}.
\end{equation}
As in Chap.~\ref{Chap:cak} we can assume that $\tilde{x}$ is close to
$\hat{x}$ and therefore use
\begin{equation}
\tilde{x} \approx \hat{x} - \frac{v_r^\prime(r)}{\vth}(\tilde{r}-r).
\end{equation}
This leads to
\begin{equation}\label{deltaglB}
\frac{\delta\gL}{\delta v_r(r)} = \frac{ik\Lsobo\gthin\bar\tauL}{\vth}
  \int_{-\infty}^\infty \phi(\hat{x})e^{-\tau(\hat{x},r)}
  \int_{\hat{x}}^\infty \phi(\tilde{x}) e^{-ik\Lsobo(\tilde{x}-\hat{x})}\,
  d\tilde{x}\,d\hat{x}.
\end{equation}
For a strong line $(\bar\tauL\gg1)$ Eq.~\ref{deltaglB} can be approximated
analytically. We shift the integration from frequencies $(x)$ to optical depths
$(\tau)$ using
\begin{equation}
\diff{\tau(x,r)}{x} = -\bar\tauL\phi(x).
\end{equation}
The $\exp\tau(\hat{x},r)$ term restricts any significant contribution to the
integral to the frequency range where $\hat{\tau}=\bar\tau(\hat{x},r)$ is close
to one. So we define $x_b$ by
\begin{equation}
\bar\tau(x_b,r)\equiv 1.
\end{equation}
We can now express $\hat{x}$ and $\tilde{x}$ to first order by
\begin{equation}
\tilde{x}-\hat{x} = -\frac{\tilde{\tau}-\hat{\tau}}{\bar\tauL\phi(x_b)}.
\end{equation}
This leads to
\begin{eqnarray}\label{deltaglC}
\frac{\delta\gL}{\delta v_r(r)} &=& \frac{ik\Lsobo\gthin}{\vth\bar\tauL}
  \int_{\bar\tauL}^0 e^{-\hat\tau}
  \int_{\hat\tau}^0 \exp
  \left(ik\Lsobo\frac{\tilde{\tau}-\hat{\tau}}{\phi(x_b)\bar\tauL}\right)\,
  d\tilde{\tau}\,d\hat{\tau}\\
&=& \frac{\gthin\phi(x_b)}{\vth}\int^0_{\bar\tauL}
    \exp\left(-\left(\frac{ik\Lsobo}{\phi(x_b)\bar\tauL}+1\right)
    \hat\tau\right)-e^{-\hat\tau}\,d\hat\tau\\
\label{deltaglD}
&\approx& \omega_b\frac{ik}{\chib+ik},
\end{eqnarray}
where we used the convention of Owocki \& Rybicki \cite{Owocki:Rybicki:84}
\begin{eqnarray}
\omega_b &=& \frac{\gthin\phi(x_b)}{\vth}\\
\chib &=& \rho\kappaL\phi(x_b).
\end{eqnarray}
The radiative acceleration of a hot star wind is not properly described by the
acceleration due to a single line. Therefore we can not expect that
Eq.~\ref{deltaglD} properly describes the perturbation in the radiative
acceleration. It is rather necessary to sum $\delta\gL$ for an ensemble of
lines using the same recipe as CAK used for the unperturbed wind. The CAK model
takes the large number of weak lines into account, as well. Therefore we can
not use Eq.~\ref{deltaglD}, when we integrate over the line distribution
function. But for a Doppler line profile
\begin{equation}
\phiD(x) = \frac{1}{\sqrt{\pi}}e^{-x^2}
\end{equation}
Eq.~\ref{deltaglB} can be integrated to
\begin{eqnarray}
\frac{\delta\gL}{\delta v_r(r)} &=& 
  \frac{ik\Lsobo\gthin\bar\tauL}{2\vth\sqrt{\pi}}\int_{-\infty}^\infty
  e^{-2\hat x^2}e^{-\tau(\hat x,r)}W\left(i\hat x-\frac{k\Lsobo}{2}\right)\,
  d\hat x\\
W(ix) &=& \exp(x^2)\erfc(x),
\end{eqnarray}
where 
\begin{equation}
\erfc(x) = \frac{2}{\sqrt{\pi}}\int^\infty_x e^{-y^2}\,dy
\end{equation}
is the complementary error function.  We find then for the perturbed line
acceleration analogous to Eq.~\ref{gCAKA}
\begin{eqnarray}
\frac{\delta\gcak}{\delta v_r(r)}
&=& \int_0^\infty N(\nuL,\kappaL)
    \frac{\delta\gL(r,\nuL,\kappaL)}{\delta v_r(r)}\,d\kappaL\\
&\approx& \frac{ik\Lsobo F}{2\sqrt{\pi}c^2}\kappa_0^{1-\alphacak}
          \left(\frac{v_r^\prime}{\rho\vth}\right)^\alphacak
          \Gamma(\alphacak+1)\times\nonumber\\
&&        \int_{-\infty}^{\infty}
          \left(\int_{\hat x}^\infty e^{-x^2}\,dx\right)^{-\alphacak-1}
          W\left(i\hat x -\frac{k\Lsobo}{2}\right)e^{-2{\hat x}^2}\,d\hat x\\
&\approx& 2^\alphacak \alphacak(1-\alphacak) \frac{\gcak}{\vth}
          \frac{ik\Lsobo}{\sqrt{\pi}}\times\nonumber\\
\label{deltagCAKA}
&&        \int_{-\infty}^\infty e^{(\alphacak-1){\hat x}^2}
          \frac{W\left(i\hat x -\frac{k\Lsobo}{2}\right)}
               {W(i\hat x)^{\alphacak+1}}
          \,d\hat x.
\end{eqnarray}
Owocki \& Rybicki \cite{Owocki:Rybicki:84} analyzed this integral
numerically. From this analysis they found that Eq.~\ref{deltagCAKA} can nicely
be approximated by the analytical expression
\begin{equation}
\frac{\delta\gcak}{\delta v_r(r)} = \OmegaOR\frac{ik}{\chiOR+ik},
\end{equation}
which is just a modified version of the analytical expression for a single
strong line (Eq.~\ref{deltaglD}). The combined amplification rate of all lines
is given by
\begin{equation}
\OmegaOR=\sqrt{\frac{2c^{\alphacak-1/2}}{1-\alphacak}}\frac{\vrbar}{\Lsobo}.
\end{equation}
And
\begin{equation}
\chiOR=\sqrt{\frac{2c^{\alphacak-1/2}}{1-\alphacak}}\frac{1}{\Lsobo}=
        \frac{\Omega}{\vrbar}
\end{equation}
is the mean blue-edge absorption strength. Here $c$ is not the speed of light
but an empirical parameter: $c\approx 1.6$.

Owocki \& Rybicki \cite{Owocki:Rybicki:84} were criticized by Lucy
\cite{Lucy:84} for the neglect of damping due to diffuse radiation. But Owocki
\& Rybicki \cite{Owocki:Rybicki:85} showed that this effect reduces the
amplification rate only by approximately 50\% at one stellar radius and 20\% at
infinity. In this initial analysis we emphasize the effect of the magnetic
field. Therefore we choose the simple description of Owocki \& Rybicki
\cite{Owocki:Rybicki:84} instead of the more exact but also more involved
description of Owocki \& Rybicki \cite{Owocki:Rybicki:85} or the description of
Gayley \& Owocki \cite{Gayley:Owocki:95}, who analyzed the instability in
optically thick winds.
\section{The dispersion relation}
\label{Sec:waveeq}
To analyze waves in the wind of hot stars we start from the equations of
magnetohydrodynamics for a compressible, nonviscous, perfectly conducting fluid
as described in Jackson \cite{Jackson:Book} and add the analytic description of
Owocki \& Rybicki \cite{Owocki:Rybicki:84} for the influence of the stellar
radiation on the plasma. We start with the time dependent equations for a
nonviscous, perfectly conducting gas:
\begin{eqnarray}\label{basiceqn1}
0 &=& \dt{\rho} + \nabla\cdot(\rho\vvec)\\ 
0 &=& \rho\dt{\vvec} + \rho(\vvec\cdot\nabla)\vvec + \nabla p + 
\frac{\Bvec}{4\pi}\times(\nabla\times\Bvec) - \fradvec\\
0 &=& \dt{\Bvec} - \nabla\times(\vvec\times\Bvec)\label{basiceqn3}
\end{eqnarray}
where $\fradvec$ is the force term due to radiation pressure. Now we express
the pressure $p$ by $\vsvs\rho$, where $\vs$ is the speed of sound, and
replace $\rho$, $\vvec$, and $\Bvec$ by sums of a equilibrium value and a
perturbation. In the comoving reference frame $(\vbarvec=0)$ this is:
\begin{eqnarray}
\rho  & = & \rhobar  + \deltarho(\xvec,t) \\
\vvec & = & \deltavvec(\xvec,t) \\ 
\Bvec & = & \Bbarvec + \deltaBvec(\xvec,t)
\end{eqnarray}
Equations \ref{basiceqn1}--\ref{basiceqn3} give to first order in small
quantities:
\begin{eqnarray}\label{lineqn1}
0 &=& \dt{\deltarho(\xvec,t)} + \rhobar\nabla\cdot\deltavvec(\xvec,t) \\
0 &=& \rhobar\dt{\deltavvec(\xvec,t)} + \vsvs\nabla\deltarho(\xvec,t) + 
\frac{\Bbarvec}{4\pi}\times(\nabla\times\deltaBvec(\xvec,t))\nonumber\\
&& -{}\delta\fradvec\\
0 &=& \dt{\deltaBvec(\xvec,t)}-\nabla\times(\deltavvec(\xvec,t)\times\Bbarvec)
\label{lineqn3}
\end{eqnarray}
We assume now, that the radiative force acts only in radial direction and
depends only on the radial velocity.  These equations can be reduced to a
single equation for $\deltavvec$:
\begin{eqnarray}
0 &=& \frac{\partial^2\deltavvec(\xvec,t)}{\partial t^2} - 
      \vsvs\nabla(\nabla\cdot\deltavvec(\xvec,t)) +\nonumber\\
&&\vAbarvec\times\nabla\times[\nabla\times(\deltavvec(\xvec,t)\times\vAbarvec)]
\nonumber\\
&&-{}\frac{1}{\rhobar}\frac{\partial\delta v_r(\xvec,t)}{\partial t} 
\frac{\partial \frad}{\partial v_r}\ervec
\label{waveeqn}
\end{eqnarray}
with the vectorial Alfv\'en velocity $\vAbarvec =
\Bbarvec/\sqrt{4\pi\rhobar}$. If we use the result of Owocki \& Rybicki
\cite{Owocki:Rybicki:84} for the linear perturbation of the radiative force, we
can get a dispersion relation from this wave equation for plane waves:
\begin{equation}
\deltavvec(\xvec,t) =\deltavvec e^{\complexi\kvec\cdot\xvec-\complexi\omega t}.
\end{equation}
Equation \ref{waveeqn} then becomes:
\begin{eqnarray}\label{vecdisp}
0&=&-\omega^2\deltavvec + (\vsvs + \vA^2)(\kvec\cdot\deltavvec)\kvec + 
(\vAbarvec\cdot\kvec)[(\vAbarvec\cdot\kvec)\deltavvec\nonumber\\
&&-{}(\vAbarvec\cdot\deltavvec)\kvec - (\kvec\cdot\deltavvec)\vAbarvec]
-\frac{\OmegaOR k_r}{\chiOR + \complexi k_r}\omega (\deltavvec\cdot\ervec)
\ervec
\end{eqnarray}

Equation \ref{vecdisp} is a vector equation, which is linear in
$\deltavvec$. We can think of it as a generalized eigenvalue problem:
\begin{equation}\label{eigeneqn}
(\mathsf{A}(\kvec,\vAbarvec,\vsvs) - \omega \mathsf{B}(k_r,\OmegaOR,\chiOR) -
\omega^2\bbbone)\deltavvec =0,
\end{equation}
where $\mathsf{A}$ and $\mathsf{B}$ are tensors. We can find $\omega$ and
$\deltavvec$ numerically.  Then $\deltaBvec$ and $\deltarho$ follow from
Eqs. \ref{lineqn1} \& \ref{lineqn3}:
\begin{eqnarray}
\deltarho  & = & \frac{\rhobar}{\omega}(\kvec\cdot\deltavvec)\\
\deltaBvec & = & \frac{1}{\omega}[(\kvec\cdot\deltavvec)\Bbarvec - 
(\Bbarvec\cdot\kvec)\deltavvec\,]
\end{eqnarray}
Although Eq.~\ref{eigeneqn} is very involved in the general case, we can
find an analytical solution for a simplified situation:
\begin{eqnarray}
\kvec(r) &=& k \ervec\\
\vAbarvec(r) &=& \vAbar \ephivec
\end{eqnarray}
The latter approximation is quite accurate far away from the star. In this
limit we find
\begin{equation}
\omega=-\frac{\OmegaOR k}{2(\chiOR+\complexi k)}\pm\sqrt{\left(
\frac{\OmegaOR k}{2(\chiOR+\complexi k)}\right)^2 + (\vsvs+\vAbar^2)k^2}.
\end{equation}
In the long wavelength limit $(k \ll \chiOR)$, where the waves are stable,
this leads to
\begin{equation}
\omega=\left[-\frac{\vrbar}{2}\pm\sqrt{\frac{\vrbar^2}{2}+\vsvs+\vAbar^2}
       \,\right] k.
\end{equation}
In the case of a weak magnetic field with $\vrbar \gg \vs \gtrsim \vAbar$, this
resembles Abbott's \cite{Abbott:80} result for stable radiative-acoustic
waves with a fast inward and a slow outward mode. In the case of a strong
magnetic field we have $\vrbar\approx\vAbar\gg\vs$.  This leads to higher phase
velocities for both modes and reduces the relative difference between inward
and outward waves. Additionally inward running waves are not advected outward
by the average wind motion any more, because their phase velocity is higher
than the velocity of the unperturbed wind. In the short wavelength limit $(k
\gg \chiOR)$ we find
\begin{equation}
\omega=\complexi\frac{\OmegaOR}{2}\pm\sqrt{-\frac{\OmegaOR^2}{4}+
(\vsvs+\vAbar^2)k^2}.
\end{equation}
These waves propagate, if $\OmegaOR$ is less than $2k\sqrt{\vsvs+\vAbar^2}$,
with the same phase velocity inward and outward. The amplification rate is also
the same for both modes.

In the case of no magnetic field Eq.~\ref{eigeneqn} leads to
\begin{equation}\label{dispnB}
0=\omega^3+\frac{\OmegaOR k_r}{\chiOR + \complexi k_r}\omega^2-\vsvs\kvec^2
\omega-\frac{\OmegaOR k_r}{\chiOR + \complexi k_r}\vsvs(k_\phi^2 + k_\theta^2).
\end{equation}
For $k_\theta=k_\phi=0$ this reproduces the result for isothermal waves found
by Owocki \& Rybicki \cite{Owocki:Rybicki:84}. For oblique waves there is a
third wave mode.
\section{The unperturbed wind models}
\label{Sec:unpermodels}
To analyze the effect of these waves we construct three wind models for a
standard massive star. Our model has $M=23\Msun$, $L=1.7\times10^5\Lsun$,
$R=8.5\rsun$, $T=60000\Kelvin$, $\alphacak=0.56$, and $\kcak=0.28$. The first
model (model~A) is the standard analytic CAK wind for a star without magnetic
field, rotation, or pressure. In this model we reproduce the previous result of
dominantly inward running waves found by Owocki et
al.~\cite{Owocki:etal:88}. In the second model (model~B) we add a radial
magnetic field. Since this magnetic field is parallel to the natural stream
lines of the wind of a nonrotating star, this field does not change the
velocity profile of the unperturbed wind. But it changes the microphysics for
waves. The last model (model~C) is a luminous fast magnetic rotator model.  The
rotating magnetic field provides an additional driving force, which changes the
properties of the wind drastically. The terminal velocity decreases and the
mass loss rate increases. The wind efficiency $(\Mdot\vinf)/(L/c)$ is 3.8,
which is much higher than for a purely radiatively driven wind in the single
scattering limit, where the efficiency can not be higher than unity. In spite
of the known spin-down problem we choose a strong magnetic field and a high
rotation rate in order to emphasize the influence of these parameters. For this
model we use the model from Chap.~\ref{Chap:cak:wd}
(Eq.~\ref{CAKWDWind}--\ref{CAKWDWindC}).  Table \ref{t:mvalues} gives the
magnetic field, rotation rate, and the resulting values for the above mentioned
unperturbed wind models. In Sect.~\ref{Sec:consec} we will discuss how these
values and their observation can be influenced by waves. In this work we do not
discuss the influence of the waves back on the unperturbed wind as Koninx
\cite{Koninx:PhD} did. A model with rotation but without magnetic field is not
included, because this model has the same microphysics for waves as
model~A\@. Just $\Mdot$ and the velocity dependence on $r$ are
different. Model~C differs from model~B in the local conditions by the fact
that, due to the rotational twist, the magnetic field and the direction of the
radiative force are not parallel anymore. Even at the base of the wind $B_\phi$
is approximately $-1.67B_r$.
\begin{table}[ht]
\caption[]{\label{t:mvalues}Results for the unperturbed wind models}
\begin{center}
\begin{tabular}{lrrrrr}
\noalign{\smallskip}
\hline 
\noalign{\smallskip}
Model                                  &    A &    B &    C \\
\noalign{\smallskip}
\hline 
\noalign{\smallskip}
$B_{r0}\ [{\rm G}]$                    &    0 &  500 &  500 \\
$\Omega/\Omega_{\rm crit}$             &    0 &    0 & 0.92 \\
$\vinf\ [{\rm km\ s^{-1}}]$            & 1146 & 1146 &  784 \\
$\Mdot\ [10^{-6}\Msun\ {\rm yr}^{-1}]$ &  0.6 &  0.6 &   17 \\
$R_{\tau=2/3}\ [\rsun]$                &  8.5 &  8.5 &   11 \\
$(\Mdot\vinf)/(L/c)$                   &  0.2 &  0.2 &  3.8 \\
\noalign{\smallskip}
\hline 
\end{tabular}
\end{center}
\end{table}
\section{Numerical results for waves}
\label{Sec:numres}
Since our unperturbed wind model is limited to the equatorial plane we limit
our discussion for waves to the same plane. We have still two free parameters
then: The wavelength and the azimuthal angle of $\kvec$. First we will discuss
radial waves. This will show all important properties of this wave model. At
the end of this section we will briefly discuss the influence of the azimuth
angle of $\kvec$.
\begin{figure}[p]
\begin{center}
\begin{picture}(80,170)
\put(0, 90){\epsfig{figure=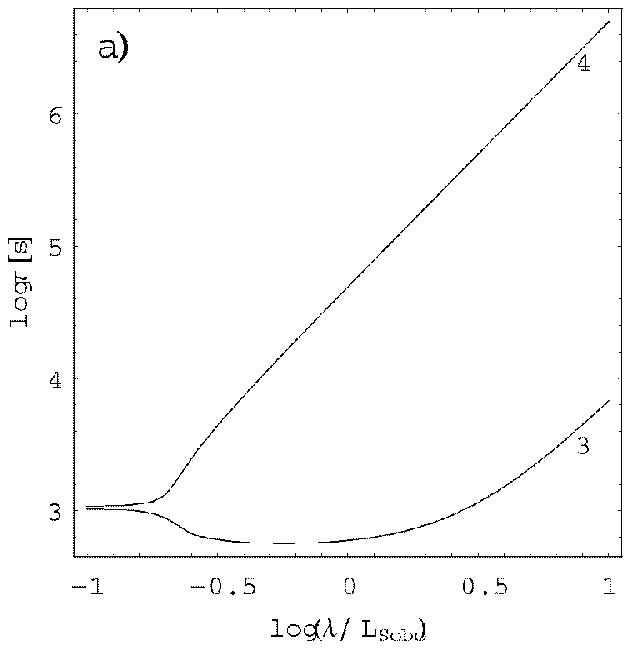,width=8cm,clip=,}}%
              %      bbllx=3.6cm,bburx=13.8cm,bblly=16.8cm,bbury=27.1cm}}
\put(0,  0){\epsfig{figure=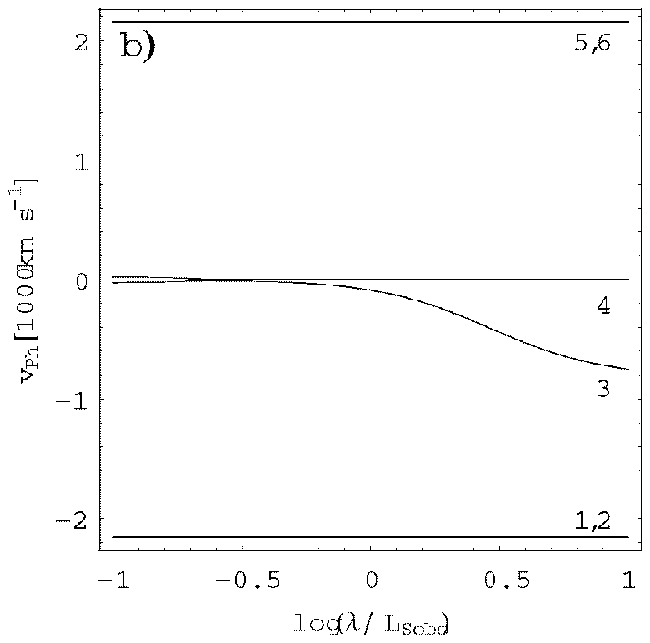,width=8cm,clip=,}}%
              %      bbllx=3.6cm,bburx=13.8cm,bblly=16.8cm,bbury=27.1cm}}
\end{picture}
\caption[]{\label{f:moda:a}Amplification timescales (a) and phase velocities 
(in the frame of the unperturbed wind) (b) versus
wavelength for model~A\&B and $r=2R$. In model~A only modes~3\&4 exist. Modes 
missing in Fig.~a) are stable. Mode~3 with $\lambda\approx\Lsobo$ has the 
shortest amplification timescale and therefore will dominate the wind. These 
inward running waves are advected outward with $\vrbar=811\rm km\ s^{-1}$ and 
steepen into reverse shocks.}
\end{center}
\end{figure}
\begin{figure}[p]
\begin{center}
\begin{picture}(80,170)
\put(0, 90){\epsfig{figure=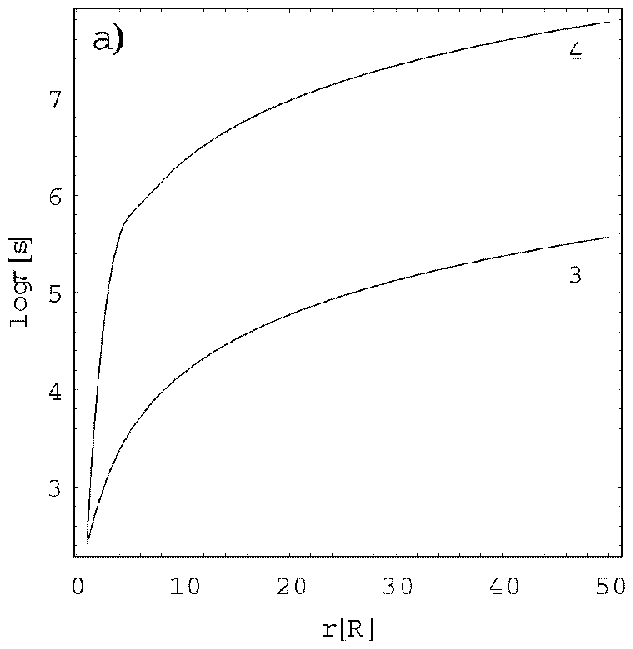,width=8cm,clip=,}}%
             %       bbllx=3.6cm,bburx=13.8cm,bblly=16.8cm,bbury=27.1cm}}
\put(0,  0){\epsfig{figure=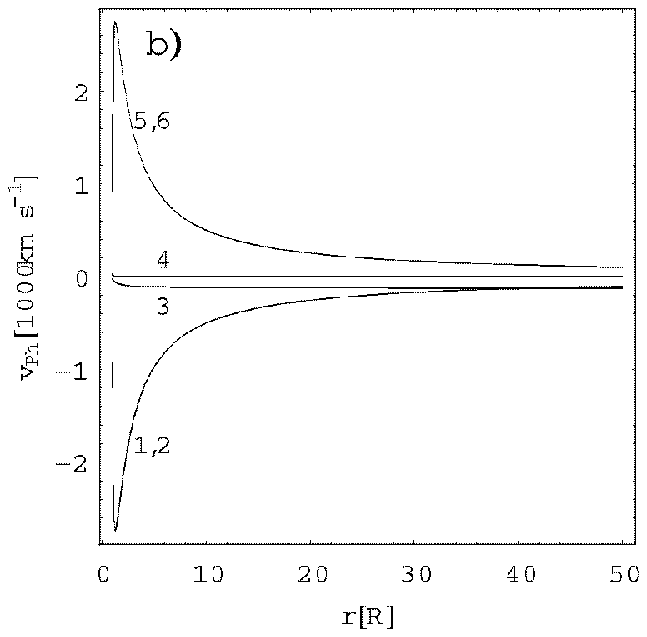,width=8cm,clip=,}}%
              %      bbllx=3.6cm,bburx=13.8cm,bblly=16.8cm,bbury=27.1cm}}
\end{picture}
\caption[]{\label{f:moda:b}Amplification timescales (a) and phase velocities 
(in the frame of the unperturbed wind) (b) versus radius 
for model~A\&B and $\lambda=\Lsobo$. In model~A only modes~3\&4 exist. 
Modes missing in Fig.~a) are stable. The wave amplification is strongest close
to the star, where the radiation field and wind acceleration are strong. Inward
running waves originating there will be advected outward and steepen into
reverse shocks.}
\end{center}
\end{figure}
Figures \ref{f:moda:a}\&\ref{f:moda:b} show numerical results for wind model
A\&B. The sonic wave modes\footnote{Ordered by the radial phase velocity we
denote in our figures the fast magnetosonic modes with 1\&6, the Alfv\'enic
modes with 2\&5, and the slow magnetosonic modes with 3\&4. The sonic modes of
model~A, which are identical to the slow magnetosonic modes of model~B, are
denoted with 3\&4 as well.} found for model~A are identical to the slow
magnetosonic modes of model~B\@. For model~B we find additionally four fast
wave modes. These modes, the fast magnetosonic and the Alfv\'enic modes, have
the same phase velocity $\vph \approx \vAbar$, because both, $\Bvec$ and
$\kvec$, are parallel to $\ervec$.  But these wave modes are stable and show no
dependence on wavelength. Therefore we concentrate our discussion for model
A\&B on the slow waves: Fig.~\ref{f:moda:a} shows the dependence of the waves
on the wavelength for $r=2R$, plotted relative to the Sobolev length, which is
the relevant length scale for radiative wave amplification in the model of
Owocki \& Rybicki \cite{Owocki:Rybicki:84}, we use here. $R=8.5\rsun$ is the
stellar radius. In the long wavelength limit we find stable waves with a high
phase velocity inward and a low phase velocity outward. This reproduces
Abbott's \cite{Abbott:80} result of radiative-acoustic waves. In the short
wavelength limit we find the same amplification timescales and phase velocities
(except the direction) for both modes. In this limit we would expect to see
both wave modes in the wind. But the most interesting case is the bridging case
$\lambda \approx\Lsobo$. Here we find the shortest amplification timescale for
all wavelengths. Inward waves are two orders of magnitude faster in
amplification than outward waves. They have also a higher phase velocity. This
resembles the result of Owocki et al.\ \cite{Owocki:etal:88}, who did a
nonlinear calculation and found that primarily the inward running waves steepen
into reverse shocks, which are advected outward. The aim of this chapter is to
argue that this scenario changes if a magnetic field {\em and} rotation are
involved. Figure~\ref{f:moda:b} shows the radial dependence of the wave with
$\lambda = \Lsobo(r)=\vth/v_r'$. The phase velocity for the slow magnetosonic
modes is approximately the velocity of sound, so that inward running waves are
advected outward in the supersonic part of the wind. The amplification of the
waves is strongest close to the star, where the Sobolev length is short. The
phase velocity of the fast modes goes with $r^{-1}$ for large radii since $B =
B_r \sim r^{-2}$. This might lead to a small velocity for outward running
shocks at large radii.

\begin{figure}[p]
\begin{center}
\begin{picture}(80,170)
\put(0, 90){\epsfig{figure=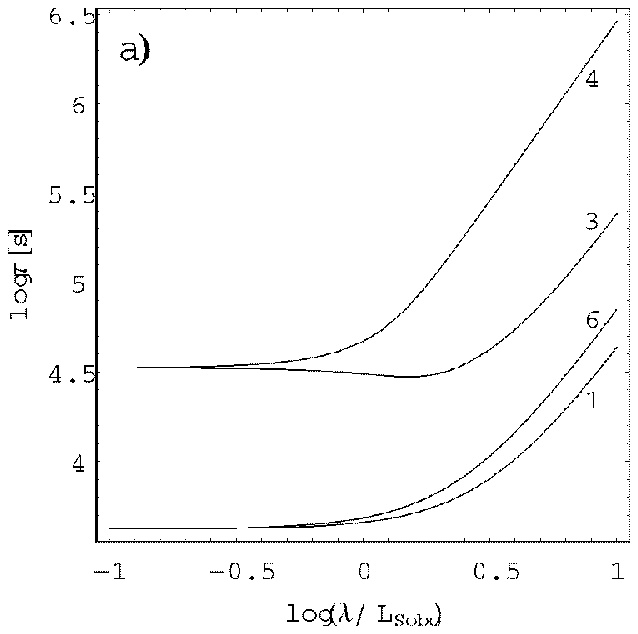,width=8cm,clip=,}}%
                 %   bbllx=3.6cm,bburx=13.8cm,bblly=16.8cm,bbury=27.1cm}}
\put(0,  0){\epsfig{figure=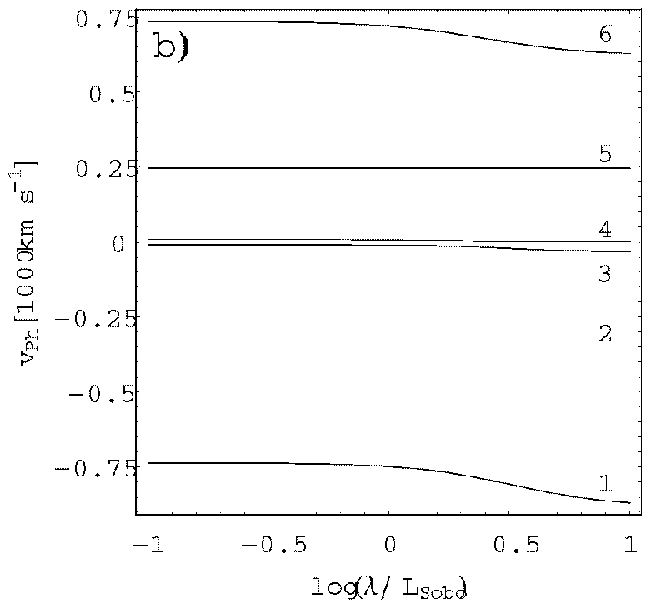,width=8cm,clip=,}}%
                  %  bbllx=3.6cm,bburx=13.8cm,bblly=16.8cm,bbury=27.1cm}}
\end{picture}
\caption[]{\label{f:modb:a}Amplification timescales (a) and phase velocities
(in the frame of the unperturbed wind) (b) versus wavelength for model~C and
$r=2R$. Modes missing in Fig.~a) are stable.  In the long wavelength limit all
modes are stable. In the short wavelength wavelength limit the fast
magnetosonic modes (1\&6) grow approximately one order of magnitude faster as
the slow magnetosonic modes (3\&4). We expect that fast magnetosonic waves in
both directions are equally dominant in the wind of model~C\@. Since the
magnetosonic waves are much faster than the unperturbed wind ($\vrbar=294\rm
km\ s^{-1}$), the inward running waves will not be advected outward. The stable
Alfv\'enic modes (2\&5) show no dependence on wavelength.}
\end{center}
\end{figure}
\begin{figure}[p]
\begin{center}
\begin{picture}(80,170)
\put(0, 90){\epsfig{figure=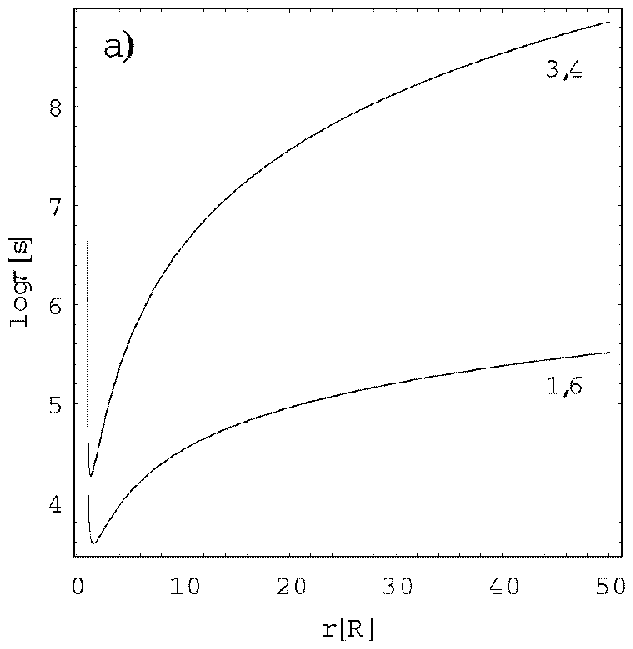,width=8cm,clip=,}}%
             %       bbllx=3.6cm,bburx=13.8cm,bblly=16.8cm,bbury=27.1cm}}
\put(0,  0){\epsfig{figure=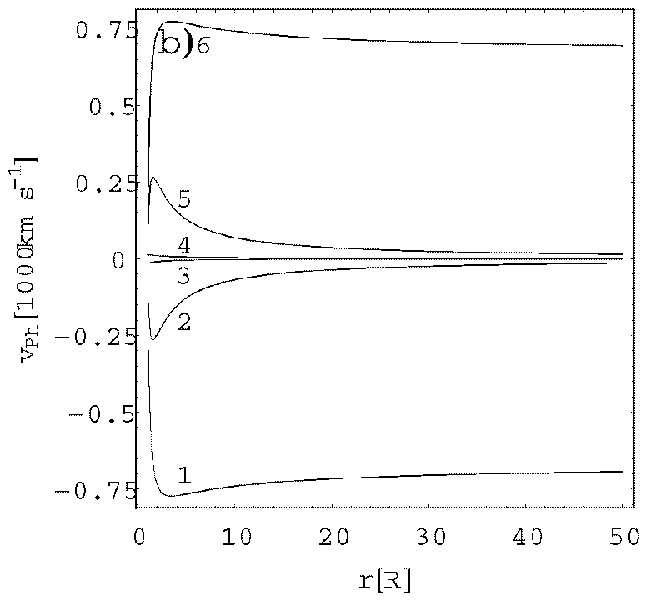,width=8cm,clip=,}}%
              %      bbllx=3.6cm,bburx=13.8cm,bblly=16.8cm,bbury=27.1cm}}
\end{picture}
\caption[]{\label{f:modb:b}Amplification timescales (a) and phase velocities 
(in the frame of the unperturbed wind) (b) versus radius for model~C and
$\lambda=0.1\Lsobo$. The Alfv\'enic modes (2\&5) missing in Fig.~a) are
stable. The wave amplification is strongest close to the star, where radiation
field and wind acceleration are strong. Fast magnetosonic waves originating
here will dominate the wind. They have the shortest amplification timescale for
all radii and a high phase velocity even at large radii. This may lead to fast
shocks running outward, which have a strong influence on observation.}
\end{center}
\end{figure}
Figures \ref{f:modb:a}\&\ref{f:modb:b} show the same plots for model~C -- with
magnetic field and rotation. The crucial point is that the magnetic field and
the amplifying stellar radiation are not parallel anymore. For large radii they
are even perpendicular. Therefore the fast magnetosonic wave modes, which are
most interesting for us, are amplified as well. Figure~\ref{f:modb:a}.b shows
six modes with different phase velocities. Two of them, the Alfv\'enic modes,
show no dependence on wavelength. They are unaffected by the radiation field
and therefore stable. Figure~\ref{f:modb:a}.a shows the amplification
timescales for the magnetosonic waves. The fast magnetosonic waves grow
approximately one order of magnitude faster than the slow waves. We can
therefore expect that these waves will dominate. The crucial point is that they
are much faster than the unperturbed wind -- especially close to the
star. Inward running fast waves will therefore not be advected away from the
star. Figure~\ref{f:modb:b}.b shows that the phase velocity for the fast
magnetosonic waves remain high for large radii. This velocity is a lower limit
for the velocity of outward running shocks. Rybicki et al.\
\cite{Rybicki:etal:90} showed that non-radial perturbations in the stellar
wind are damped close to the wind. But the magnetic field of our model~C is
mostly tangential already at the stellar surface. For the fast magnetosonic
modes $|\delta v_\phi/\delta v_r|$ is 0.5 at the stellar surface and 0.35 at
$r=2R$.  We expect therefore, that the effects found by Rybicki et al.\ will
influence but not completely dampen the magnetosonic waves. We emphasize that
fast magnetosonic waves propagate fastest perpendicular to the magnetic field
with $\vph=(\vAbarvec^2+\vsvs)^{0.5}$.

It is very speculative to draw conclusions for nonlinear waves and shocks from
a linear stability analysis. But we showed that a magnetic field has a strong
influence on the linear analysis. Therefore we expect a strong influence of the
magnetic field on nonlinear perturbations as well.  Our speculative scenario
for waves in the wind of hot stars is the following: Waves are predominantly
generated close to the star where the stellar radiation field is strong and the
Sobolev length is short. Waves with a wavelength short compared to the Sobolev
length are generated predominantly, because they have the shortest
amplification timescale. Inward running waves run into the star and disappear
because their phase velocity is higher than the unperturbed wind
velocity. Outward running waves, which have, for short wavelengths, the same
growth timescale than inward running waves, can run over many stellar radii and
grow. For the outward running waves $\deltarho$ and $\delta v_r$ are in phase.
Therefore they can steepen into forward shocks with regions of high density at
high velocity and influence the observation of the terminal velocity and mass
loss rate.

The picture drawn in the last section does not change significantly when
non-radial waves are taken into account. There are two new effects compared to
the radial case. A third wave mode appears in model~A as predicted by
Eq.~\ref{dispnB}. This inward running mode has a long amplification timescale
and a relatively low phase velocity. Therefore we do not expect a significant
contribution of this mode to the situation in the wind of model~A\@. For
model~B we find that Alfv\'en and fast magnetosonic waves have different phase
velocities now.  The fast magnetosonic waves are now amplified, too. But their
amplification timescale is much longer than for slow magnetosonic waves.
\section{Consequences of our model}
\label{Sec:consec}
In this chapter we do a linear stability analysis and show that the waves in
the stellar wind can help to understand the observations. In order to calculate
quantitative results, which can be compared with observations, it would be
necessary to analyze the detailed properties of the shocks resulting from the
waves found in this chapter. But we can derive some estimates from our
calculations.

From Fig.~\ref{f:modb:b}.b we see that the phase velocity of the outward
running waves and possibly shocks in the rest frame of the star is at least
twice the terminal velocity of the unperturbed wind. In the outward running
waves the oscillations of $v_r$ and $\rho$ are in phase. Therefore a
significant amount of matter, but presumably not all matter will escape at this
or a higher velocity, if the outward running waves steepen into forward
shocks. The terminal velocity will then be overestimated at least by a factor
of two in the observation. For our model~C this would be $\vinfobs \approx
1500{\rm km\ s^{-1}}$. Krolik \& Raymond \cite{Krolik:Raymond:85} found that
in a nonmagnetic wind shock shells are running much faster than the phase
velocity of the waves. In an unperturbed magnetic wind model such a high value
for $\vinf$ combined with an reasonable high value for $\Mdot$ can only be
obtained with a very high magnetic field and fast rotation, which leads to a
spin-down problem.

The influence of our model on $\Mdot$ is more difficult to estimate. We gain a
real factor on 28 in $\Mdot$ between our model A\&B and model~C even in the
unperturbed wind due to the driving force of the rotating magnetic
field. Furthermore the observation of $\Mdot$ is influenced by the clumping of
the wind matter. But to calculate the clumping factor
$<\!\rho^2\!>/<\!\rho\!>^2$ at large radii, where radio observations are made,
it would be necessary to do a nonlinear calculation including large radii,
because the waves steepen into shocks very rapidly due to the short
amplification timescales. This is beyond the linear model presented here.
\section{Conclusions}
\label{Sec:waveconclu}
In this chapter we analyzed the interaction between a magnetic field and linear
waves induced by the radiative instability. We found both models complement
each other. The magnetic field suppresses the inward running waves, which
dominate in nonmagnetic winds. This may allow the outward running waves to
support the unperturbed wind as described by Koninx \cite{Koninx:PhD} and to
form high density shock shells running out at a high speed. These shock shells
may explain the high terminal velocities measured in winds of massive
stars. The outward running waves will also lead to an overestimation of $\Mdot$
due to wind clumping and the overestimated $\vinf$.  Wind clumping also occurs
without a magnetic field. But in this case the resulting shock shells will run
inward; and the argument about the overestimated $\vinf$ would not apply. The
overestimated $\Mdot$ and $\vinf$ put unnecessarily strong restrictions on fast
magnetic rotator wind models. We argued that even for a magnetic field with
$B_{r0} = 500\rm G$ the spin-down time is consistent with the lifetime of the
star inferred from the mass loss rate considering the uncertainties in the
stellar structure. From the observation of nonthermal radio emission in many OB
and Wolf-Rayet stars we know that these stars have a non-negligible magnetic
field. Further direct observations are necessary to infer the actual strength
of these fields. Previous observations using the Zeeman effect were due to the
strong Doppler broadening of the lines not sensitive enough to measure reliably
the magnetic field strength in the winds of O and Wolf-Rayet stars
\cite{Landstreet:82}. It might be possible in the near future to measure these
magnetic field strengths using the Hanle effect
\cite{Ignace:etal:95,Ignace:etal:97}.
 
We showed that a luminous fast magnetic rotator model plus
wind perturbations by waves or shocks can help to explain the observed high
values for $\Mdot$ and $\vinf$ without being ruled out by the spin-down
problem. Further observation of the magnetic field and further theoretical work
on the evolution of stellar rotation are necessary to evaluate to role of
magnetic fields in winds of massive hot stars.
\begin{figure}
\begin{center}
\begin{picture}(80,85)
\put(0,0){\epsfig{figure=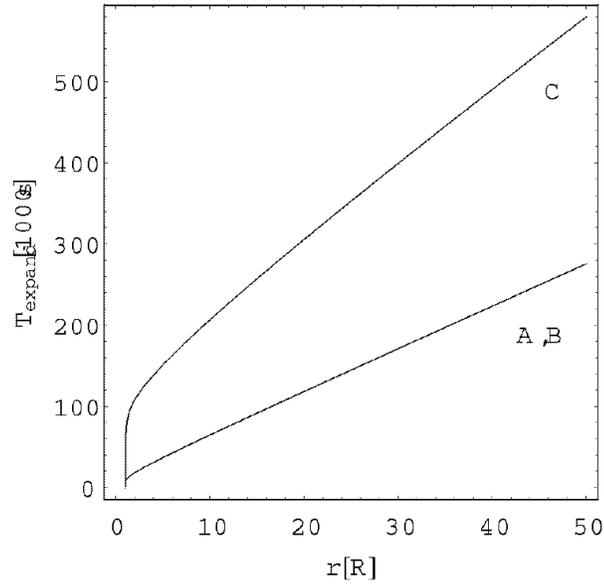,width=8cm,clip=,}}%
            %      bbllx=3.6cm,bburx=13.8cm,bblly=16.8cm,bbury=27.1cm}
\end{picture}
\caption[]{\label{f:exp}Expansion time versus radius. 
Model~C has a slower 
acceleration close to the star and a lower terminal velocity. A fast
magnetosonic wave with an amplification timescale of about $2500\rm s$
(cf. Fig~\ref{f:modb:b}.a) grows by a factor of $\approx {\rm e}^{40}$, while
the wind expands from $r=1R$ to $r=1.5R$. The electron scattering opacity is
$^2/_3$ at $r=1.3R$. Therefore we can expect shocks already at this radius,
where line observation starts. The objections of Lucy \cite{Lucy:84} do not
change this situation qualitatively.}
\end{center}
\end{figure}
%

%-*-LaTeX-*-
% This is the 7th chapter for the PhD-thesis of Henning Seemann.
% (c) 1997-98 by Henning Seemann
%
\chapter{The two dimensional fluxsheet model}
\label{Chap:fluxtube}
\section{Introduction}
In Chap.~\ref{Chap:cak:wd} we discussed an equatorial wind driven by
radiation and magnetic fields. In this chapter we extend our discussion to the
non-equatorial region. This is necessary to get better quantitative results for
mass and angular momentum loss (cf.\ Sect.~\ref{Sec:spindown}). It has also
been argued that the observed high mass loss rates and terminal velocities are
not created in the same regions of the wind \cite{Poe:Friend:86,
Poe:etal:89}. Furthermore it is possible to apply a non-equatorial model to
winds from accretion disks. This can be important for some types of stars and
for Active Galactic nuclei (AGN). It is obvious that disks rotate close to the
critical (Keplerian) rotation rate.  And it is widely assumed that the magnetic
field is a major mechanism to extract angular momentum from the disk and to
drive winds and jets \cite{Blandford:Payne:82}. But the role of radiation from
the central object or from a hot disk is often underestimated.

In this chapter we derive a model for a non-equatorial wind driven by radiation
and magnetic fields. Our model is again based on the fluxsheet concept and
therefore essentially one dimensional. But we generalize the model of
Chaps.~\ref{Chap:wd}\&\ref{Chap:cak:wd} by dropping the restriction that the
fluxsheet has to lie in the equatorial plane. In App.~\ref{App:limit} we
will show the equations of Chaps.~\ref{Chap:wd}\&\ref{Chap:cak:wd} follow
from the generalized equations developed here. At the end of the chapter we
will show results from first numerical exploration of the model.

In the first part we follow closely the derivation of Lovelace et al.\
\cite{Lovelace:etal:86} to reduce the equations for a rotational symmetric,
rotating, stationary, and perfectly conducting MHD flow to the fluxsheet
concept. Since we are primarily analyzing winds from stars and not from disks
we shifted the derivation from cylindrical to spherical coordinates. The goal
of Lovelace et al.\ is to derive an ordinary differential equation for the flow
in the fluxsheet from Bernoullis equation of energy conservation. This is
reasonable for a pure MHD model. But we can not go this way because, opposite
to Lovelace et al., we want to include radiation as a major wind driving
mechanism. We saw in Chap.~\ref{Chap:cak} that we have no simple theory
for the energy exchange between radiation an matter. The Thompson and the CAK
theory, both give us only a source term for momentum.  Therefore we leave the
path of Lovelaces model and derive a final ordinary differential equation from
Eulers equation of momentum balance. This equation is a vector equation.  We
take only the component parallel to the a priori given fluxsheet into
account. The component perpendicular to the fluxsheet can be used to improve
the shape of the fluxsheet. 

In Sect.~\ref{Sec:basicnoneq} we follow to derivation of Lovelace et al. for
the basic equations in the fluxsheet description. In Sect.~\ref{Sec:fluxcoord}
we introduce our fluxsheet coordinate system with its a priori given shape. In
Sect.~\ref{Sec:oneEuler} we use the results of the two previous sections to
derive our final wind equation as a single ordinary differential equation. In
Sect.~\ref{Sec:transf} we derive the transverse component of the Euler
equation, which was neglected in the previous section. It allows an improvement
of the chosen fluxsheet shape. and finally we will give some preliminary
numerical results in Sect.~\ref{Sec:numnoneq} to show the capabilities of this
model. 
\section{The basic model equations}
\label{Sec:basicnoneq}
As in Chap.~\ref{Chap:wd} we start from the equations of ideal
magneto-hydrodynamics (MHD) in a quasi stationary situation
(Eq.~\ref{MassConservationMHDc}--\ref{Fextdef}).
Again we assume that the wind is rotational symmetric and neglect therefore all
dependencies on the azimuthal angle $\phi$. But now we do not set $v_\theta$
and $B_\theta$ equal to zero as we did in Chap.~\ref{Chap:wd}. Rather we use
the perfect conductivity equation
\begin{equation}
\Evec+\frac{1}{c}\vvec\times\Bvec=0
\end{equation}
and the axisymmetry condition $E_\phi=0$ to find
\begin{eqnarray}
\vpvec &\equiv& v_r \ervec + v_\theta \ethetavec\\
       &=& \kappa(r,\theta)\Bpvec,
\end{eqnarray}
where the subscript ``p'' denotes the poloidal component of the vector. From $0
= \nabla\cdot(\rho\vvec) = \nabla\cdot(\rho\vpvec) =
\nabla\cdot(\rho\kappa(r,\theta)\Bpvec)$, we know that
\begin{equation}\label{kappaconst}
\Bpvec\cdot\nabla(\rho\kappa(r,\theta))=0,
\end{equation}
because $\nabla\cdot\Bpvec$ equals $\nabla\cdot\Bvec=0$. The poloidal magnetic
field can be expressed in terms of the flux function
\begin{equation}\label{fluxfunction}
\Psi = r \sin\theta A_\phi(r,\theta), 
\end{equation}
where $A_\phi$ is the azimuthal component of the vector potential. $\Psi$ is 
constant on every fluxsheet, i.e.
\begin{equation}\label{BpperdPsi}
\Bpvec\cdot\nabla\Psi=0.
\end{equation}
From Eq. \ref{kappaconst} we get
\begin{equation}
4\pi\rho(r,\theta)\kappa(r,\theta)=F(\Psi),
\end{equation}
where $F(\Psi)$ is a function whose only relevant feature is that $F$ is
constant along a fluxsheet.
Using $|\nabla\Psi|/(r\sin\theta)$ $=\Bp$ and Eq.~\ref{BpperdPsi} we get
\begin{equation}\label{btv}
\vvec\times\Bvec=\frac{1}{r\sin\theta}(\vphi-\kappa\Bphi)\nabla\Psi.
\end{equation}
From Eqs.~\ref{fluxfreezinga}\&\ref{btv} we know that
\begin{equation}
\frac{1}{r\sin\theta}(\vphi-\kappa\Bphi)=\omegabar(\Psi),
\end{equation}
where $\omegabar(\Psi)$ is another arbitrary function of $\Psi$. From
Eq.~\ref{btv} we get the electric field
\begin{equation}
\Evec=-\frac{1}{c}\omegabar(\Psi)\nabla\Psi
\end{equation}
We can now express the velocity field by
\begin{eqnarray}
\vvec &=& \frac{F(\Psi)}{4\pi\rho}\Bpvec+\left[\frac{F(\Psi)}{4\pi\rho}\Bphi+
          r\sin\theta \omegabar(\Psi)\right]\ephivec\\
\label{replaceB}
      &=& \frac{F(\Psi)}{4\pi\rho}\Bvec + \omegabarvec(\Psi)\times\rvec,
\end{eqnarray}
where $\omegabarvec$ equals $\omegabar(\Psi)\ez$. And $\ez$ is the unit vector
parallel to the rotation axis. At the base of the wind, where $\rho$ is large,
$\vphi$ equals $\omegabar r\sin\theta$. Therefore $\omegabar$ is the angular
velocity of the rotating fluxsheet. Using the toroidal component of
Eq.~\ref{SheetWinda} we find
\begin{equation}
\Bpvec\cdot\nabla(r\sin\theta(\Bphi-F\vphi))=0,
\end{equation}
which implies that 
\begin{equation}
H(\Psi)=r\sin\theta(\Bphi-F\vphi)
\end{equation}
is another arbitrary function of $\Psi$. This corresponds to the conservation
of angular momentum for matter moving on a given flux surface, $\Psi=\rm
 const$. Now we can express $\vphi$ and $\Bphi$ by
\begin{eqnarray}
\label{Bphiequ}
r\sin\theta\Bphi &=& \frac{H+r^2\sin^2\theta F\omegabar}{1-\Mp^2}\\
\label{vphiequ}
r\sin\theta\vphi &=& \frac{1}{1-\Mp^2}
     \left(r^2\sin^2\theta \omegabar+\frac{FH}{4\pi\rho}\right),
\end{eqnarray}
where we have used the poloidal Alfv\'en Mach number
\begin{eqnarray}
\Mp &=& \sqrt{\frac{4\pi\rho\vpvec^2}{\Bpvec^2}}\\
    &=& \frac{F}{\sqrt{4\pi\rho}}.
\end{eqnarray}
On the Alfv\'en surface $(r = \rAc, \theta = \thAc,\Mp=1)$ the
denominator of Eqs.~\ref{Bphiequ} \& \ref{vphiequ} vanishes. Since $\Bphi$ and
$\vphi$ must be finite on the Alfv\'en surface, we get the conditions
\begin{eqnarray}\label{defH}
H                 &=& -\rAc^2 \sin^2\thAc F\omegabar\\
\rAc^2 \sin^2\thAc \omegabar &=& -\frac{FH}{4\pi\rhoAc}.
\end{eqnarray}
From l'Hospital's rule we find
\begin{eqnarray}
(r\sin\theta\Bphi)_\mathrm{A} &=& -H\frac{d\left[\ln(r^2\sin^2\theta)\right]_
                               \mathrm{A}/ds}
                               {d(\ln\rho)_\mathrm{A}/ds}\\
(r\sin\theta\vphi)_\mathrm{A} &=& (r\sin\theta)^2_\mathrm{A}\omegabar\left[1+
                               \frac{d\left[\ln(r^2\sin^2\theta)\right]/ds}
                               {d(\ln\rho)_\mathrm{A}/ds}\right]\\
                           &=& (r\sin\theta)^2_\mathrm{A}\omegabar\left[1-
                               \frac{(r\sin\theta\Bphi)_\mathrm{A}}{H}\right],
\end{eqnarray}
where $s$ is the arc length along the poloidal projection of the field line,
i.e. $d\Psi/ds=0.$
\section{The fluxsheet coordinate system}
\label{Sec:fluxcoord}
Up to this point our model is exact besides the approximation of rotational
symmetry, stationarity, no viscosity, and perfect conductivity. But therefore
we have still a system of two partial differential equations with two dependent
$(v_r,v_\theta)$ and two independent $(r,\theta)$ variables. Such a system was
solved by Sakurai for a stellar wind without radiation \cite{Sakurai:85}.

\begin{figure}
\begin{center}
\begin{picture}(50,80)
%\put(0,0){\framebox(50,80){}}
\put(0,0){\epsfig{file=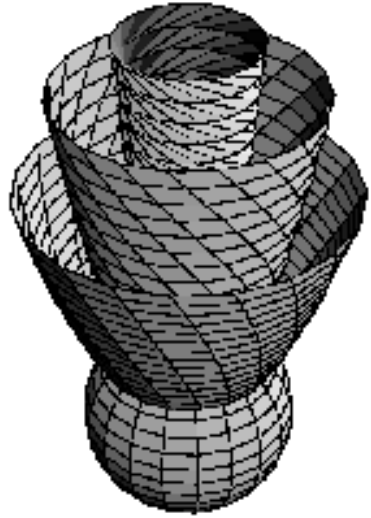,width=5cm}}
\end{picture}
\end{center}
\caption[The geometry of the fluxsheets]{\label{Fig:FluxSheetDemo} shows the
stellar surface and three members of the infinite set of fluxsheets. Every
fluxsheet has a different rotation rate $\omegabar(\eta)$. Whether $\omegabar$
is higher at the poles or at the equator is not clear yet.}
\end{figure}

In order to reduce the problem from two to one dimension we split the three
dimensional space into a one-parameter family of two dimensional fluxsheets.
Figure~\ref{Fig:FluxSheetDemo} shows how these fluxsheets fit into each other
like the leaves of a blossom. Every fluid element will flow along a single
fluxsheet. The magnetic field lines are confined in the individual fluxsheets
as well, because we assume flux freezing and a quasi stationary model. Due to
the rotational symmetry of the whole wind, the fluxsheets are rotational
symmetric as well. Fluxsheets exist even in a model which solves the full MHD
equations. In this case we would find simultaneously the shape of the
fluxsheets and the motion of the matter within the fluxsheets. We already
reduced the problem from three to two spatial dimensions by assuming rotational
symmetry. Now we reduce it again by one spatial dimension by using an a priori
given shape for the fluxsheets and only asking for the motion within the
fluxsheets. Falcke \cite{Falcke:Msc} integrated the equations of an radiation
driven nonmagnetic wind simultaneously for the wind velocity and the shape of
the fluxsheet. This is much more difficult for a magnetic wind, since our
solution has to pass through three instead of one critical point. And this will
only happen once we found the two eigenvalues of the equations. The wind
solution has to pass through three critical point if it should extend from the
subsonic base of the wind to arbitrary large radii.

We require that the shape of the fluxsheet is given by $r(u,\eta)$ and
$\theta(u,\eta)$, where $\eta$ is constant along the fluxsheet and $u$ is an
arbitrary parameter along the fluxsheet. Since the parameter $\eta$ is constant
along a fluxsheet, it depends only on the flux function $\Psi$ introduced in
Eq.~\ref{fluxfunction} and vice versa. The reason for using $\eta$ instead of
the already introduced parameter $\Psi$ is that we can choose it arbitrary in
the same way as $u$. It just distinguishes different fluxsheets, while $\Psi$
has a deeper physical meaning. For our purpose it is not necessary to know
$\Psi$ explicitly. Therefore it is numerically simpler to introduce
$\eta$. $\eta$ and $u$ are a curved coordinate system for the meridional
$(r,\theta)$ plane which is adapted to the shape of the fluxsheets. It is
therefore more appropriate for solving the wind equations than the $(r,\theta)$
system. We do not give explicit formulas for $r(u,\eta)$ and $\theta(u,\eta)$
here, so that the shape of the fluxsheets remains yet unspecified. This has the
advantage that the wind equations, we derive now, remain independent of the
explicit shape of the fluxsheet. Later, when we start our numerical
computations, we have to specify the shape of the fluxsheets through explicit
formulas for $r(u,\eta)$ and $\theta(u,\eta)$. But then we can easily compute
and compare models for different fluxsheet shapes by using different explicit
formulas for $r(u,\eta)$ and $\theta(u,\eta)$.

\begin{figure}
\begin{center}
\begin{picture}(90,50)
%\put(0,0){\framebox(90,50){}}
\put(0,0){\epsfig{file=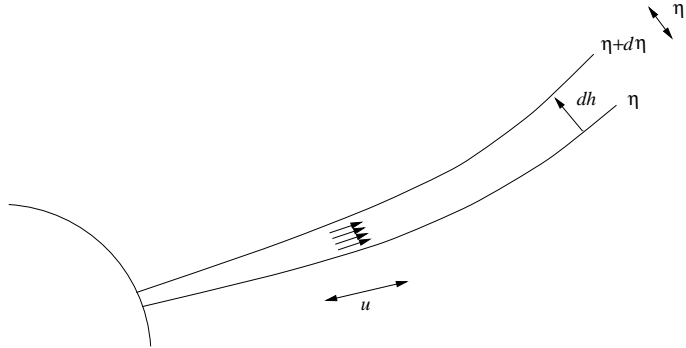,width=9cm}}
\end{picture}
\end{center}
\caption[The fluxsheet cross section factor]{\label{Fig:SheetSep}A
non-equatorial fluxsheet in the $(r,\theta)$ plane. In this plane the thickness
of the sheet is $dh(u)$. The rotational symmetry contributes a factor of $2\pi
r\sin\theta$ to the total cross section of the fluxsheet.}
\end{figure}

Now we can define a unit vector $\epvec$ tangential to the fluxsheet in
poloidal direction given by
\begin{eqnarray}
\epvec(u,\eta) &=& \frac{s\drdu}{\sqrt{\drdu^2+r^2\dthdu^2}}\ervec + 
                     \frac{sr\dthdu}{\sqrt{\drdu^2+r^2\dthdu^2}}\ethetavec\\
               &=& \epr(u,\eta)\ervec+\ept(u,\eta)\ethetavec,
\end{eqnarray}
where the dot $\dot{ }$ refers to the partial derivative with respect to $u$.
The constant $s=|\drdu|/\drdu=\pm1$ fixes for later convenience the orientation
of $\epvec$ so that $\epvec$, per definition, always points outward.  We can
now express $\vvec$ and $\Bvec$ as functions of the fluxsheet coordinates $u$
and $\eta$:
\begin{eqnarray} 
\vvec    &=& \vp(u,\eta)\epvec + \vphi(u,\eta)\ephivec\\
\Bvec    &=& \Bp(u,\eta)\epvec + \Bphi(u,\eta)\ephivec\\
         &=& \frac{4\pi\rho(u,\eta)}{F(\eta)}(\vvec-
               \omegabar(\eta)\times\rvec)
\end{eqnarray}
In order to describe $\rho$ and $\Bp$ in the new coordinates we need an
expression for the cross section of a fluxsheet. Then we can use the
conservation of matter and magnetic flux as we did in
Eqs.~\ref{FluxConservationa}\&\ref{MassConservationMHDa}. In
Fig.~\ref{Fig:SheetSep} we show the wind between two infinitesimal separated
fluxsheets. 
The cross section $dA$ between the two fluxsheets is 
\begin{equation}
dA = 2\pi r\sin\theta |dh|.
\end{equation}
The separation between the two fluxsheets $dh$ varies along the fluxsheets. But
we can relate it to $d\eta$, which is constant along the fluxsheets:
\begin{equation}
d\eta = \diff{\eta}{h}\,dh(u)
\end{equation}
Since $\eta$ is constant along every fluxsheet and $dh$ is measured
perpendicular to the fluxsheets, we can express $\partial\eta/\partial h$ by
\begin{equation}
\diff{\eta}{h} = |\nabla\eta(r,\theta)|.
\end{equation}
Finally we find for the cross section between the fluxsheets
\begin{equation}
dA = 2\pi r\sin\theta\frac{|d\eta|}{|\nabla\eta|}.
\end{equation}
Now we use $\Bp dA=\const$, $\rho\vp dA=\const$, and $d\eta=\const$ between the
two fluxsheets to express
\begin{eqnarray}
\label{rhoequ}
\rho &=& \frac{\rAc\sin\thAc}{r\sin\theta}\frac{\vpAc}{\vp}
             \frac{|\nabla\eta|}{|\nabla\eta|_\mathrm{Ac}}\rhoAc(\eta)\\
\Bp &=& \frac{R\sin\theta_0}{r\sin\theta}\frac{\AbsGradeta}{\AbsGradeta_0}\Bpo.
\end{eqnarray}
The poloidal magnetic field strength at the base of the wind $\Bpo$ is an
input parameter of our models. The matter density at the Alfv\'enic critical
point $\rhoAc$ will be derived from model input parameters later. Alternatively
we can relate the matter density to the mass loss rate in the fluxsheet
\begin{equation}
\label{masslossrate}
\diff{\Mdot}{\eta} = \rho\vp \frac{2\pi r\sin\theta}{|\nabla\eta|}.
\end{equation}
If we want to calculate the real mass loss rate for our model star we must
solve the wind equation for every fluxsheets from the equator to the poles and
then integrate the values found for $\partial\Mdot/\partial\eta$ over all flux
sheets:
\begin{equation}
\label{defMdotreal}
\Mdot_\mathrm{real} = 2\int_{\eta(\mathrm{pole})}^{\eta(\mathrm{equator})}
                      \diff {\Mdot}{\eta}\,d\eta
\end{equation}
In order to get a better intuition for the mass loss at different latitudes it
is helpful to assume for a moment that the density and the wind velocity are
constant at the base of the wind. Under this approximation we can integrate
Eq.~\ref{defMdotreal} neglecting the oblateness of the rotating star and find
\begin{equation}
\label{defMdotapprox}
\Mdot_\mathrm{approx}(\eta) = 4\pi R^2 \rho_0(\eta) v_{\mathrm{p}0}(\eta).
\end{equation}
This quantity is analogous to mass loss rate used in Chap.~\ref{Chap:cak:wd}
as one of the eigenvalues we have to find numerically. 
\section{The one dimensional Euler equation}
\label{Sec:oneEuler}
The standard way in disk wind physics is to derive $\vp$ from Bernoullis
equation, which essentially describes the conservation of energy in the
fluxsheet. We will use Eulers momentum equation (Eq.~\ref{SheetWinda}) in order
to include the momentum terms for radiation pressure. Since we have already
expressed $v_\phi$ in terms of $u$, $\eta$, and $\vp$ we can neglect the $\phi$
component of Eq.~\ref{SheetWinda}. Theoretically it would be sufficient to
integrate either the $r$ or the $\theta$ component of Eq.~\ref{SheetWinda}. But
this would lead to large numerical errors when the flow direction $\epvec$ has
a significant component perpendicular to the chosen component of
Eq~\ref{SheetWinda}. Therefore we multiply Eq.~\ref{SheetWinda} with $\epvec$
and obtain a single ordinary differential equation in $\vp$. This equation
allows us to integrate $\vp$ along the fluxsheet. To obtain a solution for the
complete wind we have to integrate $\vp$ along several fluxsheets.

Although we neglect for a while the force balance perpendicular to the
fluxsheet, we can not neglect that $\vp$ depends on $u$ and $\eta$. We solve
this problem by separating the variables:
\begin{equation}
\vp(u,\eta) = V(\eta)v(u),
\end{equation}
where we use an a priori given function $V(\eta)$. This is the analogy of the
\emph{self similar} solutions of Blandford \& Payne,
who a priori assumed certain powerlaw dependencies of $\rho$ and $B$ on the
cylindrical radius. Now we have a single differential equation with $v(u)$ as
the only dependent variable.

But this equation still contains the differential operators $\partial_r$ and
$\partial_\theta$. We transform these into our fluxsheet coordinates by
\begin{eqnarray}
\diff{}{r}     &=& \dudr \left(\diff{}{u}+\dvdu\diff{}{v}\right)+
                   \detadr\diff{}{\eta}\\
\diff{}{\theta} &=& \dudth\left(\diff{}{u}+\dvdu\diff{}{v}\right)+
                     \detadth\diff{}{\eta},
\end{eqnarray}
where we take into account, that $\rho$, $\vphi$, and $\Bvec$ depend on $u$,
$\eta$, and $y$. For brevity we will use the following convention for
partial derivatives:
\begin{equation}
\parbox{3.5cm}{
\begin{eqnarray*}
\frac{\partial f}{\partial r}      &=& f^\prime\\
\frac{\partial f}{\partial u}      &=& \dot{f}\\
\frac{\partial f}{\partial x}      &=& \tilde{f}
\end{eqnarray*}
}
\parbox{3.5cm}{
\begin{eqnarray*}
\frac{\partial f}{\partial \theta} &=& \breve{f}\\
\frac{\partial f}{\partial \eta}   &=& \check{f}\\
\frac{\partial f}{\partial y}      &=& \hat{f}.
\end{eqnarray*}
}
\end{equation}
Here we used the already the dimensionless quantities $x=r/r_*$ and $y=v/v_*$.
We can now express Eq.~\ref{SheetWinda} multiplied by $\epvec$:
\begin{equation}\label{SheetWindb}
0= \epvec\cdot\left[\left(\vvec\cdot\nabla\right)\vvec+
     \frac{\nabla p}{\rho}+\nabla\Phi-\gradvec\right]+
     \left(\epvec\times\frac{\Bvec}{4\pi\rho}\right)\cdot
     \left(\nabla\times\Bvec\right)
\end{equation}

Doing all these substitutions and simplifying the result to a readable form
includes a lot of complicated algebra. We used therefore the software package
\textsc{Mathematica} \cite{Mathematica} to do most of the steps. With a small
\textsc{Mathematica} program it is possible to show that
Eq.~\ref{SheetWindb} is equivalent to
\begin{equation}\label{SheetWindc}
A_1(u,v)\dvdu+B_1(u,v,\dvdu)=0
\end{equation}
with
\begin{eqnarray}
A_1(u,y)&=&(\drdu\dudr+\dthdu\dudth)
\left[v V^2-\frac{\vsvs}{v}+\frac{4\pi\rho}{F^2}\vphibar
           \left(\dvphidv-\frac{\vphibar}{v}\right)\right]\\
B_1(u,y,\dydu)&=&(\drdu\dudr+\dthdu\dudth)
             \left[\vsvs\frac{\drhodu}{\rho}+\dvsvsdu+\frac{4\pi\rho}{F^2}
             \vphibar\left(\frac{\drhodu}{\rho}\vphibar+\dvphidu\right)\right]+
             \nonumber\\
&&           (\drdu\detadr+\detadth\dthdu)\left[V\dVdeta v^2 + 
             \vsvs\left(\frac{\drhodeta}{\rho}-\frac{\dVdeta}{V}\right)+
             \partial_\eta\vsvs+\right.\nonumber\\
&&           \left.\frac{4\pi\rho}{F^2}\vphibar\left(\dvphideta-
             \domegabardeta r \sin\theta +
             \vphibar \left(\frac{\drhodeta}{\rho}-
             \frac{\dVdeta}{V}-\frac{\dFdeta}{F}\right)\right)\right]-
             \nonumber\\
&&           \left(\frac{\drdu}{r}+\frac{\dthdu}{\tan\theta}\right)
             \left[\frac{4\pi\rho}{F^2}
            \vphibar\left(r\sin\theta\omegabar-\vphibar\right)+\vphi^2\right]+
             \nonumber\\
&&           \drdu\left(\nabla\Phi-\grad(u,v,\dvdu)\right)\\
\vphibar &=& \vphi - r \sin\theta\omegabar.
\end{eqnarray}
We can make this equation dimensionless by multiplying it with
$1/(x^2\nabla\Phi)=r_*^2/(GM)$ and introducing the length and velocity scales
we used already in Chap.~\ref{Chap:wd}. Here we need additionally the special
rules
\begin{eqnarray}
vV     &=& v_* y Y\\
\omega &=& \frac{r_*}{v_*}\omegabar,
\end{eqnarray}
where $Y(\eta)$ is just a dimensionless version of $V(\eta)$. We get then
\begin{equation}
\label{SheetWindd}
A_2(u,y)\dydu+B_2(u,y,\dydu) = 0
\end{equation}
\begin{eqnarray}
A_2(u,y)&=&(\dxdu\dudx+\dthdu\dudth)
           \left[yY^2-\frac{\ysys}{y}+\frac{\yphibar}{\Mp^2}
           \left(\dyphidy-\frac{\yphibar}{y}\right)\right]\\
B_2(u,y,\dydu)&=&(\dxdu\dudx+\dthdu\dudth)\left[\ysys\frac{\drhodu}{\rho}+
           \dysysdu+\frac{\yphibar}{\Mp^2}\left(\frac{\drhodu}{\rho}\yphibar
           +\dyphidu\right)\right]+\nonumber\\
&&      (\dxdu\detadx+\dthdu\detadth)\left[Y\dYdeta y^2 + 
           \ysys\left(\frac{\drhodeta}{\rho}-\frac{\dYdeta}{Y}\right)+
           \dysysdeta+\right.\nonumber\\
&&      \left.\frac{\yphibar}{\Mp^2}\left(\dyphideta-
           \domegadeta x \sin\theta+
           \yphibar \left(\frac{\drhodeta}{\rho}-
           \frac{\dYdeta}{Y}-\frac{\dFdeta}{F}\right)\right)\right]-
           \nonumber\\
&&      \left(\frac{\dxdu}{x}+\frac{\dthdu}{\tan\theta}\right)
           \left[\frac{\yphibar}{\Mp^2}\left(x\sin\theta\omega-
           \yphibar\right)+\yphi^2\right]+\nonumber\\
&&      \dxdu\left(\frac{1}{x^2}-\arad(u,y,\dydu)\right).
\end{eqnarray}
If we use the line-driving mechanism of Castor, Abbott, and Klein
\cite{Castor:etal:75} plus electron scattering as description for the radiative
force, $\arad$ is given by
\begin{eqnarray}
\arad(u,y,\dydu) &=& \frac{r_*^2}{GM}\grad\\
       &=& \frac{\Gamma}{x^2} + \frac{\kcak\Gamma}{x^2}
             \left(\frac{v_*}{\kappaTh\vth\rho r_*}\right)^\alphacak
             \left|\diff{y_x}{x}\right|^\alphacak\\
       &=& \frac{\Gamma}{x^2} + \frac{C(u,y)}{\dxdu} 
             \left|\diff{y_x}{x}\right|^\alphacak\\
\label{dyxdxdef}
\dyxdx = \diff{y_x}{x} &=& 
          \left(\dydu\epr+y\deprdu\right)Y\dudx+
          \left(\dYdeta\epr+Y\deprdeta\right)y\detadx.
\end{eqnarray}
The derivation of the CAK force in Chap.~\ref{Chap:cak} shows that the CAK
force depends on the gradient of the velocity in the direction of the driving
radiation. We want to analyze the wind from a star. Therefore the photons
propagate in radial direction. If this model would be applied to systems where
a disk dominates the radiation field, the equation have to be modified at this
point. Outside the equatorial fluxsheet the poloidal flow $\vpvec$ and the
poloidal gradient $\partial_u$ are in general not in radial direction. This
leads to the complicated relation between the two spatial derivatives of the
wind velocity, $|\dyxdx|$ and $\dydu$. Using $|\dyxdx|$ instead of $\dydu$ is
the simplest approximation for the fact that in a curved fluxsheet geometry the
radiation passes through different fluxsheets. In order to compare the results
from this chapter with the results from Chap.~\ref{Chap:cak:wd} we rewrite
Eq.~\ref{SheetWindd} analogous to Eq.~\ref{CAKWind}. This allows us to use the
same arguments about critical points and reduces the necessary modifications in
our software. Since the CAK force term is now proportional to
$|\dyxdx|^\alphacak$ instead of $\dydu^\alphacak$ we use $|\dyxdx|$ as the
local unknown quantity. This leads to
\begin{equation}\label{SheetWinde}
A(u,y)\left|\dyxdx\right|+B(u,y)  = C(u,y) \left|\dyxdx\right|^\alphacak.
\end{equation}
Before we can use this equation we have to transfer $A_2$ and $B_2$ into $A$
and $B$. Once we know $|\dyxdx|$, we need a formula which gives us $\dydu$ as
function of $|\dyxdx|$. This requires to distinguish the cases $\dyxdx>0$ and
$\dyxdx<0$. Since we are only interested in expanding winds in reasonable
fluxsheets we can savely assume that the first case is always true. Then we
find
\begin{eqnarray}
A(u,y) &=& \frac{A_2}{\epr Y \dudx}\\
B(u,y) &=& B_3 - A_2 \frac{y\deprdu Y\dudx+
             \left(\dYdeta\epr+Y\deprdeta\right)y\detadx}{\epr Y\dudx}\\
\label{dydudef}
\dydu &=& \frac{\dyxdx-y\deprdu Y\dudx -
             \left(\dYdeta\epr+Y\deprdeta\right)y\detadx}{\epr Y\dudx},
\end{eqnarray}
where $B_3(u,y)=B_2+C\dyxdx^\alphacak$ is just $B_2$ without the line
acceleration term, which appears separately on the right hand side of
Eq.~\ref{SheetWinde}.  The deaccelerating solution $(\dyxdx<0)$ differs from
the result above only by a few signs.  The auxiliary quantities of
Eq.~\ref{SheetWinde} are given by
\begin{eqnarray}\label{auxa}
\mu=\frac{\drhodu}{\rho} &=& \frac{\dAbsGradetadu}{\AbsGradeta}-
                       \frac{\dthdu}{\tan\theta}-
                       \frac{\dxdu}{x}\\
\gamma=\frac{\drhodeta}{\rho}&=& \frac{\dAbsGradetadeta}
          {|\nabla\eta|}-\frac{\dthdeta}{\tan\theta}-\frac{\dxdeta}{x}-
          \frac{\dYdeta}{Y}+\frac{\drhoAcdeta}{\rhoAc}\\
\frac{\dFdeta}{F} &=& \frac{1}{2}\frac{\drhoAcdeta}{\rhoAc}\\
\Mp^2 &=& \frac{\rhoAc}{\rho}M^2_\mathrm{pAc} 
       = \frac{x\sin\theta}{\xAc\sin\thAc}\frac{y}{\yAc}
          \frac{\AbsGradeta_\mathrm{Ac}}{\AbsGradeta}\\
\label{auxe}
\yphi &=& \frac{\yrot^2-\Mp^2\yrotAc^2}{\yrot(1-\Mp^2)}\\
\frac{\partial\yphi}{\partial\Mp^2} &=& 
          \frac{\yrot^2-\yrotAc^2}{\yrot(1-\Mp^2)^2}\\
\dyphidu &=& \frac{\yrot^2+\Mp^2\yrotAc^2}{\yrot^2(1-\Mp^2)}\dyrotdu-
           \frac{\partial\yphi}{\partial\Mp^2}\mu\Mp^2\\
\label{defdyphideta}
\dyphideta &=& \frac{\yrot^2+\Mp^2\yrotAc^2}{\yrot^2(1-\Mp^2)}\dyrotdeta-
           \frac{\partial\yphi}{\partial\Mp^2}\gamma\Mp^2-
           \frac{2\yrotAc\dyrotAcdeta\Mp^2}{\yrot(1-\Mp^2)}\\
\dyphidy &=& \frac{\partial\yphi}{\partial\Mp^2}\frac{\Mp^2}{y}\\
C(u,y) &=& \dxdu\frac{\bar{C}}{x^2}\left(\frac{xy\sin\theta}
           {\vth\AbsGradeta}\right)^\alphacak\\
\label{auxz}
\bar{C} &=& \kcak\Gamma\left(\frac{v_*\AbsGradeta_\mathrm{Ac}}
            {\kappaTh r_*\xAc\sin\thAc\yAc\rhoAc}\right)^\alphacak
\end{eqnarray}
\section{The transverse force component}
\label{Sec:transf}
The model developed in this chapter so far allows us to compute the wind in a
fluxsheet of predefined shape. This is not very satisfying, because if we
choose by chance a fluxsheet whose shape differs too much from the real shape
we will get wind solutions without physical meaning. On the other hand the
numerical effort for finding the shape of the fluxsheet and the wind solution
simultaneously is much greater. In this case we have essentially a
2-dimensional problem. The compromise between these two extreme alternatives is
to assume a priori a shape for the fluxsheet, to calculate the wind solution
and finally to use the transverse component of the Euler equation to check
whether we have chosen a reasonable shape of the fluxsheet. These two steps
could even be iterated to get a consistent overall solution. Alternatively we
could use the tangential and the transverse component of the Euler equation to
find the shape of the fluxsheet and the wind velocity inside the fluxsheet
simultaneously, while we integrate from the stellar surface outward. The
difficulty to pass through all three critical points on the way would make a
completely new computer code necessary. And it would not be clear from the
beginning, whether this concept would work in the numerical praxis. For our
initial calculations we choose a less ambitious approach
by specifying shape of the fluxsheet a priori. This allows us to use our code
for the equatorial wind with few modifications. Afterwards we use the
transverse component of the Euler equation to check, whether the chosen shape
for the fluxsheet was reasonable.  This allows us to explore the raw shape of
the real fluxsheet and leads to a much higher confidence in our solutions.
For this approach we need a dimensionless form of the transverse component of
the Euler equation.

We start our derivation from the Euler equation (Eq.~\ref{SheetWinda}), whose
poloidal component (Eq.~\ref{SheetWindb}) we used to derive our wind equation.
If we choose the right fluxsheet, the transverse component of
Eq.~\ref{SheetWinda}
\begin{equation}\label{TransverseEulera}
0=\dgt\equiv \etvec\cdot\left[\left(\vvec\cdot\nabla\right)\vvec+
     \frac{\nabla P}{\rho}+\nabla\Phi-\gradvec\right]+
     \left(\etvec\times\frac{\Bvec}{4\pi\rho}\right)\cdot
     \left(\nabla\times\Bvec\right)
\end{equation}
with
\begin{equation}
\etvec = \ept\ervec-\epr\ethetavec
\end{equation}
should be fulfilled, too. But in general we do not know the exact shape of the
real fluxsheet. Therefore $\dgt$ will not be exactly zero for our solutions. In
order to get an estimate how seriously Eq.~\ref{TransverseEulera} is violated
we can compare $\dgt$ with the poloidal or transverse component of
$(\vvec\cdot\nabla)\vvec$. Therefore we use now the same \textsc{Mathematica}
code, we used for Eq.~\ref{SheetWindc}, to derive dimensionless versions of
$\dgt$ and $(\vvec\cdot\nabla)\vvec$. We find for the dimensionless version of
$\dgt$
\begin{eqnarray}
\label{defdat}
\dat &=& \frac{r_*}{v_*^2}\left[\dgt-
           \epvec\cdot((\vvec\cdot\nabla)\vvec)\right]\\
     &=& A_4(u,y)\dydu+B_4(u,y,\dydu)+\dbt
\end{eqnarray}
with
\begin{eqnarray}
A_4    &=& -\left(\ept\dudx-\frac{\epr\dudth}{x}\right)
            \left[\frac{\ysys}{y}-\frac{\yphibar}{\Mp^2}
            \left(\dyphidy-\frac{\yphibar}{y}\right)\right]\\
B_4    &=& \left(\ept\dudx-\frac{\epr\dudth}{x}\right)
            \left[\frac{\drhodu}{\rho}\ysys+\dysysdu+
            \frac{\yphibar}{\Mp^2}\left(\frac{\drhodu}{\rho}\yphibar+
            \dyphidu\right)+\frac{y^2Y^2}{\Mp^2}\frac{\drhodu}{\rho}
            \right]+\nonumber\\
&&         \left(\ept\detadx-\frac{\epr\detadth}{x}\right)
            \left[\ysys\left(\frac{\drhodeta}{\rho}-
            \frac{\dYdeta}{Y}\right)+\dysysdeta+\left.\frac{\yphibar}{\Mp^2}
            \right(\dyphideta-\domegadeta x\sin\theta+\right.\nonumber\\
&&         \left.\left.\yphibar\left(\frac{\drhodeta}{\rho}-
            \frac{\dYdeta}{Y}-\frac{\dFdeta}{F}\right)\right)+
            \frac{y^2Y^2}{\Mp^2}\left(\frac{\drhodeta}{\rho}-
            \frac{\dFdeta}{F}\right)\right]+
            \left(\frac{\epr}{x\tan\theta}-\frac{\ept}{x}\right)
            \times\nonumber\\
&&         \frac{\yphibar}{\Mp^2}(\yphi-2\yphibar)+
            \ept\left(\frac{1}{x^2}-\arad(u,y,\dydu)\right)+\nonumber\\
&&         \frac{y^2Y^2}{\Mp^2}\left(\deptdeta\detadx+\deptdu\dudx-
            \frac{\deprdeta\detadth+\deprdu\dudth}{x}+\frac{\ept}{x}\right).
\end{eqnarray}
For the tangential and transverse components of $(\vvec\cdot\nabla)\vvec$ we
find
\begin{eqnarray}
\label{defdbp}
\dbp &=& \frac{r_*}{v_*^2}\,\epvec\cdot((\vvec\cdot\nabla)\vvec)\\
     &=& \left(\epr\dudx+\frac{\ept\dudth}{x}\right)y\dydu Y^2+
          \left(\epr\detadx+\frac{\ept\detadth}{x}\right)y^2 Y\dYdeta-
          \nonumber\\
&&       \left(\frac{\epr}{x}+\frac{\ept}{x\tan\theta}\right)\yphi^2\\
\label{defdbt}
\dbt &=& \frac{r_*}{v_*^2}\,\etvec\cdot((\vvec\cdot\nabla)\vvec)\\
     &=& \left(\frac{\epr}{x\tan\theta}-\frac{\ept}{x}\right)\yphi^2-
          y^2Y^2\left[\left(\deptdeta\epr-\deprdeta\ept\right)
          \left(\epr\detadx+\frac{\ept\detadth}{x}\right)+\right.\nonumber\\
&&       \left.\left(\deptdu\epr-\deprdu\ept\right)
          \left(\epr\dudx+\frac{\ept\dudth}{x}\right)+
          \frac{\ept}{x}\right].
\end{eqnarray}
Now we can use $|\dat/\dbt|$ as relative measure how much the transverse
component of the Euler equation is violated. But in order to estimate how
relevant this violation is for the overall solution we should compare the
transverse component of the Euler equation with the tangential component. This
is given by the quantity $|\dat/\dbp|$. 
\section{First numerical models}
\label{Sec:numnoneq}
In this section we
show some initial calculations, which try to develop further some
ideas of Poe et al.\ \cite{Poe:etal:89} for winds from Wolf-Rayet stars. They
developed a wind model for Wolf-Rayet stars based on the assumption that a fast
and thin polar wind, driven only by the CAK line force, is responsible for the
high observed terminal velocities. In the equatorial plane they assumed a thick
and slow wind, additionally driven by a rotating magnetic field. Here they used
a model similar to ours. This wind should be responsible for the high radio
flux, which is otherwise explained as large spherical mass loss. Although this
model is theoretically appealing it was ruled out as general concept for
Wolf-Rayet stars by observations which showed that these stars in general do
not have a strong density gradient between the polar and equatorial regions
\cite{SchulteLadbeck:97,SchulteLadbeck:etal:92}. In this section we want to
argue that the model of Poe et al.\ might be modified to fit this observational
fact. The basic idea is that the fluxsheets are bent towards the poles and
not towards the equator. The latter is assumed in the wind compressed zone and
disk models \cite{Cassinelli:etal:95}. These models assume a nonmagnetic but
rotating wind. In this case the centrifugal force will cause the deviation from
radial outflow. But if we introduce a significantly strong magnetic field, the
Lorentz force will dominate the centrifugal force. From the physics of jets we
know that a rotating magnetic field can very efficiently collimate polar
outflows. Additionally Owocki et al. have shown that the radiation force has a
strong non-radial component \cite{Owocki:etal:96}. This will even enhance the
effect of the Lorentz force described in this section. For simplicity we do not
include this radiation effect in our calculation at the current stage.  In any
case the equatorial outflow will automatically be dispersed. This will have
several consequences: (1) The wind density in the polar region will be enhanced
due to the geometrical compression. (2) The terminal wind velocity in the polar
region will be reduced due to the fact that the polar fluxtube is collimated in
the supersonic region. The reduced terminal wind velocity increases the polar
wind density as well. (3) Exact the opposite happens in the equatorial
region. There the terminal wind velocity is increased and the density is
decreased. This reduces or maybe even eliminates the discrepancies between the
polar and the equatorial wind predicted by the model of Poe et al.\ and
criticized   by the observations of Schulte-Ladbeck et al.\
\cite{SchulteLadbeck:97,SchulteLadbeck:etal:92}.

For our numerical models we use here the ``Wolf-Rayet test model'' of Poe et
alii. This model star has a mass of $M=13\Msun$ and a luminosity of $L=3\times
10^5\Lsun$. For the equatorial radius we use a prelimnary value of $R=5\rsun$
for the base of the wind. For strong winds (high rotation rate and strong
magnetic field) this will lead approximately to a radius of $8\rsun$ for
$\taues=1$ as used by Poe et alii. For the CAK parameter we choose
$\alphacak=0.61$ and $\kcak=0.18$. The electron scattering opacity for an
Wolf-Rayet star $\kappaTh=0.2\cm^2/\gram$ due to the missing hydrogen in the
wind. For the wind temperature we choose a value of $T=20000\Kelvin$. Poe et
al.\ quote a mass loss rate of $\Mdotobs=3\times 10^{-5}\Msun/\yr$ and a
terminal velocity of $\vinfobs=2900\,\km/\s$ as observational results for the
wind. These values should be directly or indirectly be explained.

Our wind equations contain the necessary terms to include differential
rotation. And from our sun we know that stars can rotate differential. It might
even be that differential rotation is the normal case for stars. Nevertheless
we limit our discussion here to the case where the rotation rate of the star
does not depend on latitude $(\domegadeta=0)$ to reduce the number of free
parameters in this initial analysis.

We can not reproduce the numerical results of Poe et al.\ exactly as we did for
the O star models of Friend \& MacGregor~\cite{Friend:MacGregor:84} in
Chap.~\ref{Chap:cak:wd}, because for simplicity we do not include the
correction for the finite disk of the star \cite[Eq.~50]{Castor:etal:75} in our
equation at the current stage. Additionally Poe et al.\ do not print their wind
equations explicitly. Including the finite disk correction will be
neccesary later when this model is developed further to explain Wolf-Rayet star
winds quantitatively.

In a first step we calculated a set of equatorial wind models to get a first
feeling for the dependence of $\Mdot$ and $\vinf$ on the wind parameter
$\alpharot$ and $\Bpo$. These calculations confirmed the results of Poe et al.\
that in a simple equatorial model we need a high rotation rate to explain the
high mass loss rates and a very strong magnetic field to explain the high
terminal velocities. We agree with Poe et al.\ that such a model would be ruled
out by the spin-down problem.
\begin{table}
%\begin{sideways}
%\begin{minipage}{14cm}
\caption{
\label{Tab:WREqResults}Equatorial wind models for the Wolf-Rayet test star
without any extra wind compression or dilution $(a(u)=1)$.}
\begin{center}
\begin{tabular*}{13cm}{*{11}{r@{\extracolsep{\fill}\hspace{1mm}}}}
\hline
\hline
\rule[-3mm]{0mm}{8mm}
no. & $\Bpo\atop(\Gauss)$ & $\alpharot$ & $\frac{r_{c1}}{R}$ & 
$\frac{\rAc}{R}$ & $\frac{r_{c2}}{R}$ & $\vAc\atop(\km/\s)$  & 
$\vinf\atop(\km/\s)$ & ${10^6\,\Mdot}\atop(\Msun/\yr)$  & 
$\frac{\Mdot\vinf}{L/c}$ & $\frac{v_\infty}{v_{\mathrm{A}\infty}}$\\
\hline
% Dies sind meine Ergebnisse in Analogie zu Poe et al..
%no    Bp0  arot    rcn      rc      rcf    vac   vinf    Mdot     eff   y/yA
01 &   500 & 0.3 & 1.11 &  1.43 &   2.90 &  498 & 1016 &  4.71 &  0.79 & 5.96\\
02 &  1000 & 0.3 & 1.08 &  2.29 &   4.04 &  772 & 1128 &  4.75 &  0.88 & 3.50\\
03 &  2000 & 0.3 & 1.07 &  3.98 &   8.60 & 1008 & 1239 &  4.80 &  1.02 & 2.16\\
04 &  5000 & 0.3 & 1.06 &  8.42 &  34.05 & 1393 & 1740 &  4.86 &  1.39 & 1.35\\
05 & 10000 & 0.3 & 1.06 & 14.40 & 111.41 & 1890 & 2409 &  4.90 &  1.94 & 1.10\\
06 &   500 & 0.5 & 1.06 &  1.44 &   5.89 &  463 &  983 &  5.02 &  0.81 & 3.52\\
07 &  1000 & 0.5 & 1.05 &  2.21 &   5.72 &  760 & 1186 &  5.19 &  1.01 & 2.37\\
08 &  2000 & 0.5 & 1.04 &  3.66 &  11.65 & 1072 & 1477 &  5.36 &  1.30 & 1.67\\
09 &  5000 & 0.5 & 1.04 &  7.26 &  43.26 & 1643 & 2191 &  5.55 &  2.00 & 1.23\\
10 & 10000 & 0.5 & 1.04 & 12.02 & 129.31 & 2358 & 3152 &  5.64 &  2.92 & 1.07\\
11 &   500 & 0.7 & 1.04 &  1.45 &  17.08 &  395 &  895 &  5.78 &  0.85 & 2.34\\
12 &  1000 & 0.7 & 1.04 &  2.10 &  10.07 &  687 & 1160 &  6.34 &  1.21 & 1.81\\
13 &  2000 & 0.7 & 1.03 &  3.28 &  16.23 & 1040 & 1540 &  6.86 &  1.74 & 1.44\\
14 &  5000 & 0.7 & 1.03 &  6.16 &  51.11 & 1720 & 2407 &  7.35 &  2.91 & 1.16\\
15 & 10000 & 0.7 & 1.03 &  9.98 & 140.27 & 2553 & 3531 &  7.56 &  4.40 & 1.05\\
16 &   500 & 0.9 & 1.03 &  1.47 &  66.74 &  233 &  655 &  9.52 &  1.03 & 1.46\\
17 &  1000 & 0.9 & 1.04 &  1.88 &  75.00 &  415 &  850 & 13.13 &  1.84 & 1.27\\
18 &  2000 & 0.9 & 1.04 &  2.57 &  70.65 &  684 & 1173 & 17.07 &  3.30 & 1.17\\
19 &  5000 & 0.9 & 1.04 &  4.26 &  84.19 & 1246 & 1905 & 21.27 &  6.67 & 1.09\\
20 & 10000 & 0.9 & 1.04 &  6.55 & 161.36 & 1933 & 2836 & 23.16 & 10.81 & 1.03\\
\hline
\end{tabular*}
\end{center}
%\end{minipage}
%\end{sideways}
\end{table}
\begin{figure}
\begin{center}
\begin{picture}(80,85)
%\put(0,0){\framebox(80,85){}}
\put(0,0){\epsfig{file=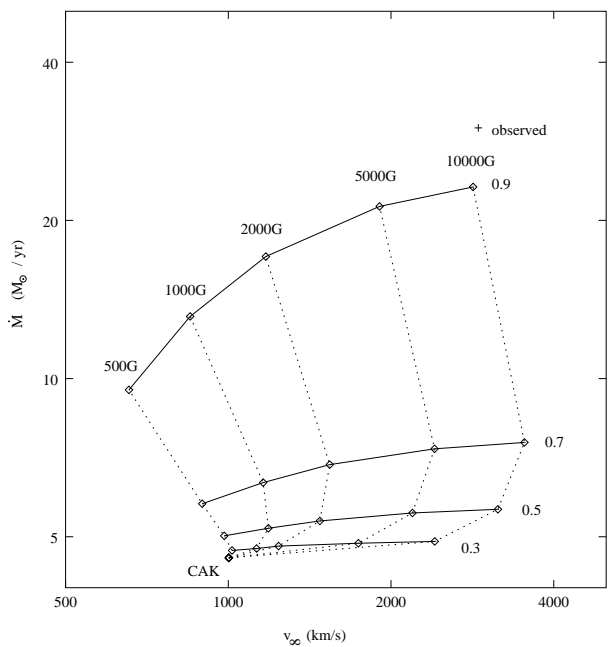,width=8cm}}
\end{picture}
\end{center}
\caption{\label{Fig:EqWinds} shows $\Mdot$ versus $\vinf$ for the models from
Tab.~\ref{Tab:WREqResults}. The solid lines connect models with a constant
value of $\alpharot$. The dotted lines connect models with a constant magnetic
field strength. Additionally the observed (+) values and the values for the
nonrotating, nonmagnetic case (CAK) are shown. Rotation basically increases
the mass loss rate. While the magnetic field basically increases the terminal
velocity rate.}
\end{figure}

\begin{figure}
\begin{center}
\begin{picture}(85,85)
%\put(0,0){\framebox(85,85){}}
\put(35,0){equatorial plane}
\put(0,35){\begin{sideways}polar axis\end{sideways}}
\put(5,5){\epsfig{file=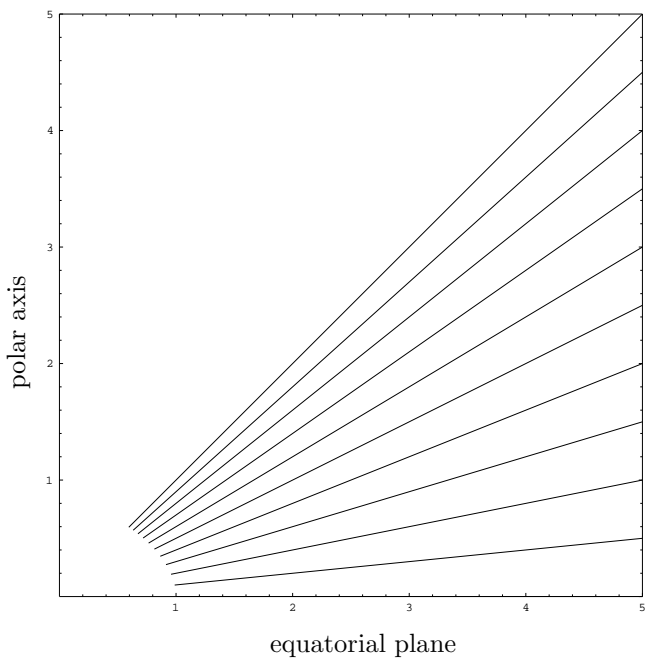,width=8cm}}
\end{picture}
\end{center}
\caption{\label{Fig:StraightSheets} shows the meridional plane ($r,\theta$ or
$u,\eta$). The fluxsheets start from the stellar surface and extend straight
outwards. This fluxsheet geometry was used for our first calculations which
should prove that the fluxsheets are actually collimated towards the poles. The
star is oblate due to its rotation $(\alpharot=0.95)$.}
\end{figure}

\begin{figure}
\begin{center}
\begin{picture}(120,80)
%\put(0,0){\framebox(120,80){}}
\put(60,0){u=R/r}
\put(0,40){\begin{sideways}$\dat$\end{sideways}}
\put(5,5){\epsfig{file=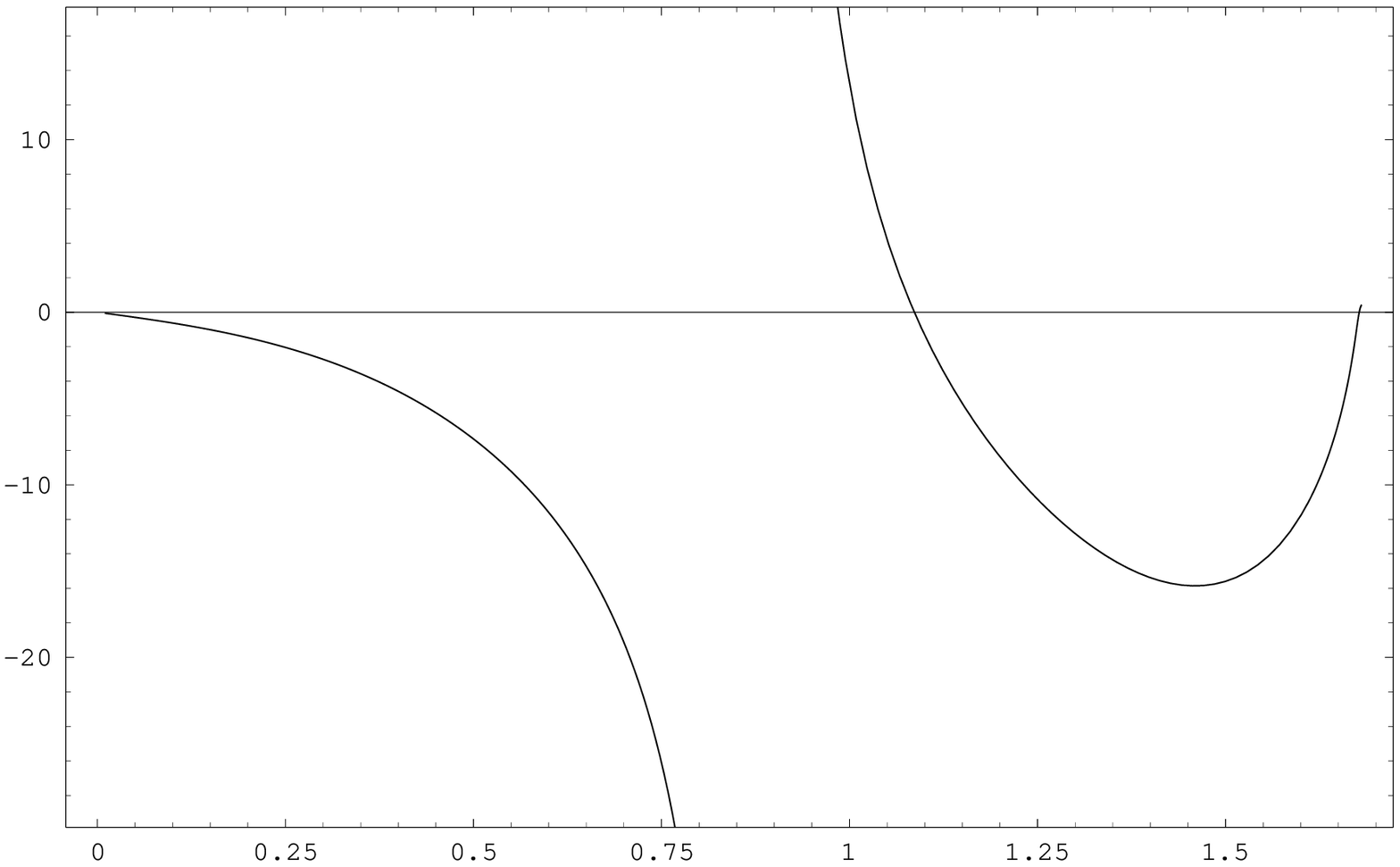,width=115mm}}
\end{picture}
\end{center}
\caption{\label{Fig:DeltaatA} shows $\dat$ as defined in Eq.~\ref{defdat} for
the $\eta=1$ solution form Tab.~\ref{Tab:WRStraightOutside}.}
\end{figure}

\begin{figure}
\begin{center}
\begin{picture}(120,80)
%\put(0,0){\framebox(120,80){}}
\put(60,0){u=R/r}
\put(0,40){\begin{sideways}$\dbt$\end{sideways}}
\put(5,5){\epsfig{file=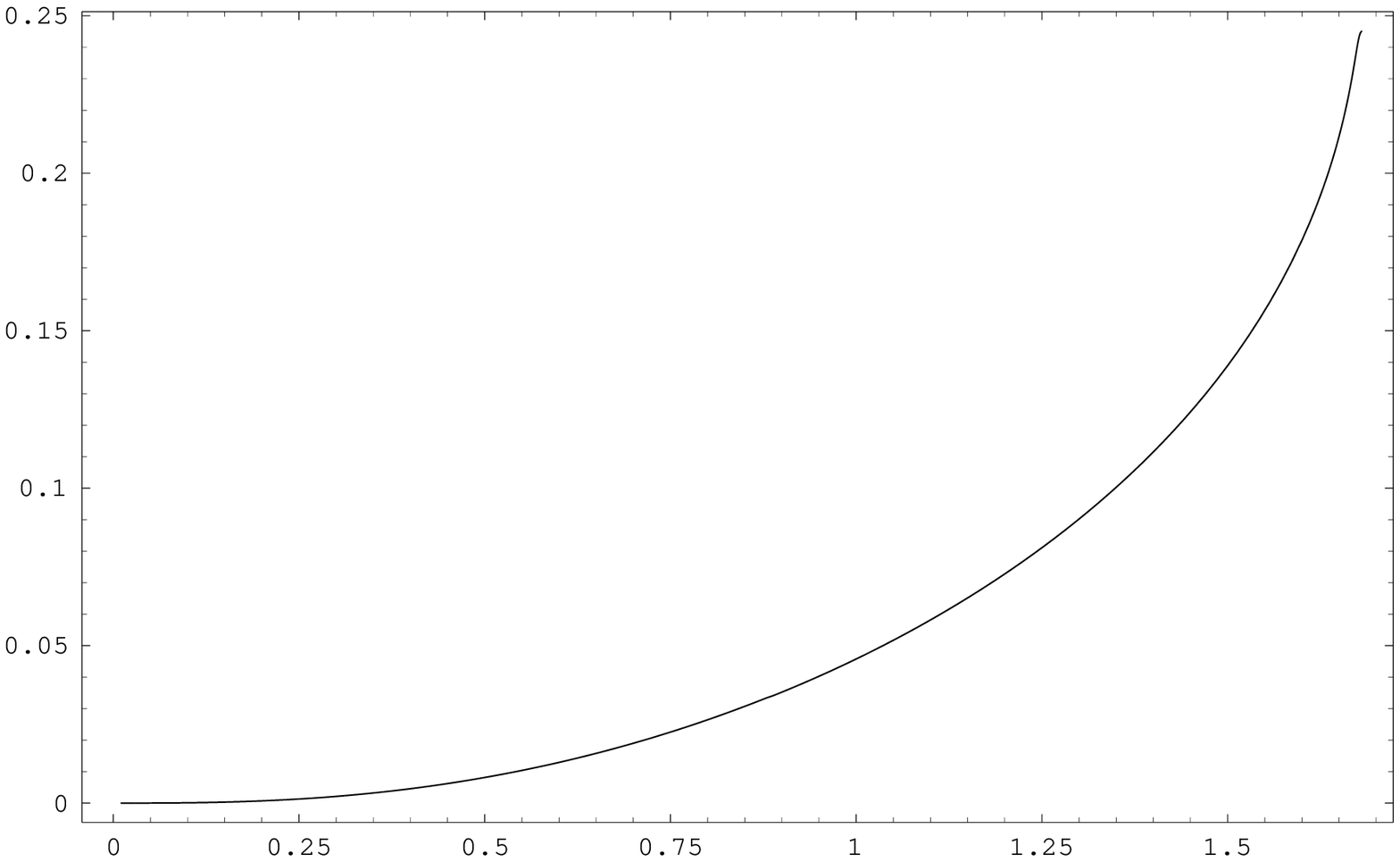,width=115mm}}
\end{picture}
\end{center}
\caption{\label{Fig:DeltabtA} shows $\dbt$ as defined in Eq.~\ref{defdbt} for
the $\eta=1$ solution form Tab.~\ref{Tab:WRStraightOutside}.}
\end{figure}

\begin{figure}
\begin{center}
\begin{picture}(120,80)
%\put(0,0){\framebox(120,80){}}
\put(60,0){u=R/r}
\put(0,40){\begin{sideways}$\dbp$\end{sideways}}
\put(5,5){\epsfig{file=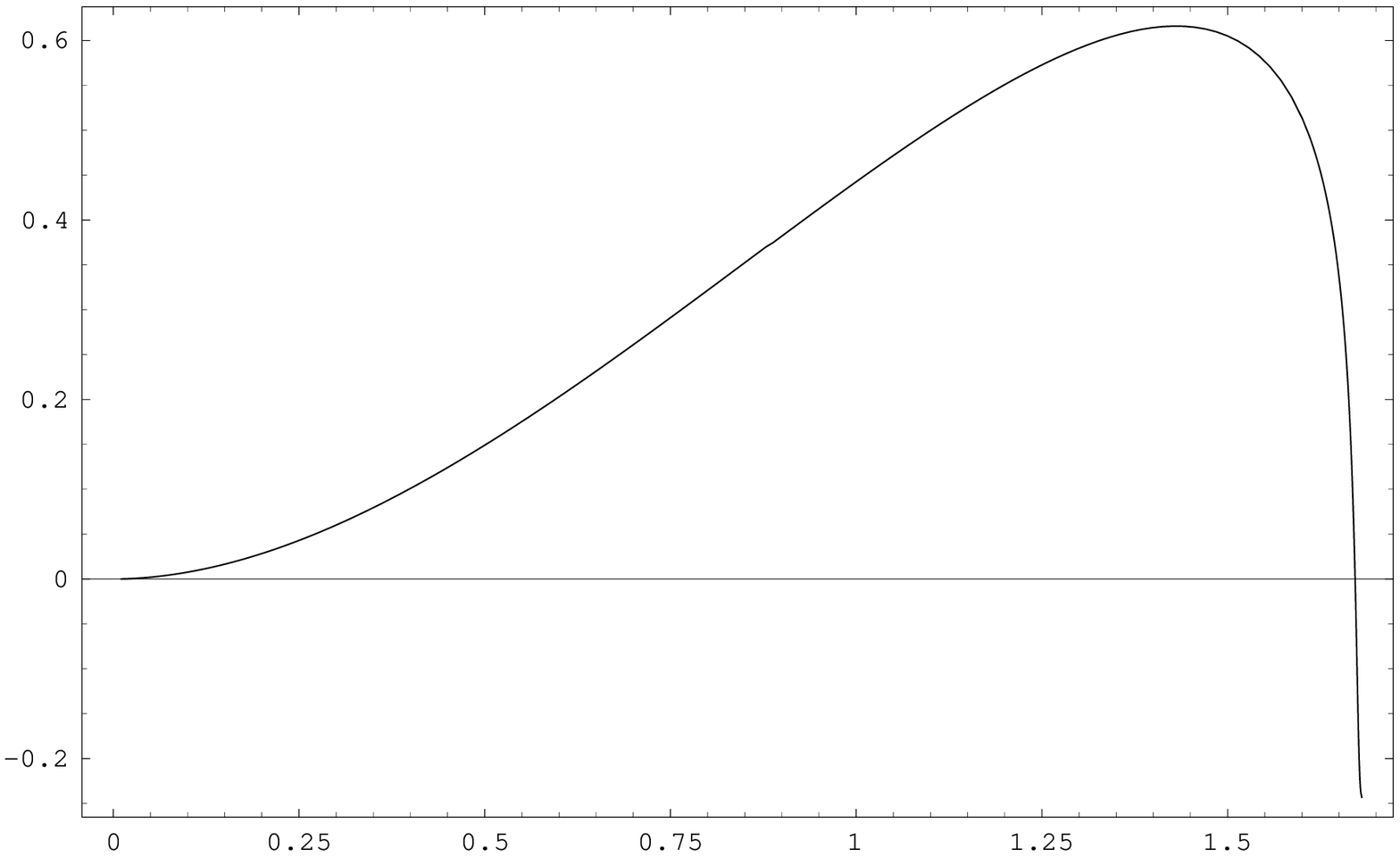,width=115mm}}
\end{picture}
\end{center}
\caption{\label{Fig:DeltabpA} shows $\dbp$ as defined in Eq.~\ref{defdbp} for
the $\eta=1$ solution form Tab.~\ref{Tab:WRStraightOutside}.}
\end{figure}

\begin{figure}
\begin{center}
\begin{picture}(80,80)
%\put(0,0){\framebox(80,80){}}
\put(0,0){\epsfig{file=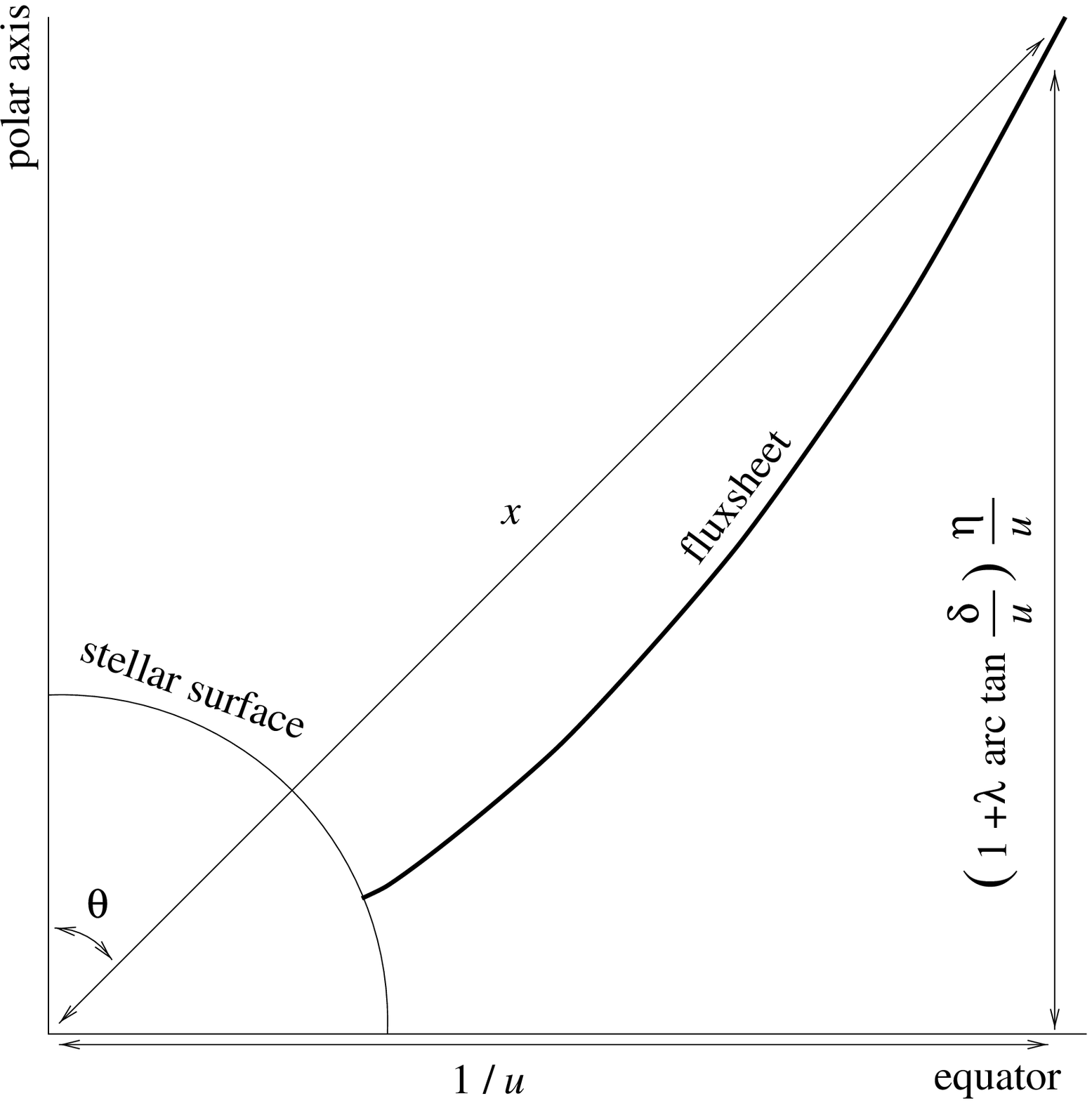,width=8cm}}
\end{picture}
\end{center}
\caption{\label{Fig:CurvedA} shows geometry of the curved fluxsheets used in
this chapter.}
\end{figure}

The next step is to check that the magnetic force really collimates the
fluxsheets towards the poles. To check this we calculated wind models for a set
of non-equatorial fluxsheets. For these first calculations we used fluxsheets
which were not bent (Fig.~\ref{Fig:StraightSheets}). This fluxsheets
configuration would be expected for a nonrotating star. This fluxsheet
configuration is mathematically given by
\begin{eqnarray}
x(u,\eta)  & = &  \frac{\sqrt{1+\eta^2}}{u}\\
\theta(u,\eta)  & = &  \mathrm{arccot}\,\eta.
\end{eqnarray}
It should be mentioned that in this special case the wind equation
(Eq.~\ref{SheetWinde}) becomes independent of $\dYdeta$, $\drhodeta$,
$\domegadeta$ and $\dFdeta$, because $\dxdu\detadx+\dthdu\detadth$ is always
zero. But this is not true for the transverse force component. We calculated 11
models with $\eta$ between 0 and 1 covering an lateral angle of $\theta$
between 90 and 45 degrees. We used a magnetic field of $1000\,\Gauss$ and a
rotation rate of $\alpharot=0.95$. The latter is more extreme than the values
given in Tab.~\ref{Tab:WREqResults} and Fig.~\ref{Fig:EqWinds}. But the idea of
this section is to propose a model with a strong equatorial wind due to rapid
rotation. The magnetic filed should be moderate, so that we do not have a
spin-down problem. But it should be strong enough to spread the mainly
equatorial mass outflow of the whole hemisphere, so that this model is not
ruled out by the lack of observed polarisation. As reference radius for the
terminal velocity we chose $100R$. For comparison with observation this
reference radius should not be chosen too large, because the P-Cygni profiles
are not too far away from the star.
\begin{table}
%\begin{sideways}
%\begin{minipage}{14cm}
\caption{
\label{Tab:WRStraightOutside}Wind solutions with three critical points for 
the fluxsheets shown in Fig.~\ref{Fig:StraightSheets}.}
\begin{center}
\begin{tabular}{rrrrrrrrrrr}
\hline \hline \rule[-3mm]{0mm}{8mm} $\eta$ & 
$\frac{r_{c1}}{R}$ & $\frac{\rAc}{R}$ & $\frac{r_{c2}}{R}$ &
$\vAc\atop(\km/\s)$ & $\vinf\atop(\km/\s)$ & ${10^6\,\Mdot}\atop(\Msun/\yr)$ &
$\frac{\Mdot\vinf}{L/c}$ & $\frac{v_\infty}{v_{\mathrm{A}\infty}}$\\ \hline
%eta   rcn      rc      rcf    vac   vinf    Mdot     eff   y/yA
0.0 & 1.03 &  1.57 & 295.91 &  132 &  455 & 59.34 &  4.44 & 1.00\\
0.1 & 1.04 &  1.71 & 173.02 &  242 &  624 & 26.80 &  2.75 & 1.10\\
0.2 & 1.03 &  1.86 &  73.01 &  418 &  860 & 12.65 &  1.79 & 1.29\\
0.3 & 1.03 &  1.92 &  37.10 &  527 & 1001 &  9.03 &  1.49 & 1.46\\
0.4 & 1.03 &  1.94 &  20.15 &  596 & 1087 &  7.47 &  1.34 & 1.63\\
0.5 & 1.03 &  1.94 &  12.68 &  641 & 1142 &  6.61 &  1.24 & 1.79\\
0.6 & 1.03 &  1.93 &   9.35 &  672 & 1176 &  6.08 &  1.18 & 1.96\\
0.7 & 1.04 &  1.93 &   7.58 &  693 & 1198 &  5.73 &  1.13 & 2.12\\
0.8 & 1.04 &  1.92 &   6.48 &  708 & 1212 &  5.49 &  1.09 & 2.28\\
0.9 & 1.04 &  1.91 &   5.74 &  719 & 1220 &  5.31 &  1.07 & 2.45\\
1.0 & 1.04 &  1.90 &   5.21 &  727 & 1225 &  5.18 &  1.04 & 2.61\\
\hline
\end{tabular}
\end{center}
%\end{minipage}
%\end{sideways}
\end{table}
The next important step is to check whether the fluxsheets should now be bent
towards the polar axis or towards the equatorial plane. For the $\eta=1$
$(\theta=45^\circ)$ solution we evaluated $\dat$, $\dbt$, and $\dbp$ as defined
in Sect.~\ref{Sec:transf}. The function $Y(\eta)$ is fitted to the wind
velocities in the different fluxsheets at $r=2R$. The result is plotted in
Figs.~\ref{Fig:DeltaatA}--\ref{Fig:DeltabpA}. The singularity we see in
Fig.~\ref{Fig:DeltaatA} is due to the Alfv\'enic point, where we yet do not use
all side conditions to keep the transverse force finite. If we mentally subtract
the singularity from the plot, we see that $\dat$ is negative everywhere, and
that we get the strongest transverse force component close to the stellar
surface. In Fig.~\ref{Fig:DeltabtA} we see the inertial contributions to the
transverse Euler equations. Since the fluxsheet is not bent we have only the
transverse component of the centrifugal force here, which of course is directed
towards the equatorial plane. The centrifugal contribution has the opposite
sign than the magnetic contribution from Fig.~\ref{Fig:DeltaatA}. And it is
close to the stellar surface more than an order of magnitude smaller. The same
is true for the poloidal component of the inertial force. We can therefore
conclude from here that the fluxsheets should be strongly bent towards the
polar axis.  We find the same result for the $\eta=0.1$ fluxsheet. Although
the force towards the polar axis is not so strong close to the equatorial
plane.
The next step is now to calculate wind solutions for fluxsheets which are bent
towards the polar axis. As an first approximation for the fluxsheets we choose
the fluxsheet shape shown in Fig.~\ref{Fig:CurvedA}.
We get again a one parameter family of fluxsheets with $\eta$ as parameter. The
additional parameters $\lambda$ and $\delta$ are constant for all fluxsheets in
a stellar wind model. The motivation for this nonobvious choice was to produce
fluxsheets which are bent towards the polar axis close to the star and rather
straight far away from the star, where the magnetic field is weak. Our
fluxsheets start with an inclination of $\theta=\mathrm{arccot}\,\eta$ for
$x\rightarrow 0$ $(u\gg\delta/2)$ and have an inclination of
$\eta(1+\lambda\pi/2)$ at $x\rightarrow\infty$ $(u\ll\delta/2)$. So $\lambda$
controls how strong the fluxsheets are bent. While $\delta$ controls how fast
they are bent. For $\lambda=0$ these fluxsheets reduce to the straight
fluxsheets used previously in this section. And for $\lambda=\eta=0$ we obtain
the equatorial fluxsheet used everywhere in this thesis. The equatorial
fluxsheet should always be straight due to the symmetry between northern and
southern hemisphere.

We started our analysis of bent fluxsheets trying to find solutions with three
critical points. But we found that the solutions extend only beyond the outer
critical point if the fluxsheets are nearly straight. A numerical analysis of
the solution around the outer critical points showed that the third condition
for the critical point (Eq.~\ref{critequCb}) is not fulfilled with the
necessary numerical precision. This might have two reasons. (1) The numerical
roundoff errors in the formulas for the critical point are too large. Therefore
the root finding algorithm of our code can not find the position of the outer
critical point with the required precision. This would require an improvement
of our numerical code in order to reduce roundoff errors. (2) There might be no
point on our fluxsheet where all three conditions for the outer critical point
are fulfilled. This would require a deeper mathematical analysis of our wind
equation. It might be that we can derive from this problem further conditions
for the shape of fluxsheets which have a complete wind solution. For this point
see also Appendix~\ref{App:noneqcritp}.  Due to this problem we calculated few
initial wind solutions without the outer critical point analogous to our models
in Chap.~\ref{Chap:cak:wd}. But these models still have fairly straight
fluxsheets. The results are still too few and inconsistent to be published
here. But we showed here that the non-equatorial fluxsheets are indeed bended
towards the poles. Future numerical computations will complete the picture we
sketched here and show how strong the effect of the bent fluxsheets really is.
%

%-*-LaTeX-*-
% This is the 8th chapter for the PhD-thesis of Henning Seemann.
% (c) 1997-98 by Henning Seemann
%
\chapter{Conclusions}
\label{Chap:conclusions}
In this thesis we analyzed new aspects in the theory of magnetic winds from
massive stars. This work was motivated by the fact that we have many indirect
hints for the existence of a significant magnetic field in the wind of massive
stars \cite{Biermann:Cassinelli:93}, although the magnetic field is not
accessible for direct observation yet \cite{Ignace:etal:95}. The development of
the theory is still at a very early stage. Therefore we can not yet produce
models, which allow to describe and understand in detail the complex structure
of the winds from massive stars we know from observation. But in this thesis we
present two new and important components in the theory of magnetic winds from
massive stars. It is important to recognize that both components are not
optional. Both effects are automatically present without further conditions, if
the star rotates with a significant magnetic field.

In Chap.~\ref{Chap:waves:shocks} we analyzed, in the linear limit, the
evolution of wind instabilities in magnetic winds. The foundation for this
analysis is the well known instability of the line driving mechanism. This
instability has been intensively studied \cite{Owocki:Rybicki:84,
Owocki:Rybicki:85, Owocki:Rybicki:86, Owocki:Rybicki:91, Owocki:etal:88}. And
it seems that this instability can explain some details in the spectra of
massive stars. But in spite of initial hopes \cite{Lucy:82} no way was found
how these waves can help to solve the basic wind momentum problem in many of
these stars, especially Wolf-Rayet stars. We showed in our analysis that the
magnetic field is the missing key. It has a strong influence of the behavior of
the waves resulting from this instability. In the nonmagnetic case the waves
are dominantly running inward, leading to reverse shocks. If a significant
magnetic field is present, the waves are running dominantly outward, leading to
forward shocks. These forward shocks will contain layers of material which
have a much higher velocity than the terminal velocity found in stationary and
smooth wind models. This is an important contribution to the explanation of the
observations. Stationary wind models need a very high magnetic field to produce
a high terminal velocity. Such a high magnetic field would cause a spin-down
problem, which is easily avoided in our wind models with waves. Additionally
our waves produce a higher \textsl{observed} mass loss rate by feigning a
higher terminal velocity and a higher wind density through clumping. This helps
to overcome the wind problem in Wolf-Rayet stars as well.

Our second model, the bending of the non-equatorial fluxsheets towards the
poles, is a direct and unavoidable consequence of significant rotation and
magnetic field strength. This effect is a ideal extension to our first model,
because its results lead into the same direction. We showed that the
non-equatorial fluxsheets are bent towards the poles. This has several
consequences. Due to the extension of the flux cross section in the equatorial
region we will get an even higher terminal velocity, because the wind is
supersonic except very close to the star. The wind material will be spread more
equally over all lateral regions. This will avoid the problems which previous
models \cite{Poe:etal:89} had with the unobserved polarization of the stellar
light \cite{SchulteLadbeck:etal:92}.

Previous wind models had to use rather slow rotation and rather strong
magnetic fields to produce the high observed terminal velocities and to avoid
conflicts with the not observed polarization of the stellar light. But such
models lead to a rapid spin-down. We can now avoid this using a wind model with
rapid rotation and moderate magnetic fields. In this case we have strong mass
loss combined with a sufficiently small angular momentum loss. In such a model
the mass loss will be concentrated in the equatorial regions. But our bent
fluxsheets will distribute the wind material in the polar regions as well, so
that we can avoid the polarization problem. Additionally our outward running
waves will reproduce the high observed terminal velocities without requiring a
stronger magnetic field. These two models in combination help a lot to fit
theoretical models for magnetic winds to the observations.

Nevertheless we are not a the end of the development. We are only at the
beginning. And there is still a lot of interesting work to do. The next, direct
step should be a deeper mathematical and numerical analysis of our fluxsheet
model in order to complete this work. In general both of our models can be
improved. For our analysis of waves in magnetic winds the next step is to look
at the nonlinear case and to demonstrate that our predictions for the behavior
of magnetic shock fronts are correct. This work could produce quantitative
predictions for the X-ray emission, for the nonthermal radio emission, and for
the cosmic ray emission. This would improve our understanding of the
microphysics of these winds. Our fluxsheet model can be improved beyond the
steps already mentioned here and in Chap.~\ref{Chap:fluxtube} in two
directions: (1) A more detailed description of the radiation driving mechanism
can give a better understanding of optical thick winds
\cite{Gayley:etal:95}. And the non-radial component of the radiation force can
bend the fluxsheets even more towards the poles \cite{Owocki:etal:96}. It
should therefore be included in our model. This might be important in order to
show, that the wind material is really equally distributed around the star, and
we therefore see no polarization.  (2) The second major direction in the future
development of the fluxsheet model is the implementation of a real two
dimensional treatment of the MHD equations to get a better understanding of the
shape of the fluxsheets and of the large scale structure of the wind.

Another large area of future work is the application of the fluxsheet model to
other winds. On purpose we did not integrate a specific geometrical shape of
the fluxsheets into the fluxsheet model. With few modifications it is possible
to use the model with a different fluxsheet geometry. This allows e.g.\ to
model the wind from accretion disks in young stellar objects or in AGNs. This
opens a wide range of future projects. 

Thus at the end of this thesis we have the exciting situation of science:
Answer one question, and you find three new ones!

\vfill
%

%-*-LaTeX-*-
% This is the appendix of the PhD-thesis of Henning Seemann.
% (c) 1997-98 by Henning Seemann
%
\begin{appendix}
\chapter{The equatorial wind as limiting case}
\label{App:limit}
In this appendix we derive the equations for the equatorial wind from
Chap.~\ref{Chap:wd} as a limiting case of the wind equations from
Chap.~\ref{Chap:fluxtube}. This is an important consistency check for the new
model. Additionally it allows to relate results calculated with the equations
of Chap.~\ref{Chap:wd} to results calculated with the equations of
Chap.~\ref{Chap:fluxtube}.  The equations of Chap.~\ref{Chap:wd} can be
obtained by using the fluxsheet shape functions
\begin{eqnarray}
x(u,\eta)      &=& \frac{1}{u}\\
\theta(u,\eta) &=& \arccos[a(u)\eta],
\end{eqnarray}
where $a(u)$ is the area function for the equatorial fluxsheet we have
introduced in Chap.~\ref{Chap:wd}. The equatorial fluxsheet is specified by
\begin{equation}
\eta=0.
\end{equation}
The symmetry between northern and southern hemisphere requires that our
self-similarity functions $\dYdeta$, $\dFdeta$, $\domegadeta$, $\drhoAcdeta$,
and $\check{B}_\mathrm{pAc}$ vanish in the equatorial plane. For $Y$ we choose
1 in the equatorial plane. We find then for
the geometry expressions
\begin{eqnarray}
\dxdu                    &=& -\frac{1}{u^2}\\
\dudx                    &=& -u^2\\
\dthdeta                 &=& -a\\
\AbsGradeta              &=& \frac{u}{ar_*}\\
\partial_u\AbsGradeta    &=& \frac{1}{ar_*}
                               \left(1-\frac{u\dadu}{a}\right)\\
s                        &=& -1\\
\epr                     &=& 1\\
\ept                     &=& 0.
\end{eqnarray}
All other relevant geometrical quantities are zero. 
We find then for the auxiliary quantities
(Eqs.~\ref{dyxdxdef},\ref{auxa}--\ref{auxz})
\begin{eqnarray}
\left|\dyxdx\right| &=& -u^2\dydu\\
\frac{\drhodu}{\rho} &=& \frac{2}{u}-\frac{\dadu}{a}\\
\frac{\drhodeta}{\rho} &=& 0\\
\Mp^2 &=& \MAr^2 = \frac{\uAc^2}{u^2}\frac{y}{\yAc}\frac{a}{\aAc}
                 = \frac{1}{1-\UM}\\ 
\yphi  &=& \omega\frac{\Mp^2-\frac{x^2}{\xAc^2}}{\Mp^2-1}\frac{\xAc^2}{x}\\
       &=& \yrot\frac{\frac{\xAc^2}{x^2}\MAr^2-1}{\MAr^2-1}\\
       &=& \yrot\frac{\UM-\Ue}{\UM}\\
\frac{\partial\yphi}{\partial\Mp^2} &=& \yrot\frac{1-\frac{\xAc^2}{x^2}}
           {(1-\MAr^2)^2}\\
       &=& \frac{\yrot}{\MAr^4}\frac{\Ue}{\UM^2}\\
\dyphidu &=& -\frac{\omega}{u^2}\frac{1+\MAr^2\frac{\xAc^2}{x^2}}{1-\MAr^2}-
           \frac{\partial\yphi}{\partial\Mp^2}\left(\frac{2}{u}-
           \frac{\dadu}{a}\right)\MAr^2\\
       &=& \frac{\yrot}{u}\left[\frac{2-\Ue-\UM}{\UM}-\frac{1}{\MAr^2}
           \frac{\Ue}{\UM^2}\left(2-\frac{u\dadu}{a}\right)\right]\\
\dyphideta &=& 0\\
\dyphidy &=& \frac{\yrot}{y}\frac{\Ue}{\UM^2}\frac{1}{\MAr^2}\\
       &=& \frac{\yrot}{y}\frac{\Ue}{\UM^2}\left(1-\UM\right)\\
\label{Cequiv}
C\left|\dyxdx\right|^\alphacak 
       &=& -\kcak\Gamma\left(\frac{4\pi GM}{\kappaTh\vth\Mdot}
           \right)^\alphacak(-y\dydu a)^\alphacak\\
\yphibar &=& \yphi-\yrot\\
       &=& -\yrot\frac{\Ue}{\UM},
\end{eqnarray}
where we have used $\yrot=\omega/u$ and $Q=y_0/M_{\mathrm{A}r0}^2$. We can now
express Eq.~\ref{SheetWinde} with
\begin{eqnarray}
A\dyxdx &=& A_2\dydu\\
  &=& \dydu\left[y-\frac{\ysys}{y}-\yrot\frac{\Ue}{\UM}(1-\UM)
        \left[\frac{\yrot}{y}\frac{\Ue}{\UM}\left(\frac{1}{\UM}-1\right)+
        \right.\right.\nonumber\\
  & & \left.\left.\frac{\yrot}{y}\frac{\Ue}{\UM}\right]\right]\\
  &=& \dydu\left[y-\frac{\ysys}{y}-(1-\UM)\frac{\yrot^2}{y}\frac{\Ue^2}{\UM^3}
        \right]\\
\label{Aequiv}
  &=& y\dydu\left[1-\frac{\ysys}{y^2}-\frac{\yroto^2}{y^2}\frac{Q}{y}
        \frac{\Ue^2}{\UM^3}\frac{a_0}{a}\right]\\
B &=& B_2\\
  &=& -1+\Gamma+\ysys\left(\frac{2}{u}-\frac{\dadu}{a}\right)+\dysysdu-
        \yrot\frac{\Ue}{\UM}(1-\UM)\times\nonumber\\
  & & \left[-\yrot\frac{\Ue}{\UM}\left(\frac{2}{u}-\frac{\dadu}{a}\right)+
        \frac{\yrot}{u}\left[\frac{2-\Ue-\UM}{\UM}\right.\right.\nonumber\\
  & & \left.\left.-\frac{1}{\MAr^2}\frac{\Ue}{\UM^2}\left(2-\frac{u\dadu}{a}
        \right)\right]\right]+\frac{1}{u}\left[-\yrot\frac{\Ue}{\UM}(1-\UM)
        \times\right.\nonumber\\
  & & \left.\left(\yrot+\yrot\frac{\Ue}{\UM}\right)
        +\yrot^2\left(1-\frac{\Ue}{\UM}\right)^2\right]\\
  &=& -1+\Gamma+\ysys\left(\frac{2}{u}-\frac{\dadu}{a}\right)+\dysysdu+
        \frac{\yrot^2}{u\UM^2}\left[-2\Ue(1-\UM)+\right.\nonumber\\
  & & \left.(\Ue-\UM)^2+\frac{\Ue^2}{\UM}
        (1-\UM)\left(2-\frac{u\dadu}{a}\right)\right]\\
  &=& -1+\Gamma+\ysys\left(\frac{2}{u}-\frac{\dadu}{a}\right)+\dysysdu-
        \frac{\yrot^2}{u\UM^2}\left(\UM-\Ue\right)\times\nonumber\\
  & &   \left[-2\frac{\Ue}{\UM}
        (1-\UM)+\UM-\Ue\right]-\frac{\yrot^2}{\UM^2}\frac{\Ue^2}{\UM}(1-\UM)
        \frac{\dadu}{a}\\
  &=& -1+\Gamma+\ysys\left(\frac{2}{u}-\frac{\dadu}{a}\right)+\dysysdu-
        \frac{\lambda}{\UM^2}\frac{u}{u_0}\left[\epsilon-\frac{Q}{y}
        \frac{a_0}{a}\right]\times\nonumber\\
\label{Bequiv}
  & & \left[\left(2\frac{\Ue}{\UM}+1\right)\frac{Q}{y}\frac{a_0}{a}
        -\epsilon\right]-\frac{\lambda}{\UM^2}\frac{u_0}{u}\frac{\Ue^2}{\UM}
        \frac{Q}{y}\frac{a_0u\dadu}{a^2},
\end{eqnarray}
where we have used $\lambda=\omega^2/u_0^3$. This result
(Eqs.~\ref{SheetWinde}, \ref{Cequiv}, \ref{Aequiv} \& \ref{Bequiv}) reproduces
our results from chapter \ref{Chap:cak:wd}
(Eqs.~\ref{CAKWDWind}--\ref{CAKWDWindC}). It might now appear that the 
equatorial version of the wind equations is superfluous since we have the
general version. But they were an important step in the development of the
generalized model. Additionally they are much simpler, so that there numerical
implementation is faster and more robust against numerical roundoff
errors. Therefore it is useful to start numerical computations with the
equatorial wind equations.
\chapter{Numerical considerations}
\label{App:noneqcritp}
Before we can integrate Eq.~\ref{SheetWinde} to find a unique wind solution, we
have to specify some numerical parameters: (1) We have to describe the
geometrical shape of the fluxsheet by specifying $r_*$, $x(u,\eta)$, and
$\theta(u,\eta)$.  This defines $\AbsGradeta$ as well. (2) The next step is to
specify the self-similarity functions $Y(\eta)$ and $\rhoAc(\eta)$. These
functions have to be guessed for the first calculations. Then we can derive
them from previous calculations until we get a self-consistent combination of
self-similarity functions and wind solutions. (3) The star itself is described
by its luminosity $L_*$, its mass $M$, and its equatorial radius
$R_\mathrm{eq}$, its (equatorial) magnetic field $B_\mathrm{p0Eq}$, and its
equatorial rotation rate $\alpharoteq$. (4) The physics in the wind depends on
the CAK parameters $\alphacak$, $\kcak$ and on the temperature stratification
$T(u)$. All models presented in this thesis assume an isothermal wind. (5)
Finally some of theses quantities may depend on latitude. For all numerical
models presented here we ignore differential rotation
$\alpharot(\eta)=\alpharoteq$ and a lateral dependence of the magnetic field.
But the lateral dependence of the stellar radius can not be ignored. Some wind
models in this thesis have a rather high rotation rate, which will lead to an
oblate star. Without going deep into the theory of stellar structure we can
describe the stellar surface as an equipotential surface of
\begin{equation}
\Phi(r,\theta) = -\frac{GM}{r}-\frac{\Omega^2 r^2\sin^2\theta}{2}.
\end{equation}
The radius $R(\eta)$ for the base of the wind is given by the point where the
fluxsheets intersects the stellar surface. We can find this point by solving
\begin{equation}
\Phi(r_* x(u_0(\eta),\eta),\theta(u_0(\eta),\eta)) = 
\Phi\left(R_\mathrm{eq},\frac{\pi}{2}\right)
\end{equation}
for $u_0(\eta)$. Finally we have to specify eigenvalues as in the equatorial
case. We chose again the position of the Alfv\'enic point $\uAc$ and the
approximate mass loss rate (Eq.~\ref{defMdotapprox}). Using
\begin{eqnarray}
\BpAc &=& \Bpo\frac{\AbsGradeta_\mathrm{Ac}}{\AbsGradeta_0}
          \frac{x_0\sin\theta_0}{\xAc\sin\thAc}\\
\yAc  &=& \frac{1}{v_*}\frac{\BpAc}{\sqrt{4\pi\rhoAc}}
\end{eqnarray}
the wind velocity $\yAc$ and the density $\rhoAc$ at the Alfv\'enic point can
be found.

In the case of a cold wind $(\vs=0)$ we integrate Eq.~\ref{SheetWinde}
from Alfv\'enic critical point $\uAc$ to the base of the wind $u_0$ and
to infinity. We can obtain $\uAc$ by fitting a reasonable initial wind velocity
$v_0$ at the base of the wind.

In the case of a warm wind $(\vs>0)$ we integrate from the inner critical point
to the base of the wind and to the Alfv\'enic critical point and from the outer
critical point to the Alfv\'enic critical point and to infinity.  We  can
now fit the initial wind velocity at the base of the wind and a continuous wind
acceleration at the Alfv\'enic critical point. These two conditions allow to
find $\uAc$ and the mass loss rate. 

One important point has yet not been taken into account properly. The plot for
the transverse component of the magnetic force shows a singularity at the
Alfv\'enic point. This is of course not physical. The reason for this is the
uncompensated $(1-\Mp^2)^{-1}$ term in the formulas for the various
derivatives of $\yphi$ (Eqs.~\ref{auxe}--\ref{defdyphideta}). These derivatives
enter the wind equation (Eq.~\ref{SheetWinde}) and the equation for the
transverse forces (Eq.~\ref{defdat}). The wind equation has a singularity at
the Alfv\'enic point. But this is already known from the equatorial theory. The
coordinates of the Alfv\'enic point are chosen so that we get a solution with
finite $y$ and $\yphi$ at the Alfv\'enic point. But nevertheless this is a
problem which has to be fixed before really reliable results for winds in
bended fluxsheets can be calculated. We might derive an additional condition
for the wind solution at the Alfv\'enic point from the proper treatment of the
transverse Euler equation.

Another challenging point is the computation of the inner and outer critical
points.  The wind equation derived in this chapter is much more complicated
than the equatorial equations derived in the previous chapters.  We derive here
the equations for the CAK-type critical points in the non-equatorial wind using
the same argument as in the case of the equatorial wind. At the CAK-type
critical points we have
\begin{eqnarray}
\label{critequAa}
0 &=& A \dyxdx + B - C \dyxdx^\alphacak\\
\label{critequBa}
0 &=& A - \alpha C \dyxdx^{\alphacak-1}\\
\label{critequCa}
0 &=& \frac{d}{du}\left(A\dyxdx+B - C\dyxdx^\alphacak\right).
\end{eqnarray}
With Eq.~\ref{critequBa} $\ddot y$ can be eliminated from Eq.~\ref{critequCa}
leading to
\begin{equation}
\label{critequCb}
0 = \left[\dAdu + \dydu\dAdy\right] \dyxdx +
    \left[\dBdu + \dydu\dBdy\right] -
    \left[\dCdu + \dydu\dCdy\right] \dyxdx^\alphacak.
\end{equation}
Eqs.~\ref{critequAa}, \ref{critequBa} \& \ref{critequCb} form a set of three
nonlinear equations with three unknowns: $u$, $y$, and $\dydu$. We can
therefore expect at least locally unique solutions. Since these equations are
very complicated and essentially non-algebraic, it is not possible to find an
analytic solution. We use a numerical method based on Newton's algorithm. This
algorithm requires the Jacobian matrix of the terms on the right hand sides of
Eqs.~\ref{critequAa}, \ref{critequBa} \& \ref{critequCb} with respect to $u$,
$y$, and $\dydu$. This requires the second derivatives of $A$, $B$, and $C$.

Since Eq.~\ref{critequCb} is very complicated it seems reasonable to evaluate
the derivatives of $A$ and $B$ numerically. But we have to consider that the
Jacobian matrix, which contains the second derivatives of $A$, $B$, and $C$, is
even more complicated and requires certainly numerical evaluation. Tests have
shown that evaluating the first and second derivatives of $A$, $B$, and $C$
numerically causes severe numerical roundoff errors, which make it impossible
to find the critical points. The reason is that numerical derivatives contain
the difference of two nearly identical quantities. This leads to a loss of
precision. In our case the situation is worse due to the fact that the right
hand sides of Eqs.~\ref{critequAa}, \ref{critequBa} \& \ref{critequCb} vanish
at the critical point as well. I.e.\ they are differences of nearly identical
quantities close to the critical points. We control these problems by using
quadruple precision for all numerical calculations. Additional we use numerical
derivatives only for the Jacobian matrix. Nevertheless our code has some times
problems finding the outer critical point numerically. But it is not clear yet,
whether this is an numerical or a mathematical problem. See
Sect.~\ref{Sec:numnoneq} for more details. Finally it should be mentioned that
in general a forth condition for the inner and outer critical point can be
found. A `perfect' wind solution should fulfill the transverse Euler equation
everywhere. In this case we can demand that analogous to Eq.~\ref{critequCa}
the total derivative of the transverse Euler equation with respect to $u$
vanishes. Further details are discussed by Falcke \cite[Sect.~5.3]{Falcke:Msc}.
Our solutions do not obey the transverse Euler equation in general. Therefore
we can not use this condition for our critical points.

The analytical expressions for the first derivatives of $A$, $B$, and $C$ are
given now. The formulas are rather complicated and offer certainly some space
for simplifications.
\begin{eqnarray}
\dAdu &=& \left(\frac{\dAzweidu}{A_2}-\frac{\deprdu}{\epr}-
                  \frac{\ddudxdu}{\dudx}\right)A\\
\dAzweidu   &=& A_2\frac{\dchidu}{\chi}+\chi
                  \left[-\frac{\dysysdu}{y}+\varrho\left(\ddyphiduy-
                  \frac{\dyphibardu}{y}\right)+\dvarrhodu\left(\dyphidy-
                  \frac{\yphibar}{y}\right)\right]\\
\dAdy       &=& \frac{\hat{A}_2}{\epr Y \dudx}\\
            &=& \frac{\chi}{\epr Y \dudx}\left[Y^2+\frac{\ysys}{y^2}+\varrho
                  \left(\ddyphidyy-\frac{\dyphidy y-\yphibar}{y^2}\right)+
                  \right.\nonumber\\ 
            & & \left.\dvarrhody\left(\dyphidy-\frac{\yphibar}{y}\right)
                  \right]\\
\dBdu       &=& \dBdreidu-\dAdu(y\deprdu Y\dudx+(\dYdeta\epr+Y\deprdeta)y
                  \detadx)-\nonumber\\
            & & A(yY(\ddeprduu\dudx+\deprdu\ddudxdu+
                  \ddeprdetadu\detadx+\deprdeta\ddetadxdu)+
                  \dYdeta y(\deprdu\detadx+\epr\ddetadxdu))\\
\dBdreidu   &=&
   \dchidu\left[\ysys\mu+\dysysdu+\varrho
     \left(\mu\yphibar+\dyphidu\right)\right]+
     \chi\left[\mu\dysysdu+\dmudu\ysys+\ddysysduu+\right.\nonumber\\
&& \left.\dvarrhodu\left(\mu\yphibar+\dyphidu\right)+
     \varrho\left(\dmudu\yphibar+\mu\dyphibardu+\ddyphiduu\right)\right]+
     \dxidu\left[Y\dYdeta y^2 +\right.\nonumber\\ 
&& \ysys\left(\frac{\drhodeta}{\rho}-\frac{\dYdeta}{Y}\right)+
     \dysysdeta+\varrho\left(\dyphideta-\domegadeta x \sin\theta+
     \dyphidy\frac{\dYdeta}{Y} y+\right.\nonumber\\
&& \left.\left.\yphibar \left(\frac{\drhodeta}{\rho}-
     \frac{\dYdeta}{Y}-\frac{\dFdeta}{F}\right)\right)\right]+
     \xi\left[\dysysdu\left(\frac{\drhodeta}{\rho}-\frac{\dYdeta}{Y}\right)+   
     \ysys\dmudeta+\ddysysdetau+\right.\nonumber\\
&& \dvarrhodu\left(\dyphideta-
     \domegadeta x \sin\theta+\dyphidy\frac{\dYdeta}{Y}
     y +\yphibar \left(\frac{\drhodeta}{\rho}-
     \frac{\dYdeta}{Y}-\frac{\dFdeta}{F}\right)\right)+\varrho\left(
     \ddyphidetadu-\right.\nonumber\\
&& \left.\left.\domegadeta(\dxdu\sin\theta+x\dthdu\cos\theta)+
     \ddyphiduy\frac{\dYdeta}{Y} y+\dyphibardu\left(\frac{\drhodeta}{\rho}-
     \frac{\dYdeta}{Y}-\frac{\dFdeta}{F}\right)+\yphibar\dmudeta\right)
     \right]-\nonumber\\
&& \left(\frac{\ddxduu x-\dxdu^2}{x^2}+\frac{\ddthduu}{\tan\theta}-
     \frac{\dthdu^2}{\sin^2\theta}\right)\left[\varrho
     \left(x\sin\theta\omega-\yphibar\right)+\yphi^2\right]-\nonumber\\
&& \left(\frac{\dxdu}{x}+\frac{\dthdu}{\tan\theta}\right)
     \left[\dvarrhodu\left(x\sin\theta\omega-\yphibar\right)\right.\nonumber\\
&& +\left.\varrho\left(\omega\left(\dxdu\sin\theta+x\cos\theta\dthdu\right)-
     \dyphibardu\right)+2\yphi\dyphidu\right]+\nonumber\\
&& \frac{\ddxduu x-2\dxdu^2}{x^3}(1-\Gamma)\\
\dBdy &=& \dBdreidy-\dAdy(y\deprdu Y\dudx+(\dYdeta\epr+Y\deprdeta)y
            \detadx)-\nonumber\\
      & & A(\deprdu Y\dudx+(\dYdeta\epr+Y\deprdeta)\detadx)\\
\dBdreidy &=&
   \chi\left[\varrho\left(\mu\dyphidy+\ddyphiduy\right)+
     \dvarrhody\left(\mu\yphibar+\dyphidu\right)\right]+\nonumber\\
&& \xi\left[2Y\dYdeta y+\dvarrhody\left(\dyphideta-\domegadeta x\sin\theta+
     \dyphidy\frac{\dYdeta}{Y} y+\yphibar\left(\frac{\drhodeta}{\rho}-
     \frac{\dYdeta}{Y}-\frac{\dFdeta}{F}\right)\right)+\right.\nonumber\\
&& \left.\varrho\left(
     \ddyphidetay+\frac{\dYdeta}{Y}\left(\ddyphidyy y+\dyphidy\right)+\dyphidy
     \left(\frac{\drhodeta}{\rho}-\frac{\dYdeta}{Y}-\frac{\dFdeta}{F}\right)
     \right)\right]\nonumber\\
&& -\left(\frac{\dxdu}{x}+\frac{\dthdu}{\tan\theta}\right)
     \left[\dvarrhody\left(x\sin\theta\omega-\yphibar\right)-\varrho\dyphidy+
     2\yphi\dyphidy\right]\\
\dCdu &=& \left[\frac{\ddxduu}{\dxdu}+(\alphacak-2)\frac{\dxdu}{x}+\right.
            \nonumber\\
      & & \left.\alphacak\left(\frac{\dthdu}{\tan\theta}-
            \frac{\dvthdu}{\vth}
            -\frac{\dAbsGradetadu}{\AbsGradeta}\right)
            \right]C\\
\dCdy &=& \alphacak \frac{C}{y}.
\end{eqnarray}
For brevity we used the following auxiliary quantities
\begin{eqnarray}
\chi       &=& \dxdu\dudx+\dthdu\dudth\\
\dchidu    &=& \ddxduu\dudx+\dxdu\ddudxdu+\dudth\ddthduu+\ddudthdu\dthdu\\
\xi        &=& \dxdu\detadx+\dthdu\detadth\\
\dxidu     &=& \ddxduu\detadx+\dxdu\ddetadxdu+
               \ddthduu\detadth+\dthdu\ddetadthdu\\
\varrho    &=& \frac{\yphibar}{\Mp^2}\\
\dvarrhodu &=& \frac{\dyphibardu+\yphibar\mu}{\Mp^2}\\
\dvarrhody &=& \frac{\dyphidy-\yphibar/y}{\Mp^2}\\
\mu        &=& \frac{\drhodu}{\rho}\\
\dmudu     &=& \frac{(\ddAbsGradetaduu)\AbsGradeta-(\dAbsGradetadu)^2}
               {\AbsGradeta^2}-
               \frac{\ddthduu}{\tan\theta}+\frac{\dthdu^2}{\sin^2\theta}-
               \frac{\ddxduu x-\dxdu^2}{x^2}\\
\dmudeta   &=& \frac{(\ddAbsGradetadetau)\AbsGradeta-\dAbsGradetadu
               \dAbsGradetadeta}{\AbsGradeta^2}-
               \frac{\ddthdetau}{\tan\theta}+\frac{\dthdu\dthdeta}
               {\sin^2\theta}-\nonumber\\
           & & \frac{\ddxdetau x-\dxdu\dxdeta}{x^2}\\
\gamma     &=& \frac{\drhodeta}{\rho}.
\end{eqnarray}
\end{appendix}
\backmatter
%
%-*-LaTeX-*-
% This is the bibliography for the PhD-thesis of Henning Seemann.
% (c) 1997 by Henning Seemann
%
\addcontentsline{toc}{chapter}{\protect\numberline{ }Bibliography}

%
%-*-LaTeX-*-
% These are the acknowledgements for the PhD-thesis of Henning Seemann.
% (c) 1997-98 by Henning Seemann
%
\chapter{Danksagungen}
\addcontentsline{toc}{chapter}{\protect\numberline{ }Danksagungen}
\label{Chap:acknowledgements}
Zu aller erst m\"ochte ich meinem Doktorvater Prof.\ Dr.\ P. L. Biermann f\"ur
das Thema dieser Arbeit und die unerm\"udliche Unterst\"utzung danken , ohne
die diese Doktorarbeit niemals zu Stande gebracht h\"atte. Er gew\"ahrte mir
die gro{\ss}e Freiheit, mich auf dem Gebiet der magnetischen Sternwinde nach
eigenem Willen auszutoben, ohne da{\ss} er das Interesse an mir und meiner
Arbeit verlor und mich mit ihr alleine gelassen h\"atte. Sein steter Rat war
f\"ur mich immer ein Ansporn und ein wichtiger Wegweiser durch das manchmal
doch recht un\"ubersichtliche Terrain der Astrophysik.

Mein Dank gilt auch Prof.\ Dr.\ H. J. Fahr f\"ur sein T\"atigkeit als
Gutachter f\"ur diese Dissertation.

Dem Max-Planck-Institut f\"ur Radioastronomie danke ich f\"ur den mir zur
Verf\"ugung gestellten Arbeitsplatz samt der komfortablen EDV-Ausstattung, die
einem Theoretiker wie mir das Leben so viel einfacher macht. Hier gilt mein
besonderer Dank H. G. Girnstein und Dr.\ P. M\"uller. Au{\ss}erdem danke ich
dem Institut f\"ur die langfristige finanzielle Unterst\"utzung, die mir die
Freiheit gegeben hat, meinem Forscherdrang ungehindert nachzugehen. Dies gilt
auch f\"ur die gro{\ss}z\"ugige Unterst\"utzung durch die Studienstiftung des
deutschen Volkes.

Ganz wichtig f\"ur das Gelingen dieser Doktorarbeit war auch die wunderbare
Arbeitsatmosph\"are, die ich meinen Arbeitskollegen Alina \& F\u anel Donea,
Torsten En{\ss}lin, Dr.\ Heino Falcke, Dr.\ Wolfram Kr\"ulls, Dr.\ Chun-yu Ma,
Dr.\ Biman Nath, Dr.\ Martina Niemeyer, Giovanna Pugliese, Dr.\ J\"org Rachen,
Iliya Roussev, Yiping Wang und Christian Zier verdanke. Sie haben daf\"ur
gesorgt, da{\ss} die letzten vier Jahre eine wundersch\"one Zeit waren, an die
ich immer gerne erinnern werde. Speziell danken m\"ochte ich Dr.\ Heino Falcke
f\"ur die Einf\"uhrung in die Geheimnisse der Unix-Welt, Dr.\ Wolfram Kr\"ulls
f\"ur die Idee, Wellen in magnetischen Sternwinden zu untersuchen, und Torsten
En{\ss}lin f\"ur die gewissenhafte Durchsicht dieser Arbeit. Mein Dank gilt
auch meinen Freunden au{\ss}erhalb der Astronomie, die mich immer wieder daran
erinnerten, da{\ss} es auch ein Leben au{\ss}erhalb der Astronomie gibt.

Ganz wichtig war f\"ur mich nat\"urlich auch die Unterst\"utzung durch meine
Eltern und meine ganze Familie, die mir erst meinen Lebensweg und damit auch
diese Doktorarbeit erm\"oglicht haben. Sie haben nie den Glauben an mich
verloren haben und immer unendliche Geduld mit mir bewiesen. 

%
%-*-LaTeX-*-
% These is the curriculum vitae of Henning Seemann.
% (c) 1997-98 by Henning Seemann
%
\chapter*{Lebenslauf}
\addcontentsline{toc}{chapter}{\protect\numberline{ }Lebenslauf}
\begin{tabbing}
xxxxxxxxxxxxxxxxxxxxxxxx\=xxxxxxxxxxxxxxxxx\= \kill
\\
\textbf{Name}               \>               \> Henning Seemann \\
\\
\textbf{Geburtsdatum/-ort}  \>               \> 28.01.1969 in L\"uneburg\\
\\
\textbf{Nationalit\"at}      \>               \> deutsch\\
\\
\textbf{Familienstand}      \>               \> ledig, kinderlos\\
\\
\textbf{Schulbildung}       \> 08/75 - 01/77 \> Grundschule in L\"uneburg \\
                            \> 02/77 - 07/79 \> Grundschule in Gelnhausen\\
                            \> 08/79 - 06/81 \> Gymnasium in Gelnhausen\\
                            \> 07/81 - 07/85 \> Gymnasium in Dortmund\\
                            \> 08/85 - 06/88 \> Gymnasium in Braunschweig\\
                            \>               \> Abschlu{\ss} mit Abitur\\
\\
\textbf{Wehrdienst}         \> 07/88 - 09/89 \> \\
\\
\textbf{Studium}            \> 10/89 - 08/92 \> Studium der Physik, Universit\"at Bonn\\
                            \> 10/91         \> Vordiplom\\
                            \> 09/92 - 05/93 \> Hauptstudium der Physik, Graduate School,\\
                            \>               \> University of Wisconsin, Madison, USA\\
                            \> 10/93 - 03/94 \> Hauptstudium der Physik, Universit\"at Bonn und\\
                            \>               \> Suche nach einer Doktorarbeit\\
\textbf{Studienabschlu{\ss}}   \> 05/93         \> Master of Science, Physics (Madison)\\
\\
\textbf{Promotion}          \> 04/94 - 03/98 \> Doktorarbeit am Max-Planck-Institut\\
                            \>               \> f\"ur Radioastronomie, Bonn\\
\\
\end{tabbing}

\end{document}